\documentclass[smallextended,runningheads,natbib]{svjour3} 
\RequirePackage{fix-cm}
\smartqed
\usepackage[colorinlistoftodos]{todonotes}

\usepackage{float}
\usepackage{graphicx}
\usepackage[multidot]{grffile}
\usepackage{subcaption}
\usepackage[final]{pdfpages}
\usepackage{amssymb,amsmath}
\usepackage{mathtools} %for dcases
\usepackage{listings} %for pretty print code
\usepackage{enumitem} %for specifying label style for enumerate
\usepackage[unicode=true,
  linktocpage,
  linkbordercolor={0.5 0.5 1},
  citebordercolor={0.5 1 0.5},
  colorlinks=true,
  linkcolor=blue]{hyperref} %must be the last one

\renewcommand{\cite}{\citep}

\DeclareMathOperator{\Tr}{Tr}

\newcommand{\pd}[2]{\frac{\partial #1}{\partial #2}}

\newcommand{\lb}{\left(}
\newcommand{\rb}{\right)}

\newcommand{\specialcell}[2][c]{%
  \begin{tabular}[#1]{@{}l@{}}#2\end{tabular}}
  
\newfloat{Program}{htbp}{lop} %for including code, usage: \begin{Program}...\end{Program}
\usepackage[left=1.0in, right=1.0in, top=1.0in, bottom=1.0in,
  includehead,includefoot]{geometry}
\journalname{Journal of Mathematical Biology}

\begin{document}

\title{Spots, strips, and spiral waves in models for static and motile cells}
\subtitle{GTPase patterns in cells}
\author{Yue Liu \and Elisabeth G. Rens \and Leah Edelstein-Keshet}
\institute{Y.\ Liu \at
          Department of Mathematics, University of British Columbia, Vancouver
          V6T~1Z2, BC, Canada\\
          \email{liuyue@math.ubc.ca}\\
             \emph{Present address:} Mathematical Institute, University of Oxford, Oxford, OX2 6GG, UK  %  if needed
          \and
          E. G. Rens \at
          Department of Mathematics, University of British Columbia, Vancouver
          V6T~1Z2, BC, Canada\\
          \email{rens@math.ubc.ca}           %  \\
          \and
          L.\ Edelstein-Keshet \at
          Department of Mathematics, University of British Columbia, Vancouver
          V6T~1Z2, BC, Canada\\
          \email{keshet@math.ubc.ca}           %  \\
}
\date{Received: date / Accepted: date}% The correct dates will be entered by the editor

\maketitle

\begin{abstract}
The polarization and motility of eukaryotic cells depends on assembly and contraction of the actin cytoskeleton and its regulation by proteins called GTPases.
The activity of GTPases causes assembly of filamentous actin (by GTPases Cdc42, Rac), resulting in protrusion of the cell edge.
%and activity of myosin motors (by Rho) which contracts the actin network.  %
Mathematical models for GTPase dynamics address the spontaneous formation of patterns and nonuniform spatial distributions of such proteins in the cell.
Here we revisit the wave-pinning model for GTPase-induced cell polarization, together with a number of extensions proposed in the literature. These include introduction of sources and sinks of active and inactive GTPase (by the group of A. Champneys), and negative feedback from F-actin to GTPase activity. We discuss these extensions singly and in combination, in 1D, and 2D static domains. We then show how the patterns that form (spots, waves, and spirals) interact with cell boundaries to create a variety of interesting and dynamic cell shapes and motion. 
\end{abstract}

\vspace{.2in}

\section{Introduction}

The dynamics of the actin cytoskeleton determines internal cell structure, cell shape, and cell motility. By accumulating at a cell edge, filamentous actin (F-actin) produces outwards protrusion. 
Actin assembly is regulated by signaling networks. Central in those networks are the small GTPases, Rac, Cdc42, and Rho. Rac promotes assembly of F-actin, whereas Rho activates myosin motors.
The interactions of Rac, Rho, Cdc42, and other molecular players has been modeled in previous work \cite{mori2008wave,Champneys,holmes2012regimes,holmes2016,zmurchok2018coupling,walther2012deterministic,edelstein2013simple,jilkine2011comparison,otsuji2007mass} both in 1D and 2D. These studies made different modelling decisions and ranged from simple \cite{mori2008wave} to detailed \cite{maree2008quantitative}.  It is challenging to determine parameter sensitivity and map out regimes of behavior of the more detailed models. This motivates studying minimal models that showcase the possible realms of predicted behavior.

It was shown previously that the biology of GTPases permits a single member of this family to spontaneously polarize (i.e. form spatial regions of high vs low activity). This idea was the basis of the wave-pinning model \cite{mori2008wave,mori2011asymptotic}, and depends on the large difference in diffusion of the active (slow) and inactive(fast) forms of a GTPase.

 Several models have been examined mathematically to describe how a single GTPase coupled to other effectors or influences could results in spatio-temporal patterns. These include a GTPase with sources and sinks \cite{Champneys}, with feedback from F-actin \cite{holmes2012regimes,mata2013model}, with mechanical tension \cite{zmurchok2018coupling} and with effects of changing cell size \cite{buttenschon2019cell}. Many of these were explored in reaction-diffusion (RD) equations within a 1D static single cell domain or with spatially uniform distribution in each of many cells \cite{zmurchok2018coupling}.
Some of the behaviors found in such models include, traveling waves, pulses, or oscillating fronts \cite{holmes2012regimes,mata2013model}, 
%Hopf bifurcations leading to cycles of high and low GTPase activity \cite{zmurchok2018coupling}, 
or localized peaks and ``solitons'' \cite{Champneys}.

 Here we have two main purposes: (1) to explore what happens when two distinct minimal models are coupled, and whether this leads to new behavior, (2) to study these systems in 2D domains to determine whether they produce spots or stripes, and (3) to simulate the same models on a deforming 2D domain depicting the shape and motility of a cell.

Biological motivation for this work comes from several sources. (A) Waves of actin are observed in a number of experimental systems \cite{inagaki2017}. In some of these, such waves are seen to cause cell edge to cyclically protrude outwards (as the waves impinge on the cell edges). We wondered whether a model for Rac interacting with F-actin could mimic this kind of behavior. (B) The GTPase model generalized by the group of Alan Champneys in \cite{Champneys} converts the polarizing cell behaviour into multiple coexisting peaks. We wondered how such peaks would interact with cell boundaries, and, in particular, whether they would be associated with smaller protrusions such as filopodia. (C) In some cells, notably the embryos of C. elegans, localized Rho-associated actin clusters are seen to ``blink'' (oscillate temporally while maintaining a fixed location) \cite{robin2016excitable}. We asked whether the combined F-actin-Rho model with localized sources could account for such behavior. 

We first briefly review the three classes of minimal models, show results for the combined model, and then demonstrate the novel 2D behaviors that are observed once these models are simulated in the deforming 2D cell.

\begin{figure}[H]
    \centering
    \includegraphics[scale=0.4]{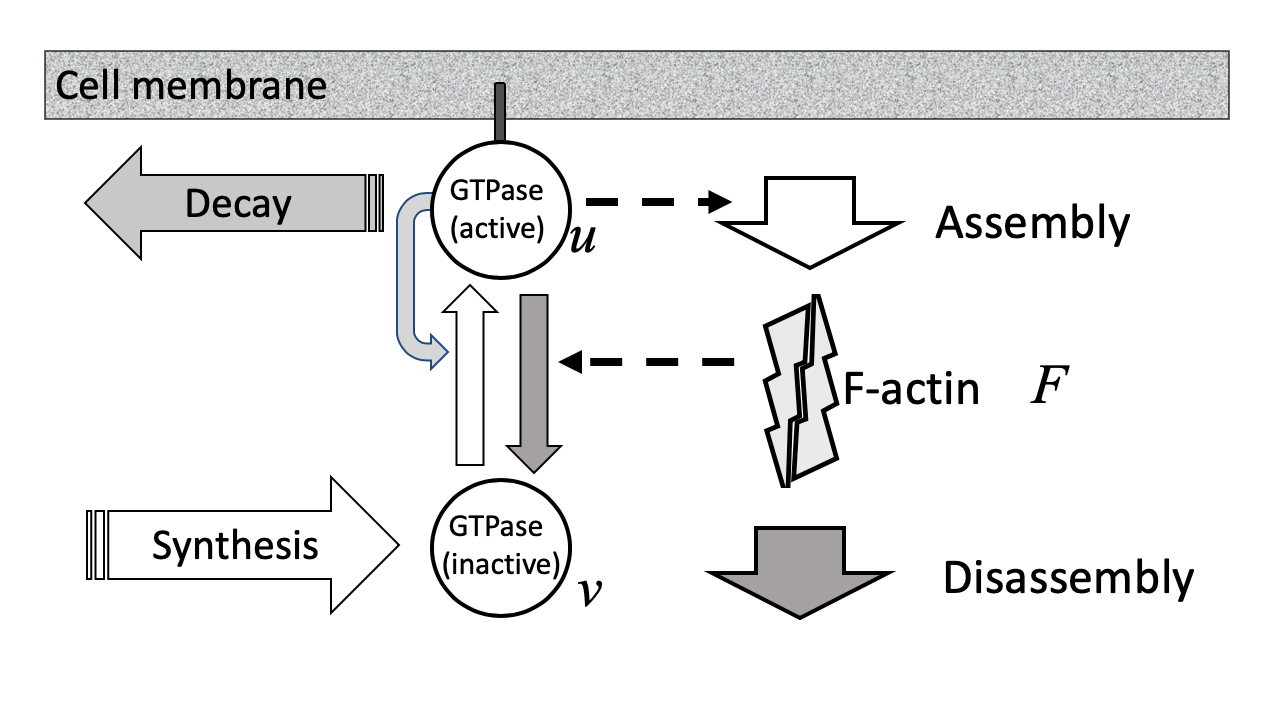}
    \caption{Schematic diagram of the models. The original wave-pinning model consists of GTPase (circles) in the active, (membrane-bound) form, $u$ and inactive form $v$, with positive feedback (curved grey arrow) from $u$ to its own activation (upwards white arrow). The F-actin extension model \cite{holmes2012regimes} includes GTPase activation of F-actin assembly and GTPase inactivation by F-actin (dashed arrow). The source-sink (nonconservative) extension by \cite{Champneys} includes removal of active GTPase and synthesis of inactive GTPase so that the total amount is no longer conserved.}
    \label{fig:model_schematic}
\end{figure}

\section{The models}

Our model is a system of reaction-diffusion partial differential equations (PDEs) based on the wave pinning model first proposed by \cite{mori2008wave}. The model is extended with a source and sink terms following \cite{Champneys}, and feedback from actin, proposed by \cite{holmes2012regimes}. 

\subsection{Model equations}
The dimensionless form of the model combining both extensions can be written as:
\begin{subequations}
\begin{align}
\pd{u}{t} &= \delta \nabla^2 u +f(u,v,F) - c\theta u ,\\
\pd{v}{t} &=  \nabla^2 v -f(u,v,F) +c \alpha ,\\
\pd{F}{t} &= \epsilon (k_n u - k_s F) ,\\
f(u,v,F) &=A(u) v - \lb \eta + s \frac{F}{1+F} \rb u , \quad A(u)= k + \gamma \frac{u^n}{1 + u^n},  \\
\pd{u}{\vec{n}} \bigg|_{\partial \Omega} &= 0, \quad \pd{v}{\vec{n}} \bigg|_{\partial \Omega} = 0,
\quad x \in \Omega, \quad t \geq 0. \notag 
\end{align}\label{sys:combined}
\end{subequations}
Here $u(x,t)$ and $v(x,t)$ represent the active and inactive GTPase, respectively. $F(x,t)$ represents filamentous actin (F-actin). $\delta \ll 1$ is the diffusion coefficient for the active form, which is slow due to attachment to the membrane. The reaction function $f(u,v,F)$ describes the net rate of GTPase activation, with $A(u)$ representing activation rate. Parameters $k, \gamma, \eta, s$ are the basal activation rate, self-feedback activation, basal inactivation and actin-feedback inactivation rates, respectively. Neumann boundary conditions are used to represent the fact that GTPases and F-actin do not leak out of the cell edges.

%The $c$ and $s$ parameters act as switches to turn the two extensions on and off: $c$ for the source and sink terms, and $s$ for feedback from actin. 
Setting $c=s=0$ reduces the system to the original wave pinning (WP) model \cite{mori2008wave}, which conserves the total $u+v$ inside the domain. The model has been analyzed in detail elsewhere \cite{mori2011asymptotic}, but we briefly mention its key property: under specific parameter settings, the WP model sustains waves that decelerate and stall in the domain, leading to a stable spatially heterogeneous steady state distribution of $u$ (the ``pinned wave'').

When $c=1, s=0$, the system corresponds to the non-conservative (NC) model of \cite{Champneys}. When $c=0, s>0$, we have the actin feedback (AF) model of \cite{holmes2012regimes}. While each of the above models has been studied previously, here we will also be concerned with their union, i.e. the so-called ``combined model'' (CM) with $c=1, s>0$. 
%We refer the system in this case as the combined model. 
The four models of interest are then (I) WP, (II) NC, (III) AF, and (IV) CM. These four models all have very distinct characteristic behaviors.
We will consider these models in several settings (A) a 1D spatial domain, as previously described in the literature, (B) a static 2D spatial domain where we can distinguish between spots and stripes, and finally (C) a deforming domain whose boundary dynamics is coupled to the evolving solution $u$ (or $F$) of the PDE. 

\subsection{Geometry}

In many previous papers, simulations were restricted to 1D \cite{holmes2012regimes,mata2013model}, but a variety of actin wave models exist in more detailed geometries, including 2D \cite{doubrovinski2011} and 3D \cite{bretschneider2009}. Here, for for simplicity in the 2D static domain case we consider a unit square. For the deforming domain, we use the Cellular Potts Model (CPM) to simulate a dynamic 2D cell. The methods and results are introduced in Sec.~\ref{sec:cpm}.

For the ease of analysis and identification of distinct patterns, we first discuss and examine results in a 1D spatial version of the models. We can interpret this 1D geometry in one of two classic ways: (1) as a cross-section along the diameter of a cell. This cross-section neglects any variation in the cell thickness and includes both the intracellular volume (cytosol) and the top and bottom membranes at every point. Neumann (no flux) boundary conditions are used for the endpoints of the interval. (2) Alternatively, another common assumption is a 1D cell perimeter. In this case, the region considered is close to the cell membrane, with periodic boundary conditions. 
%(2) Is somewhat less appropriate for the WP model, where the variable $v$ is resident in the cytosol. (The model can be adjusted to correct for this fact.) Hence, 
Here we adhere to the first approach.
%
%Since these two approaches, illustrated in Fig.~\ref{fig:1ddomain}, yielded similar patterns in numerical simulations, we will adhere to the first approach only. 
The case of 1D dynamic cell size is considered in \cite{andreassinglecell}.

%\begin{figure}[H]
 %   \centering
 %   \includegraphics[width=0.8\textwidth]{figures/1dslice_new.png}
%
%
 %   \caption{Replace with better figure. Possibly group with Fig 1}
 %   \label{fig:1ddomain}
%\end{figure}

\section{Methods of Analysis}

We briefly describe methods used to analyse the models. We use local perturbation analysis (LPA) to study the bifurcation behavior of each model, and compare with results from Turing Linear stability analysis.
A full description of these methods is found in the MSc thesis by one of us (YL) \cite{liu2019}.

\subsection{Local perturbation analysis}

Local perturbation analysis is a method for examining the evolution of a localized perturbation to a homogeneous steady state (HSS) for a fast-slow diffusion-reaction system. It provides a way to systematically detect certain forms of nonlinear instabilities that are not detectable by the more traditional Turing analysis. LPA was first developed by AFM Mar\'ee and V Grieneisen (Cardiff University)  \cite{grieneisen_LPA}, and has been used in \cite{edelstein2013simple,holmes2016,holmes2012regimes,mata2013model} and elsewhere to analyze wave pinning and related models. 

The basic idea of LPA is to take the limit where the slow diffusion coefficients goes to $0$ and the fast diffusion coefficients goes to infinity. We then consider an initial condition where the system is at HSS with a localized perturbation in the form of a spike of infinitesimal width but finite height. The behavior of the PDE can then be captured with an ODE system with ``global variables" representing the levels of the PDE variables away from the spike, and ``local variables" for the slow PDE variables at the spike. For example, using subscript $L$ to denote local variables, the LPA system for our combined model \eqref{sys:combined} is:
\begin{subequations}
\begin{align}
\pd{u}{t} &= f(u,v,F) - c\theta u ,\\
\pd{v}{t} &=  -f(u,v,F) +c \alpha ,\\
\pd{F}{t} &= \epsilon (k_n u - k_s F),\\
\pd{u_L}{t} &= f(u_L,v,F_L) - c\theta u_L ,\\
\pd{F_L}{t} &= \epsilon (k_n u_L - k_s F_L) .
\end{align}\label{sys:combined_LPA}
\end{subequations}
In the cases where $c=0$ or $s=0$, we will use mass conservation to remove irrelevant equations and eliminate degeneracy.
This allow us to easily produce bifurcation diagrams using AUTO \cite{auto} and delineate parameter regimes.
Notice that the LPA system \eqref{sys:combined_LPA} contains the well-mixed system (i.e. the system without local variables), so any features (branches and bifurcations) of the well-mixed system will also be present in the LPA system. Hence we can obtain any information that can be gained by analyzing the well-mixed system through LPA.

\subsection{Bifurcation analysis}

We refer to branches of equilibria and periodic solutions in the LPA system  that are also present in the well-mixed model as ``global" branches, as they correspond to solutions in which the local variables are equal to the global variables and the spike disappears, i.e. a homogeneous solution. The others branches are referred to as ``local" branches; they correspond to some kind of pattern.

We classify the parameter regimes into three categories: (a) stable, where only global branches are stable. In this regime, no pattern can arise from localized perturbation; (b) polarizable, where stable global and local branches coexist. In this regime, patterns can form only if the perturbation is sufficiently strong. Finally, (c) unstable, where all global branches are unstable. In this regime even infinitesimal perturbations can lead to pattern formation. In Appendix~\ref{apd:turing_vs_lpa}, we show that this is equivalent to the classical Turing regime.

The sets of parameters for each model are listed in Table~\ref{tab:parameters}. For the cases where the total amount of GTPase is conserved, we define the total mass of GTPase in the cell,
\[w = \int_\Omega (u+v) dx ,\]
as an additional constant parameter. This allow us to eliminate $v$ from the equations by writing it in terms of $u$ and $w$.

All bifurcation diagrams follow AUTO's conventions. On one-parameter diagrams, red/black curves indicate positions of stable/unstable equilibria respectively, while green/blue indicate the range of stable/unstable limit cycles. On two-parameter diagrams, red/light blue/dark blue curves trace the position of limit points (fold points)/branch points (transcritical points)/Hopf points, respectively.

\section{Results}

We next apply the methods to compare the behaviors of the four models of interest.

\subsection{Wave pinning (WP) model}

\begin{figure}
    \centering
    \vspace{-0.3cm}
    \begin{subfigure}[h]{0.5\textwidth}
        \centering
        \caption{WM, $w=2$}
        \includegraphics[width=\textwidth]{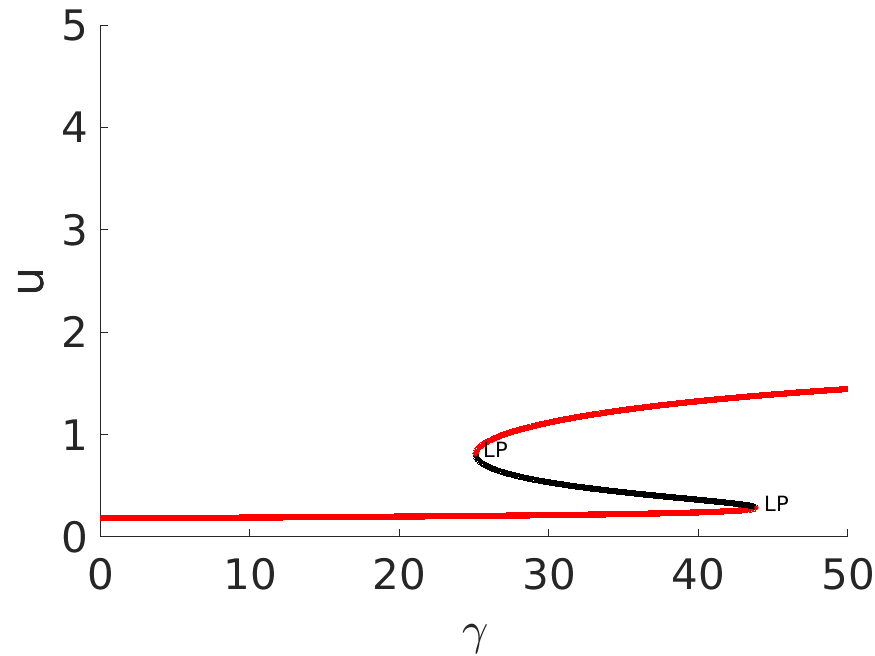}
    \end{subfigure}%
    \begin{subfigure}[h]{0.5\textwidth}
        \centering
        \caption{LPA, $w=2$}
        \includegraphics[width=\textwidth]{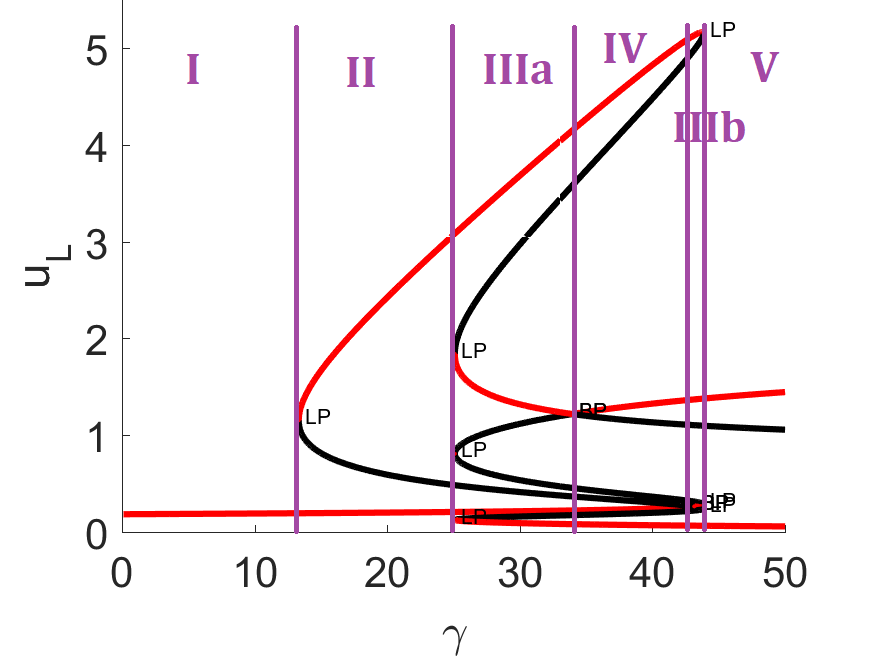}
    \end{subfigure}
    \begin{subfigure}[h]{0.5\textwidth}
        \centering
        \caption{WM, $w=3.5$}
        \includegraphics[width=\textwidth]{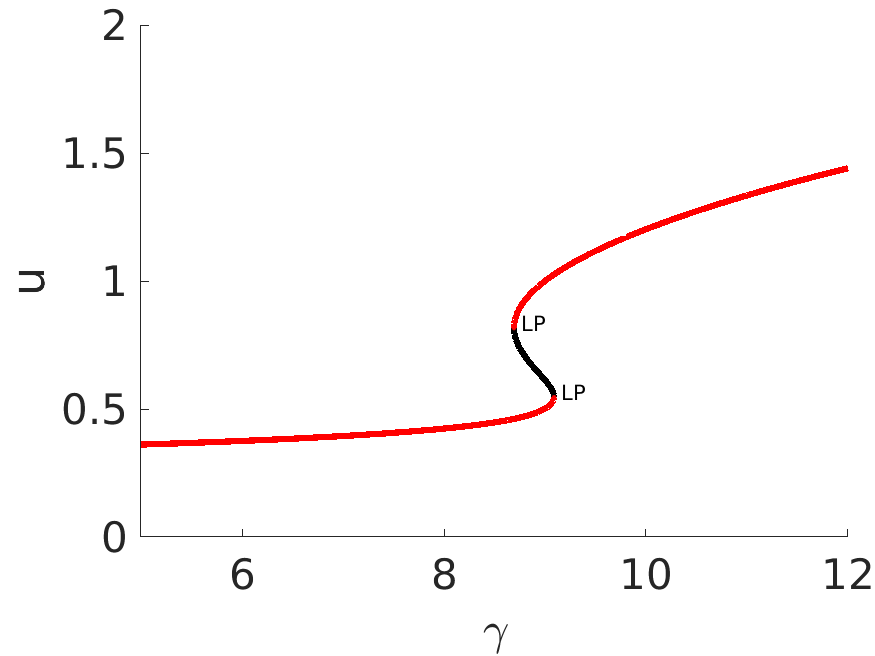}
    \end{subfigure}%
    \begin{subfigure}[h]{0.5\textwidth}
        \centering
        \caption{LPA, $w=3.5$}
        \includegraphics[width=\textwidth]{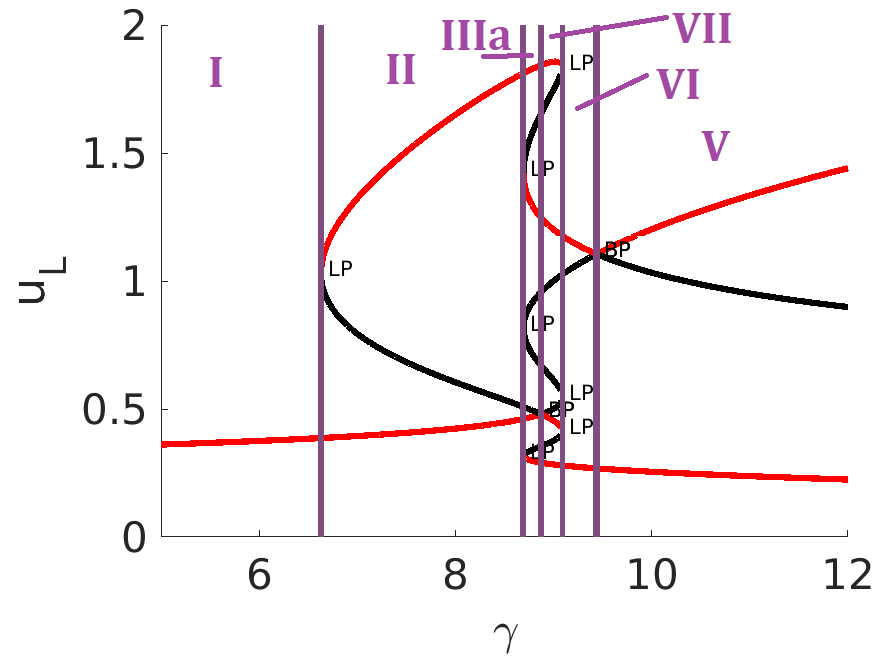}
    \end{subfigure}
    \begin{subfigure}[h]{0.5\textwidth}
        \centering
        \caption{WM, $w=4$}
        \includegraphics[width=\textwidth]{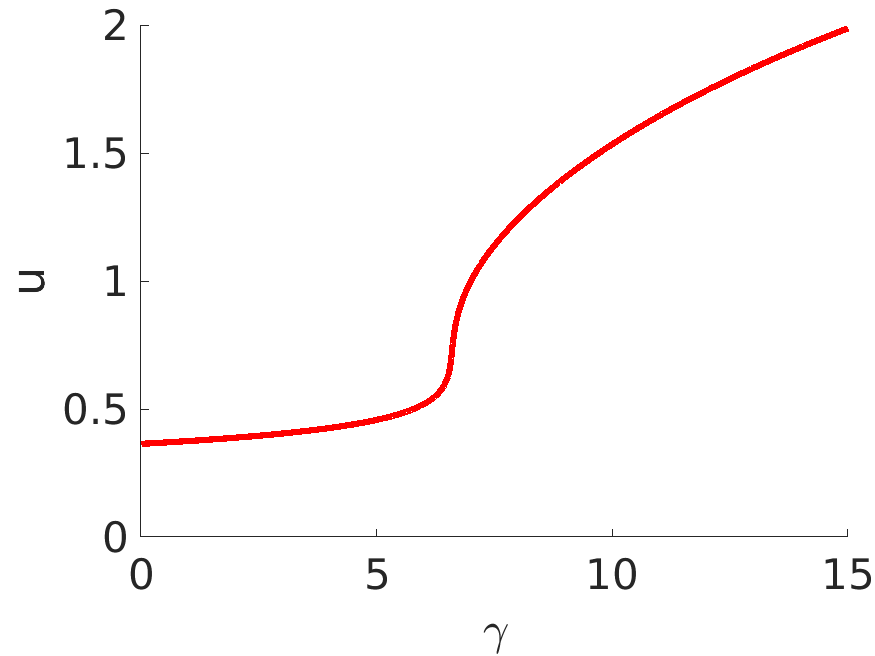}
    \end{subfigure}%
    \begin{subfigure}[h]{0.5\textwidth}
        \centering
        \caption{LPA, $w=4$}
        \includegraphics[width=\textwidth]{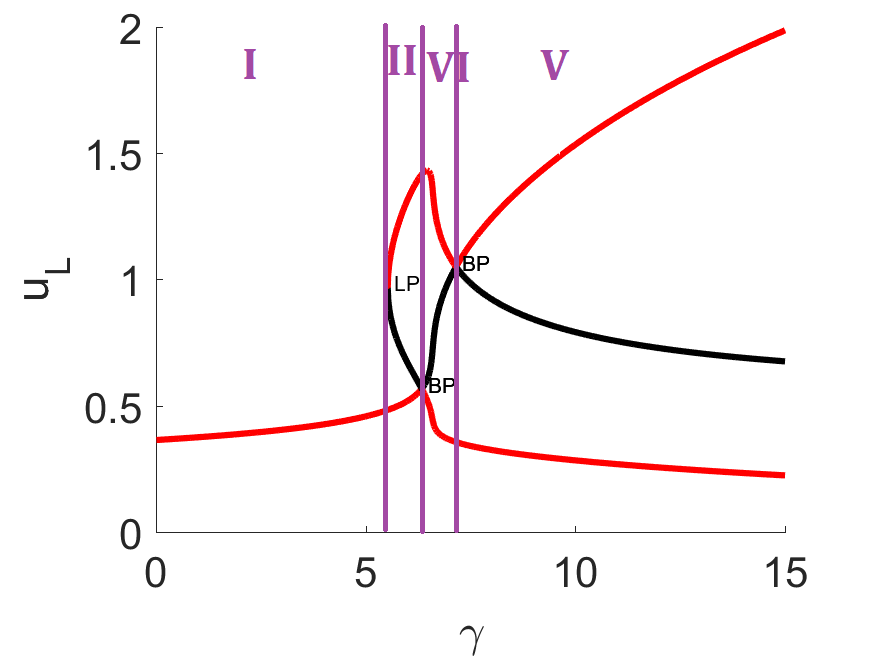}
    \end{subfigure}
    \vspace{-0.1cm}
    \caption[One-parameter bifurcation diagrams of the LPA and well-mixed wave pinning system]{Bifurcation diagrams of the  well-mixed (WM) and LPA wave pinning system %\eqref{sys:wavepin_lpa} 
    with respect to the rate of activation parameter $\gamma$. Other parameters as in Table~\ref{tab:parameters}(WP) except $w$. The purple lines are located at bifurcation points separating the distinct regimes. Note that the ``global branches'' (curves in the WM diagrams) also appear in LPA, though their stability can be different in LPA over certain intervals.
    %, as discussed in Sec.~\ref{sec:lpa_intro}. 
    }
    \label{fig:lpa_wavepin}
\end{figure}

\begin{figure}
    \centering
    \begin{subfigure}[h]{0.5\textwidth}
        \centering
        \caption{WM}
        \includegraphics[width=\textwidth]{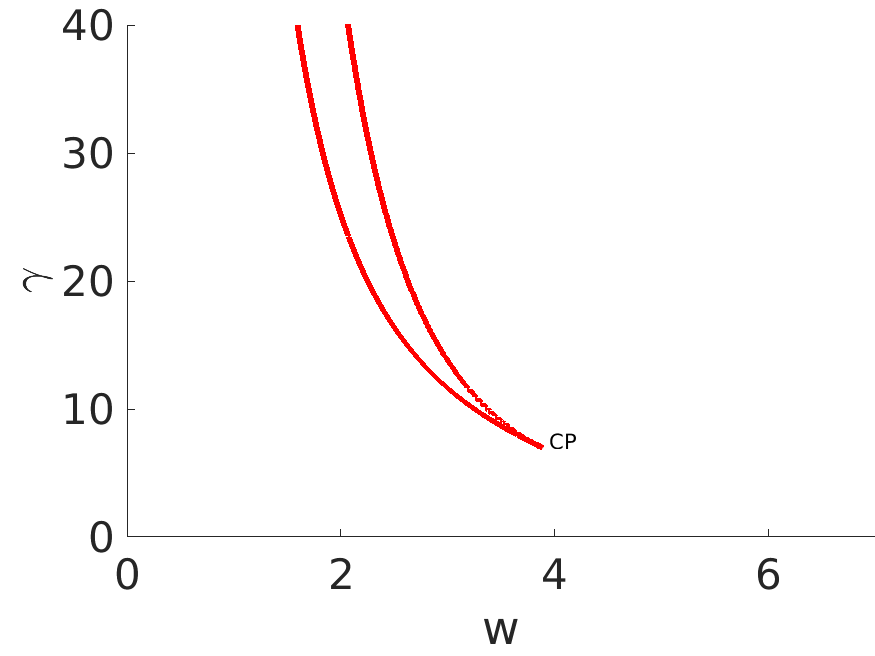}
    \end{subfigure}%
    \begin{subfigure}[h]{0.5\textwidth}
        \centering
        \caption{LPA}
        \includegraphics[width=\textwidth]{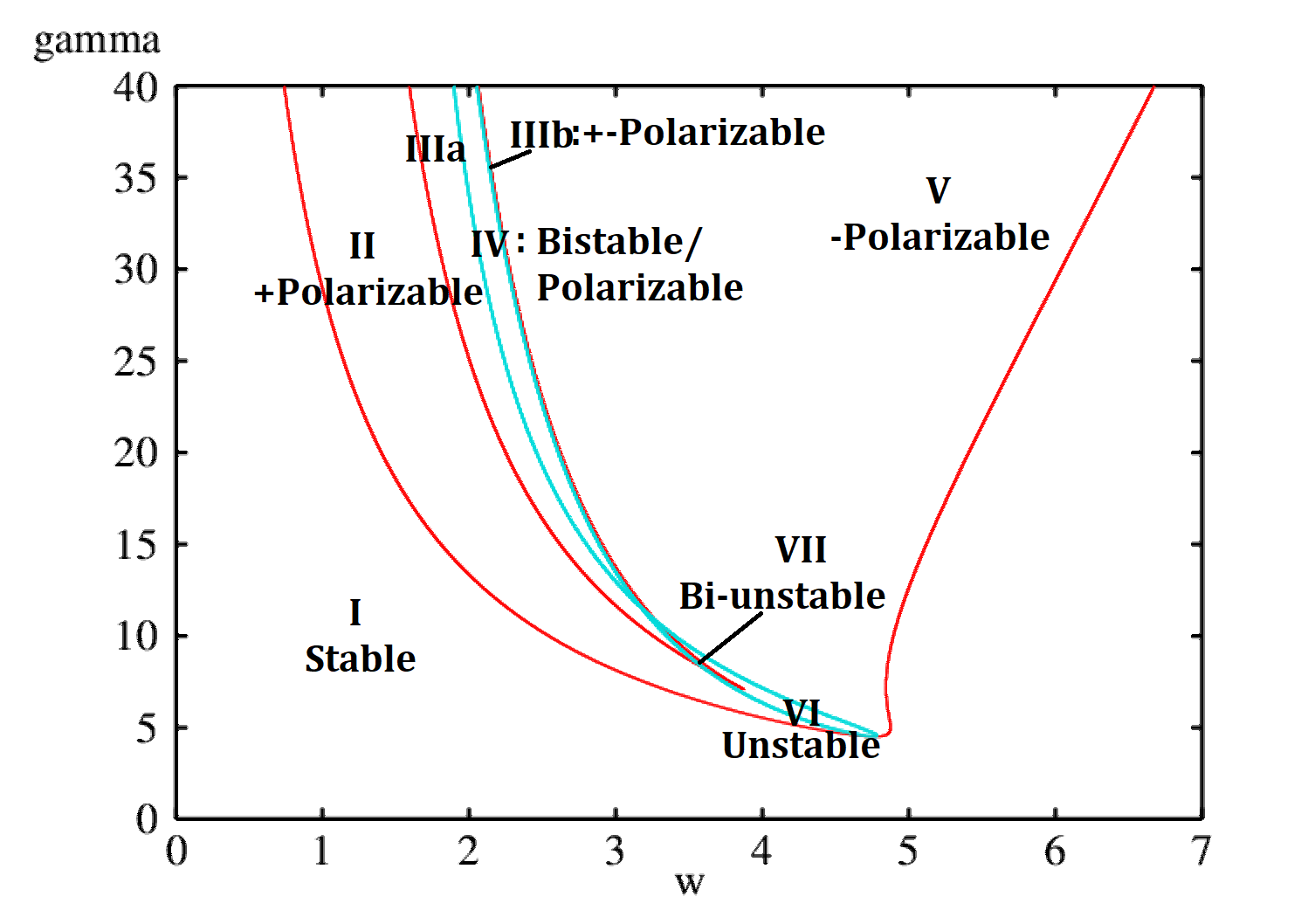}
    \end{subfigure}
    \begin{subfigure}[h]{0.8\textwidth}
        \centering
        \caption{LPA}
        \includegraphics[width=\textwidth]{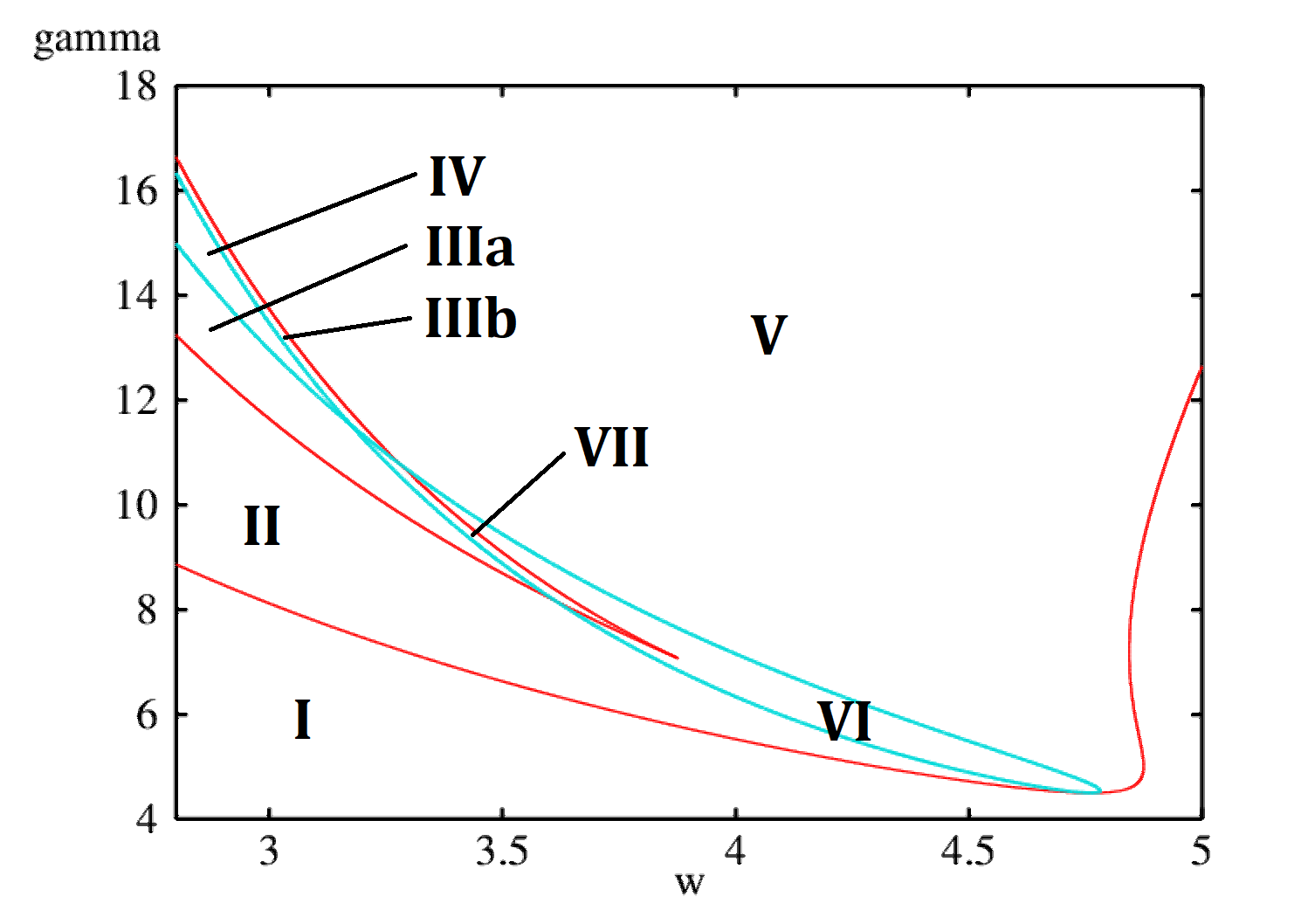}
    \end{subfigure}
    \caption[Two-parameter bifurcation plots of the LPA and well-mixed wave pinning system]{Two-parameter bifurcation plots of the wave pinning (WP) model with respect to  parameters $w, \gamma$. (a) Well-Mixed (WM) and (b,c) LPA system. Other  parameters as in Table~\ref{tab:parameters}(WP). Each curve in these diagrams traces the location of a bifurcation point shown in Fig.~\ref{fig:lpa_wavepin}, and forms the boundary of a parameter regime. The one-parameter bifurcation diagrams in Fig.~\ref{fig:lpa_wavepin} correspond to vertical cross-sections of the diagrams here. The LPA regimes I - VII match with the regimes in Fig.~\ref{fig:lpa_wavepin}(b,d,f). See summary in Table~\ref{tab:lpa_wavepin_regimes}. (c) A zoom into the cusps in (b). (Compare (b) to LPA Fig. 3(a) of \cite{holmes2016} for the same model with different parameter values: our figures agree on the (red) fold curves but ours includes an additional transcritical curve (light blue) separating several distinct regimes.)} %The dashed green lines shows the cross-sections in the parameter plane represented by (a,b).
    \label{fig:lpa_wavepin_2par}
\end{figure}

Based on extensive previous analysis \cite{mori2011asymptotic,mori2008wave,holmes2016} we highlight the results in Fig.~\ref{fig:lpa_wavepin} and Fig.~\ref{fig:lpa_wavepin_2par} merely for comparison with the extended model variants. 
Distinct regimes are summarized in Table~\ref{tab:lpa_wavepin_regimes}.
We identify $\gamma$ (the magnitude of the only nonlinear term) and $w$ (total concentration) as primary parameters of interest. Extending the earlier study \cite{holmes2016}, we also trace a branch of transcritical bifurcation in two-parameter continuation in Fig.~\ref{fig:lpa_wavepin_2par}. This allows us to identify several new regimes. We verify that LPA predictions in each regime are indeed correct with simulations of the full PDEs.

\begin{table}[h]
    \centering
    \begin{tabular}{l||l|l}
         Regime  & Classification & Description  \\
         \hline
         \hline
         I &  Stable & \specialcell{One stable GB, no LB} \\
         \hline
         II &  Polarizable & \specialcell{One stable GB,  one stable LB\\ located above the GB} \\
         \hline
         III & Polarizable & \specialcell{One stable GB,  three stable LBs \\located on both sides of the GB} \\
         \hline
         IV & Polarizable & \specialcell{Two stable GBs, three stable LBs: one above both\\ GBs, one in between, and one below both GBs} \\
         \hline
         V & Polarizable & \specialcell{One stable GB, one stable LB\\ located below the GB} \\
         \hline
         VI & Unstable & \specialcell{The only GB is unstable, two stable LBs\\ located on both sides of the GB} \\
         \hline
         VII & Unstable & \specialcell{Three GBs, all unstable, four stable LBs \\ located on both sides of the GB} \\
         \hline
    \end{tabular}
    \caption[Summary of the wave pinning model regimes]{Summary of the wave pinning (WP) regimes identified in Fig.~\ref{fig:lpa_wavepin} and \ref{fig:lpa_wavepin_2par}. GB: global branch; LB: local branch. Stable: all stable branches are global branches. Polarizable: there exist both stable global and local branches. Unstable: all global branches are unstable, so some local branches have to be stable.}
    \label{tab:lpa_wavepin_regimes}
\end{table}

\subsection{Non-conservative (NC) model}

In addition the main bifurcation parameter $\gamma$, we also take $c$, the parameter that controls the magnitude of the source/sink terms. 
This model possess a unique global equilibrium:
\[u_* = \frac{\alpha}{\theta}, \quad \ v_* = \frac{c \alpha + \eta u_*}{A(u_*)}=\frac{c \alpha + \eta u_*}{k + \gamma \frac{u_*^n}{1+u_*^n}} \,.\]
Any local branches $u_{L*}$ must satisfy $f(u_{L*},v_*)=0$. After expanding and some manipulations, we obtain
\begin{equation}
\frac{A(u_{L*})}{A(u_*)}=\frac{u_{L*}}{u_*}.
\label{eqn:champ_lpa_local}
\end{equation}
Since neither $A(u)$ nor $u_*$ involve $c$ and $\eta$, we conclude that the local branches are independent of these parameters.
Furthermore, for $\gamma \ll k$, the LHS of \eqref{eqn:champ_lpa_local} $\approx 1$ so $u_{L*}=u_*$, which means that there is no local branch for small $\gamma$. 

We will show that for $\gamma \gg k$, there are always a high and low local branches. The low branch is $u_L \approx 0$, since with $\gamma \to \infty$ and $u_L =0$, both sides of \eqref{eqn:champ_lpa_local} evaluate to 0. With a bit of further manipulation, we get (in the limit $\gamma \to \infty$):
\[h(u_{L_*}) = h(u_*), \quad \text{where \ } h(u)=\frac{u^{n-1}}{1+u^n} \,.\]
The function $h(u)$ satisfies $h(0)=0=h(u \to \infty)$, and it has a single peak at $u_{p} \geq 1$ (provided $n \geq 2$). Since we focus on parameters with $\alpha<\theta$, that is $u_* <1$, there exists a point $u_{L*}> u_p > 1$ such that $h(u_{L*})=h(u_*)$, which corresponds to the high local branch.

\begin{figure}
    \centering
    \begin{subfigure}[h]{0.5\textwidth}
        \centering
        \caption{WM}
        \includegraphics[width=\textwidth]{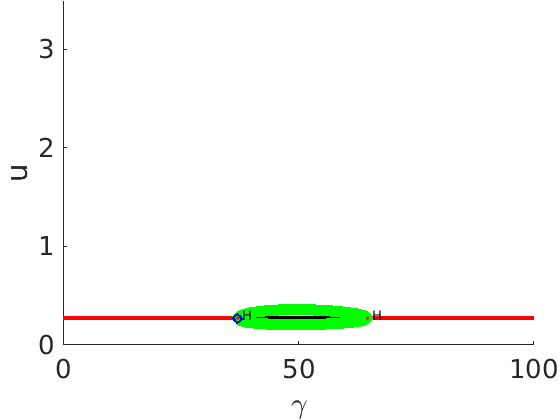}
    \end{subfigure}%
    \begin{subfigure}[h]{0.5\textwidth}
        \centering
        \caption{LPA}
        \includegraphics[width=\textwidth]{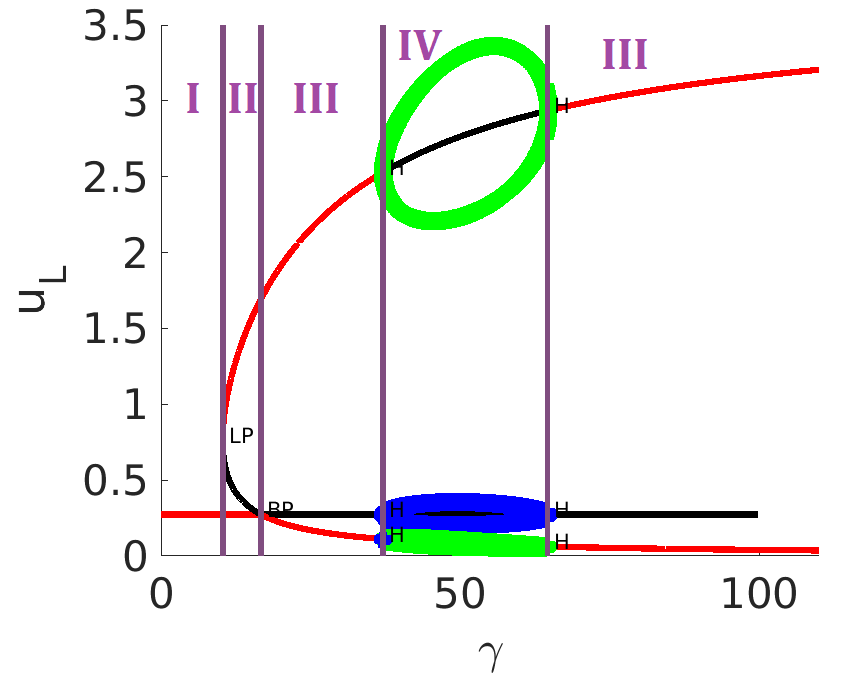}
    \end{subfigure}
    \begin{subfigure}[h]{0.8\textwidth}
        \centering
        \caption{LPA}
        \includegraphics[width=\textwidth]{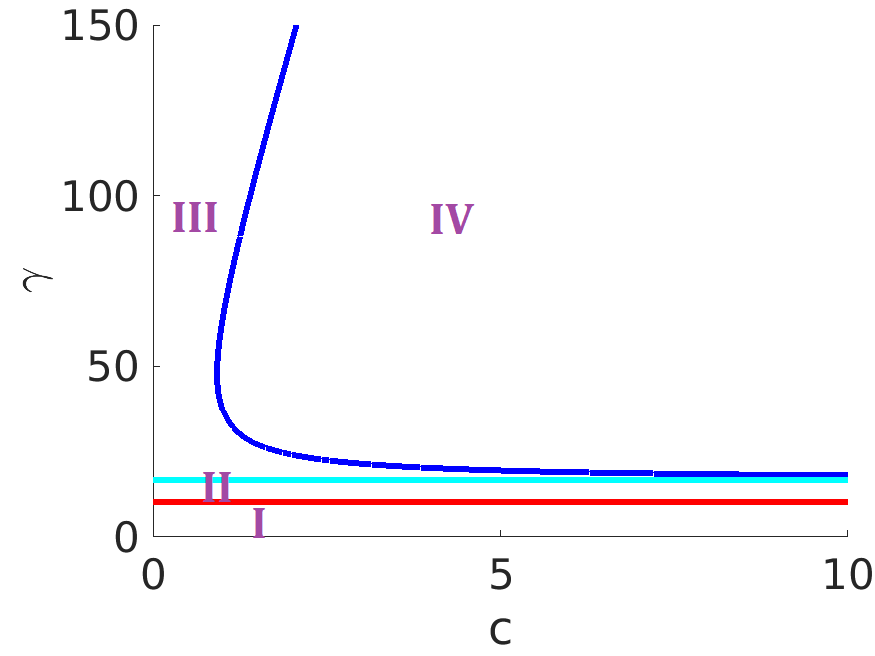}
    \end{subfigure}
    \caption[Bifurcation diagrams of the LPA and well-mixed non-conservative model]{Bifurcation diagrams for the non-conservative (NC) model, with parameter values from Table~\ref{tab:parameters}(CM2) except $\eta=5$. (a) WM, (b,c) LPA,  using bifurcation parameters (a,b) $\gamma$, with $c=1$, (c) $c$ and $\gamma$. A thin polarizable regime  II is sandwiched between the stable I and Turing III regimes. The triplet of Hopf bifurcations (not present in WP) does not show up as new behavior in the full PDE simulations. }
    %In (b), there is a fold point at $\gamma=10.31$, transcritical point at $\gamma=16.76$, and triplets of Hopf bifurcation at $\gamma=37.09, 64.68$
    \label{fig:lpa_champ}
\end{figure}

In the bifurcation diagrams in Fig.~\ref{fig:lpa_champ}, we use parameters from Table~\ref{tab:parameters} (CM2) but with $\eta=5$. (These parameters yield visually optimized bifurcation diagrams whose regimes are neither too wide nor too narrow; the same regimes are present for parameters from Table~\ref{tab:parameters}(NC) used for PDE simulations, but the resulting bifurcation diagram is harder to read.) Fig.~\ref{fig:lpa_champ} identifies four distinct regimes in the LPA system whose interpretation is as in the previous section. The location of the branches agrees with earlier analysis.
The regimes are summarized in Table~\ref{tab:lpa_champ_regimes}.

In Regime I, no pattern forms, as expected. In Regime II a small perturbation decays, but a sufficiently large perturbation will persist. In the full PDEs, such perturbation  leads to the soliton solution shown in Fig.~\ref{fig:sim_champ}(c,d). Both Regime III and IV are unstable, and any perturbation leads to a Turing-type pattern consisting of a series of evenly spaced, static spikes in the full PDE, as in Fig.~\ref{fig:sim_champ}(a,b). The limit cycles in Regime IV suggests that the spikes might oscillate, but this, in fact, does not occur: we found that the PDE behavior is qualitatively indistinguishable in Regime III and IV. This suggests that the Hopf bifurcations in the LPA diagram may not necessarily correspond to actual bifurcations for the full PDE, pointing to a limitation of LPA.

\begin{table}[h]
    \centering
    \begin{tabular}{l||l|l}
         Regime  & Classification & Description  \\
         \hline
         \hline
         I &  Stable & \specialcell{One stable GB, no LB} \\
         \hline
         II &  Polarizable & \specialcell{One stable GB,  one stable LB\\ located above the GB} \\
         \hline
         III & Unstable & \specialcell{The only GB is unstable, two LBs \\located on both sides of the GB} \\
         \hline
         IV & Unstable & \specialcell{One GB, two LBs all unstable, \\ each enclosed by a periodic orbit} \\
         \hline
    \end{tabular}
    \caption[Summary of the non-conservative model regimes]{Summary of the non-conservative (NC) model regimes identified in Fig.~\ref{fig:lpa_champ}. Abbreviations as in Table.~\ref{tab:lpa_wavepin_regimes}.}
    \label{tab:lpa_champ_regimes}
\end{table}

\subsection{Actin feedback (AF) model}
We use mass conservation to eliminate $v$ from the LPA system as before. The strength of actin feedback $s$ and the basal rate of activation $k$ were our bifurcation parameters. LPA for this model was previously discussed in \cite{holmes2012regimes,mata2013model}, but here we traced more bifurcations in greater detail.

\begin{figure}
    \centering
    \begin{subfigure}[h]{0.5\textwidth}
        \centering
        \caption{WM, $k=1.5$}
        \includegraphics[width=\textwidth]{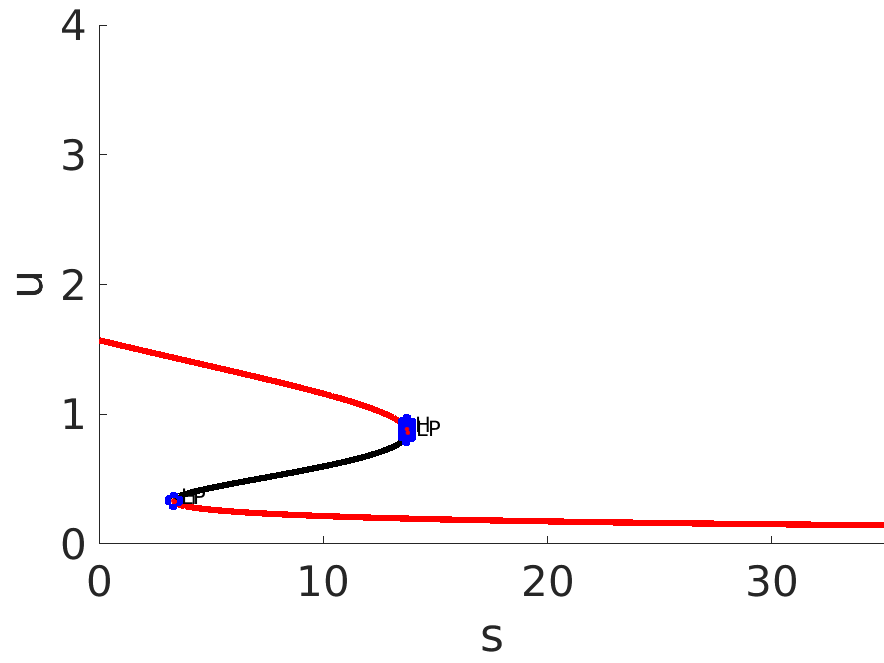}
    \end{subfigure}%
    \begin{subfigure}[h]{0.5\textwidth}
        \centering
        \caption{LPA, $k=1.5$}
        \includegraphics[width=\textwidth]{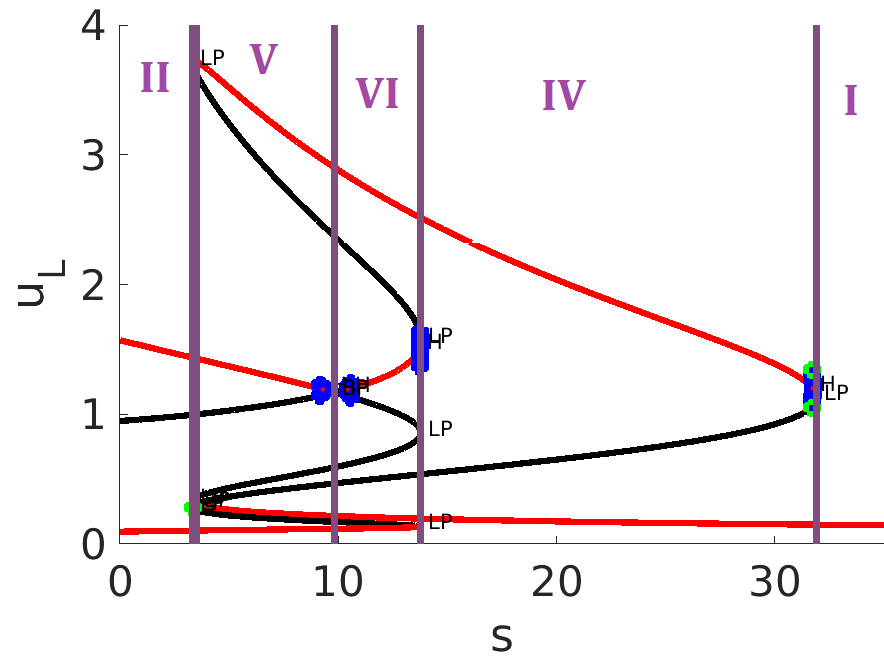}
    \end{subfigure}
    \begin{subfigure}[h]{0.5\textwidth}
        \centering
        \caption{WM, $k=6$}
        \includegraphics[width=\textwidth]{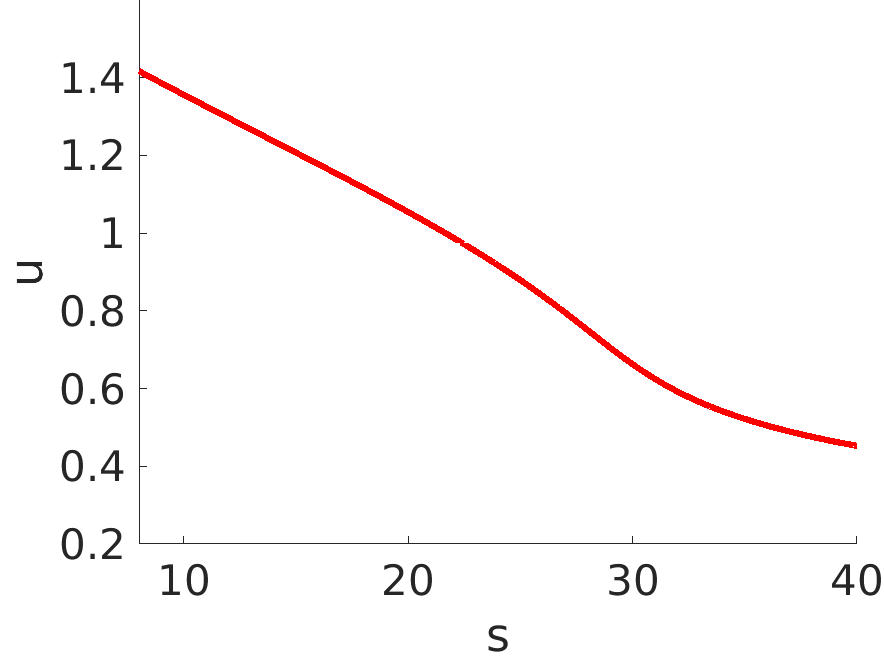}
    \end{subfigure}%
    \begin{subfigure}[h]{0.5\textwidth}
        \centering
        \caption{LPA, $k=6$}
        \includegraphics[width=\textwidth]{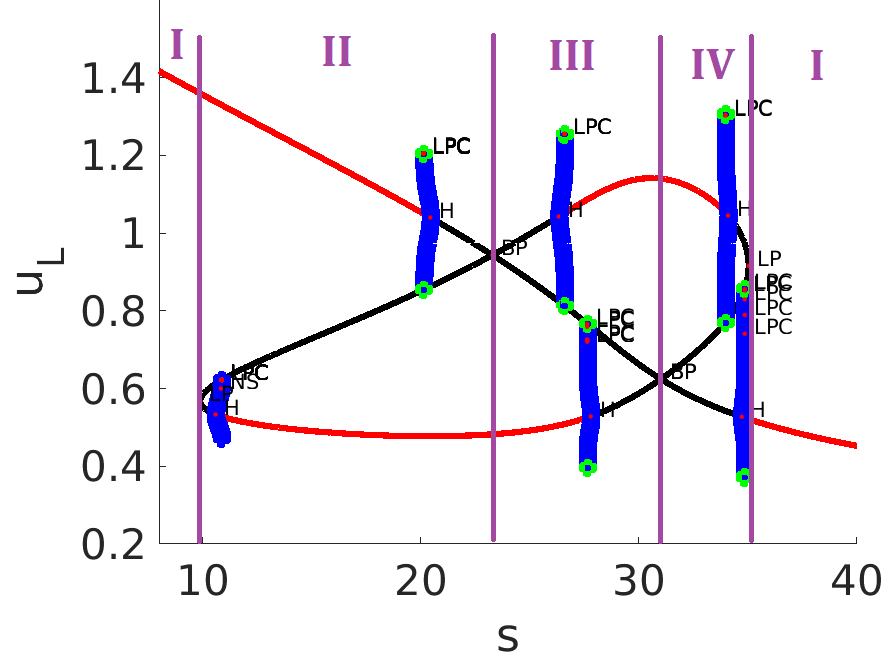}
    \end{subfigure}
    \begin{subfigure}[h]{0.45\textwidth}
        \centering
        \caption{LPA}
        \includegraphics[width=\textwidth]{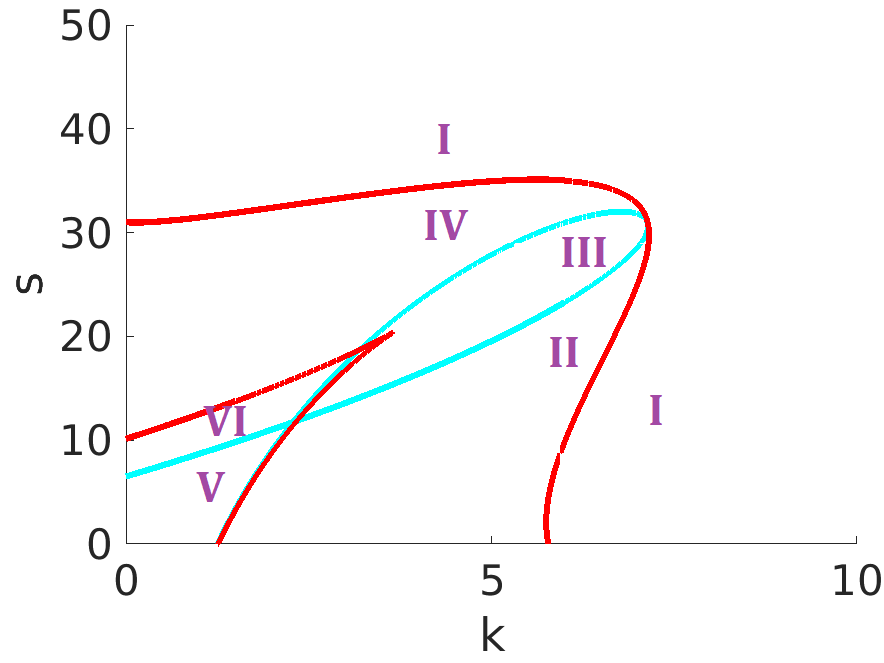}
    \end{subfigure}
    \begin{subfigure}[h]{0.45\textwidth}
        \centering
        \caption{LPA}
        \includegraphics[width=1.0\textwidth]{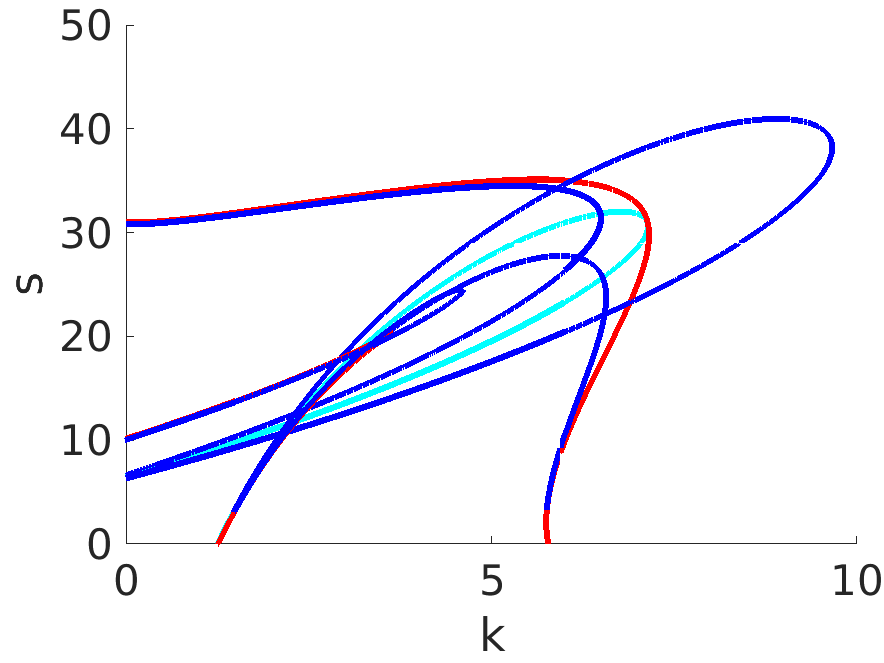}
    \end{subfigure}
    \caption[Bifurcation diagrams of the actin feedback extension]{Bifurcation diagrams of the actin feedback (AF) model with respect to parameter $s$ (a-d) and with respect to $k,s$ in (e,f). (In (e), the Hopf curves are omitted for clarity of the diagram. They are then included in (f).) The narrow regimes are not labelled. The nearly vertical blue curves indicate unstable periodic orbits.}
    \label{fig:lpa_actin}
\end{figure}

\begin{figure}
    \centering
    \includegraphics[width=1.0\textwidth]{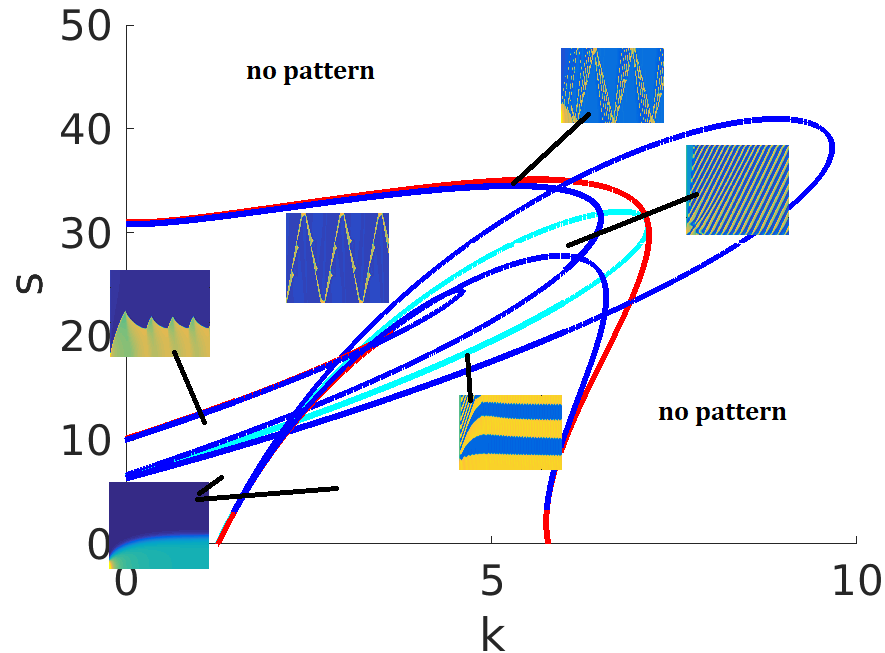}
    \caption{Same as Fig.~\ref{fig:lpa_actin}(f) with the Hopf curves included, and with an indication of patterns in several regimes. A few Hopf curves lie very close to one of the other curves for most of their length, creating some very narrow regimes. The simulation results from Fig.~\ref{fig:sim_actin} are identified with their corresponding regions on the parameter plane. %\mycomment{Do we want this picture instead of the other one? It's a bit messier}
    }
    \label{fig:lpa_actin_twoparam_annotated}
\end{figure}

The results are shown in Fig.~\ref{fig:lpa_actin} and \ref{fig:lpa_actin_twoparam_annotated}. We only distinguish between the regimes separated by fold and transcritical curves and omit the Hopf curves, as explained below. We also ignore some very narrow regimes, to concentrate on six major regimes as summarized in Table.~\ref{tab:lpa_actin_regimes}.

One interesting characteristic of these diagrams is the presence of unstable periodic orbits that emerge as subcritical Hopf bifurcations and exist for very narrow parameter ranges. The unstable cycle enlarges until it collides with a saddle point, turning into a homoclinic orbit to the saddle, and then disappearing. This is known as saddle-loop bifurcation, or homoclinic bifurcation (see \cite[Ch.6.2]{kuznetsov}). Parameter regimes where the periodic solutions exist are very narrow. Hence, while Hopf bifurcations occur, they are unlikely to be playing a major role in the biological application of this model.

We can compare our results to those of \cite{holmes2012regimes}
(Fig.~5, a LPA diagram in $k-s$ plane containing only one of the Hopf curves). The Hopf curve in \cite{holmes2012regimes} corresponds to the dark blue curve on our diagram, which traces the pair of Hopf points on the global branch in Fig.~\ref{fig:lpa_actin}(d)). Furthermore, our diagram (Fig.~\ref{fig:lpa_actin_twoparam_annotated}) traces the fold (red) and transcritical (light blue) bifurcation points and hence identifies a larger number of distinct regimes. 
%Given that the periodic solutions exist only in a very narrow range of parameters, the Hopf curve is unlikely to form the only regime boundary as discussed by \cite{holmes2012regimes}. %\mycomment{I'm not sure how much to say here. The current text sounds too 'rude'}

Interpreting the LPA diagrams (as in the WP model), we can conclude that a stable local branch in LPA corresponds to a regime of pattern formation in the PDE. Unlike WP, there are multiple possible patterns in this AF model. LPA cannot accurately predict the type of pattern. In particula, the consequence of the subcritical Hopf bifurcations to the full PDE is unclear, possibly suggesting some kind of (quasi-)periodic behavior that we did not fully characterize. In \cite{holmes2012regimes,mata2013model}, a parameter scan of the PDE system was included with the LPA diagrams. As previously noted, PDE regimes are not exactly aligned with LPA regimes since $\delta \ne 0$ in the full PDEs.

In summary, in our hands, LPA worked well in identifying no-pattern and WP regimes, but was less useful for predicting the emergence of more complex patterns. Many of those patterns involve interacting waves, which suggests that they are non-linear, non-local phenomena, explaining why LPA cannot account for them.

\begin{table}[h]
    \centering
    \begin{tabular}{l||l|l}
         Regime  & Classification & Description  \\
         \hline
         \hline
         I &  Stable & \specialcell{One stable GB, no LB} \\
         \hline
         II &  Polarizable & \specialcell{One stable GB,  one stable LB\\ located above the GB} \\
         \hline
         III & Unstable & \specialcell{The only GB is unstable, two stable LBs \\located on both sides of the GB} \\
         \hline
         IV & Unstable & \specialcell{The only GB is unstable, one stable LB\\ located above the GB} \\
         \hline
         V & Polarizable & \specialcell{Two stable GBs, three stable LBs: one above both\\ GBs, one in between, and one below both GBs} \\
         \hline
         VI & Polarizable & \specialcell{One stable GB, three stable LBs \\ located on both sides of the GB} \\
         \hline
    \end{tabular}
    \caption[Summary of the actin feedback model regimes]{Summary of the actin feedback (AF) model regimes identified in Fig.~\ref{fig:lpa_actin}. For abbreviations see caption of Table.~\ref{tab:lpa_wavepin_regimes}.}
    \label{tab:lpa_actin_regimes}
\end{table}

%\begin{figure}
 %   \centering
%    \includegraphics[width=1.0\textwidth]{bif/actin/holmes_LPA,w=2.5,k=6,gamma=30,n=3,eta=15_twoparam_autotomatlab_thick2.png}
%    \caption[Details of the two-parameter bifurcation diagrams for the actin feedback extension]{Same as Fig.~\ref{fig:lpa_actin}(e) but with the Hopf curves included. A few Hopf curves lie very close to one of the other curves for most of their length, creating some very narrow regimes.}
%    \label{fig:lpa_actin_twoparam}
%\end{figure}

\subsection{Combined model (CM)}
The LPA diagrams for the combined model are very complex, and mostly beyond the scope of interpretation (see Appendix \ref{apd:NC_lpa}.) This is unsurprising given the complex behavior exhibited by the PDE. The bifurcation diagram shown in Fig.~\ref{fig:lpa_combined}, which uses parameter values from Table~\ref{tab:parameters}(CM2), contains many limit cycle bifurcations, such as torus and period-doubling. One thing the diagram can provide is the minimum value of $s$ required for any non-static patterns (corresponding to the first triplet of Hopf bifurcation in Fig.~\ref{fig:lpa_combined}). With $s$ below this value, the system behaviour is the same as that of the $s=0$ case, which reduces back to the non-conservative (NC) model.

%**LPA for each model, especially actin feedback and source/sink models. The wave pinning models has been done by multiple existing papers, and the combined model is a mess.**

%Points to make: Turing analysis can't distinguish polarizable regimes

%%%%%%%%%%%%%%%%
%%%%%%%%%%%%%%%
\subsection{Comparison with linear (Turing) stability analysis}

Linear stability analysis (LSA) was previously applied by  \cite{mori2008wave} and \cite{Champneys} for the wave pinning and non-conservative models respectively. The relative merits of LSA and LPA have been described in \cite{mata2013model,holmes2014efficient} and we briefly summarize some of these in the Appendix.
 
LPA is only valid in the limit of $\delta \to 0$. In this limit, LPA contains the Turing stability properties: a branch that is LPA-unstable is also Turing-unstable. See Fig.~\ref{fig:lpa_turing_comparison}, where we show how the LPA regimes from Fig.~\ref{tab:lpa_champ_regimes}(b) line up with Turing regimes. 
LPA can detect instabilities that require a perturbation of sufficient magnitude (the polarizable regimes), which cannot be detected by Turing analysis. This means LPA can potentially find more types of pattern.

LPA does not predict details of the pattern. We saw this most evidently in the actin feedback (AF) model, where many possible patterns and a large number of parameter regimes exist. 
Turing analysis predicts pattern initiation, but often fails to specify the final pattern that depends on nonlinear interactions. We give an example of this type for the NC model in 
Fig.~\ref{fig:turing_zoom}. We also indicate how the ``minimal patch size" idea from \cite{painter2011spatio} can be used to help predict the final pattern using LSA.

%%%%%%%%%%%%%%
%%%%%%%%%%%%%
\section{Numerical simulations}
We simulated the model for a static cell in 1D ($0 \leq x \leq 1$) and 2D ($0 \leq x,y \leq 1$), and for a motile cell in two spatial dimensions using the Cellular Potts Model (CPM).
The four main parameter sets we used for numerical simulations are summarized in Table~\ref{tab:parameters}. The selection of values for most of these parameters is based on \cite{holmes2012regimes}, with $\alpha, \theta$ coming from \cite{Champneys}, and some modifications guided by LPA and Turing analysis. In contrast to \cite{holmes2012regimes} we use a much larger domain size $L$, corresponding to a larger cell and allowing for more complex patterns to develop.
\begin{table}[ht]
\centering
\begin{tabular}{l|l||c|c|c|c}
Parameter & Meaning & WP & NC & AF & CM2 \\
\hline
\hline
$\delta$ & Diffusion coefficient ratio & \multicolumn{4}{c}{$0.01$} \\
\hline
$L$ & Domain length & 1 & \multicolumn{3}{c}{$10$}\\
\hline
$k$ & Basal activation rate & \multicolumn{2}{c|}{$1.5L^2$} & $1L^2-6L^2$ & $1L^2$\\
\hline
$\gamma$ & Nonlinear activation rate & \multicolumn{4}{c}{$30L^2$}\\
\hline
$n$ & Hill coefficient & \multicolumn{3}{c|}{$3$} &2\\
\hline
$\eta$ & Inactivation rate & $15L^2$& $5L^2$& $15L^2$& $5.2L^2$\\
\hline
$c$ & NC terms on/off & 0& 1& 0&1\\
\hline
$\alpha$ & Source strength & \multicolumn{4}{c}{$1.5L^2$}\\
\hline
$\theta$ & Sink strength & \multicolumn{3}{c|}{$4.5L^2$} & $5.5L^2$\\
\hline
$s$ & Actin feedback strength & \multicolumn{2}{c|}{$0$}& \multicolumn{2}{c}{$0-50L^2$}\\
\hline
$\epsilon$ & Actin reaction rate & \multicolumn{4}{c}{$0.1$}\\
\hline
$k_n$ & Actin activation rate & \multicolumn{4}{c}{$24L^2$}\\
\hline
$k_s$ & Actin inactivation rate & \multicolumn{4}{c}{$7.5L^2$}\\
\hline
\end{tabular}
\caption[Parameter values for numerical simulations]{The parameters in the combined model, their meanings and values for various simulations. WP: wave pinning; %(Sec.~\ref{sec:numerics_wavepin}); 
NC: non-conservative extension; %(Sec.~\ref{sec:numerics_champ}); 
AF: actin feedback extension; %(Sec.~\ref{sec:numerics_actin}); 
CM2: one of the parameter sets used for the combined model (CM). %(Sec.~\ref{sec:numerics_combined}). 
All parameters (except $L$) are scaled to be non-dimensional.
% as in Sec.~\ref{sec:nondimlize}.
}
\label{tab:parameters}
\end{table}

%%%%%%%%%%%%%%%%
%%%%%%%%%%%%%%%%
\subsection{Simulations in a fixed 1D domain}

While 1D simulations for the WP, NC and AF appear in previous works \cite{mori2008wave,Champneys,holmes2012regimes,mata2013model}, we present them here as comparison to the combined model and the 2D case.
Results are  shown as kymographs, with time on the horizontal axis and is space on the vertical axis. Color indicates the levels of $u$ and $v$ and/or $F$ (if $s>0$). 
For most simulations, we start at a homogeneous steady state (HSS), and perturb the system either with small global noise or with a localized pulse. The first leads to Turing-type patterns, while the latter can lead to the patterns described by LPA.

\begin{figure}
    \centering
    \vspace{-0.25cm}
    \begin{subfigure}[h]{0.5\textwidth}
        \centering
        \caption{u}
        \includegraphics[width=\textwidth]{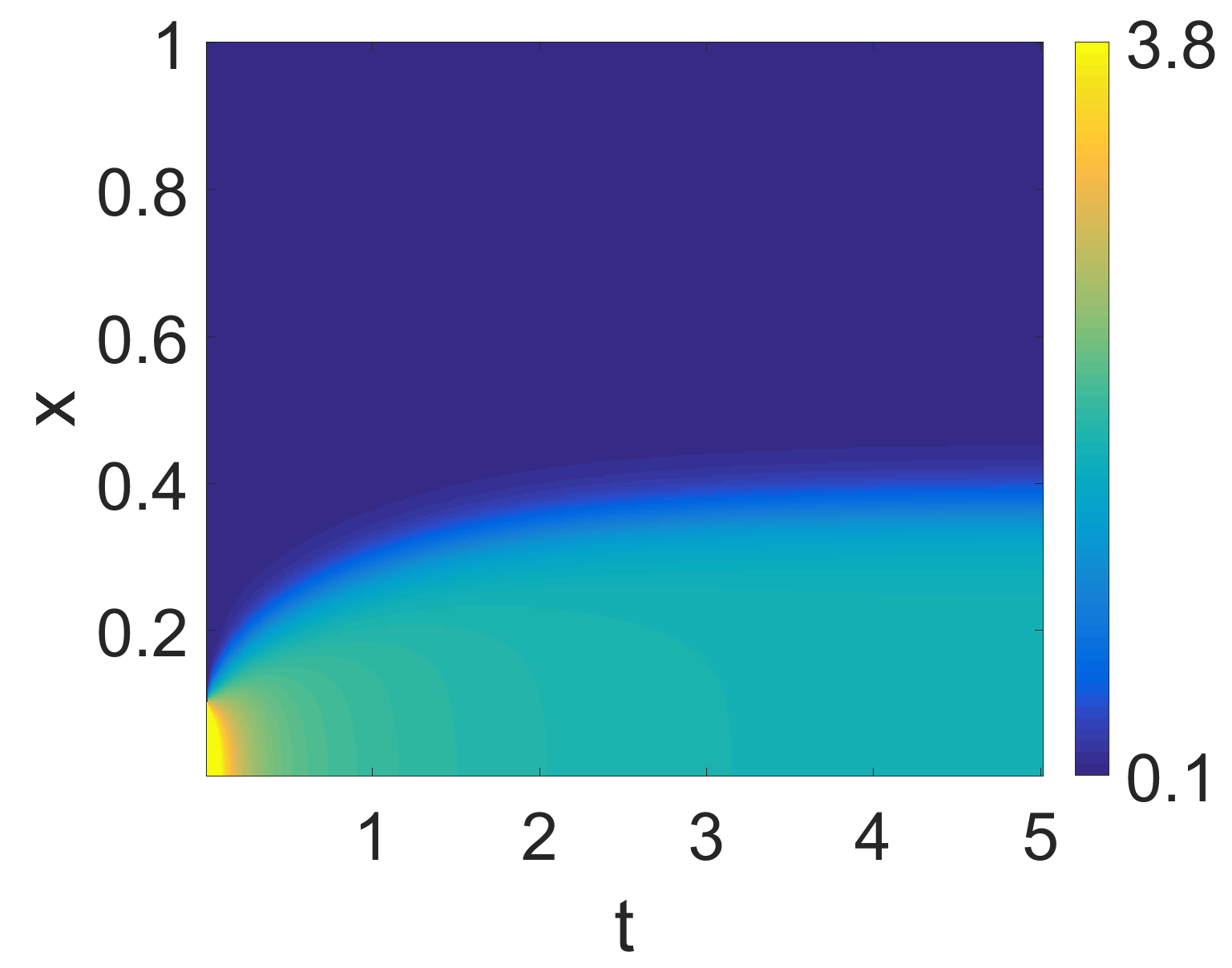}
    \end{subfigure}~
    \begin{subfigure}[h]{0.5\textwidth}
        \centering
        \caption{v}
        \includegraphics[width=\textwidth]{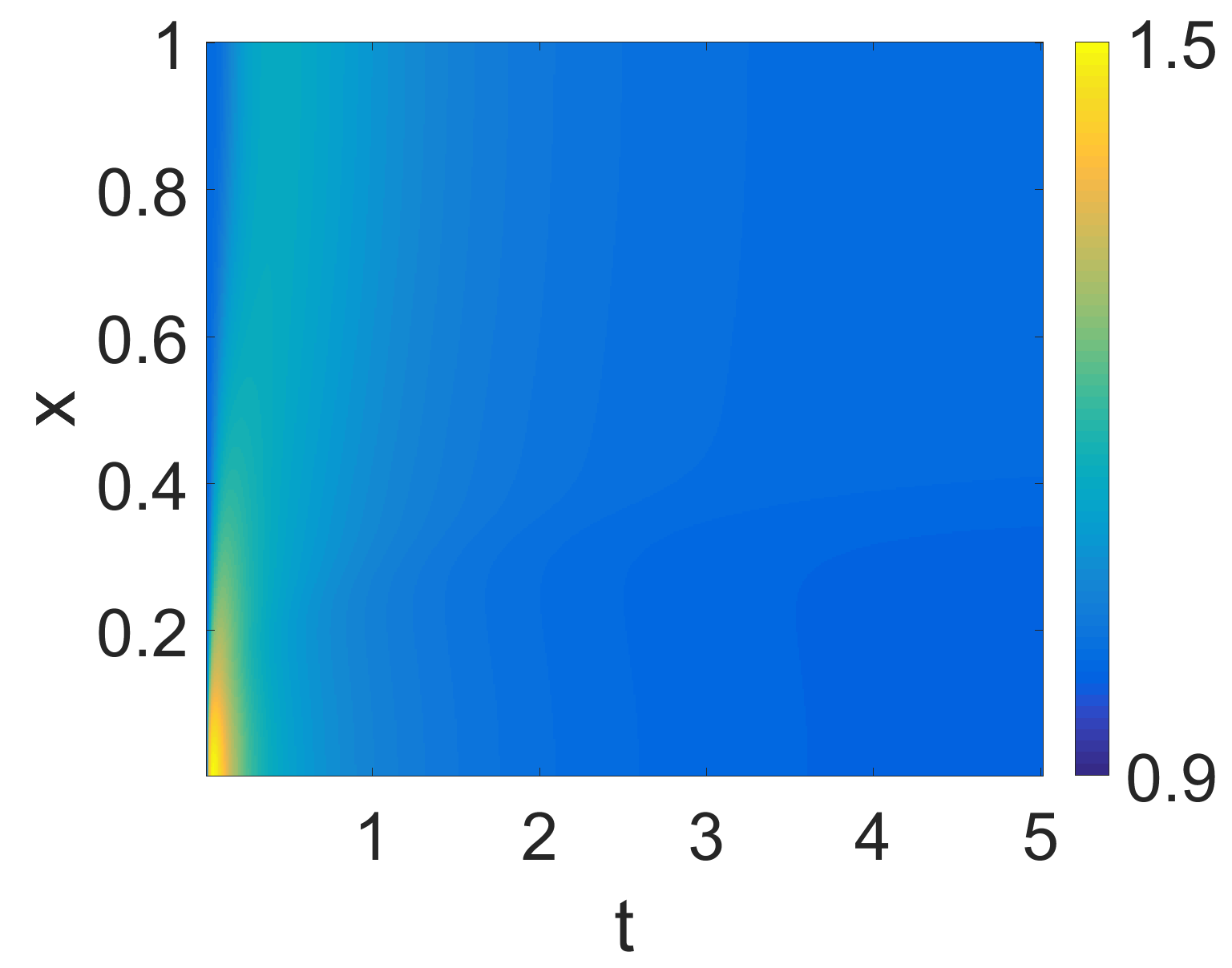}
    \end{subfigure}
    \begin{subfigure}[h]{0.5\textwidth}
        \centering
        \caption{u}
        \includegraphics[width=\textwidth]{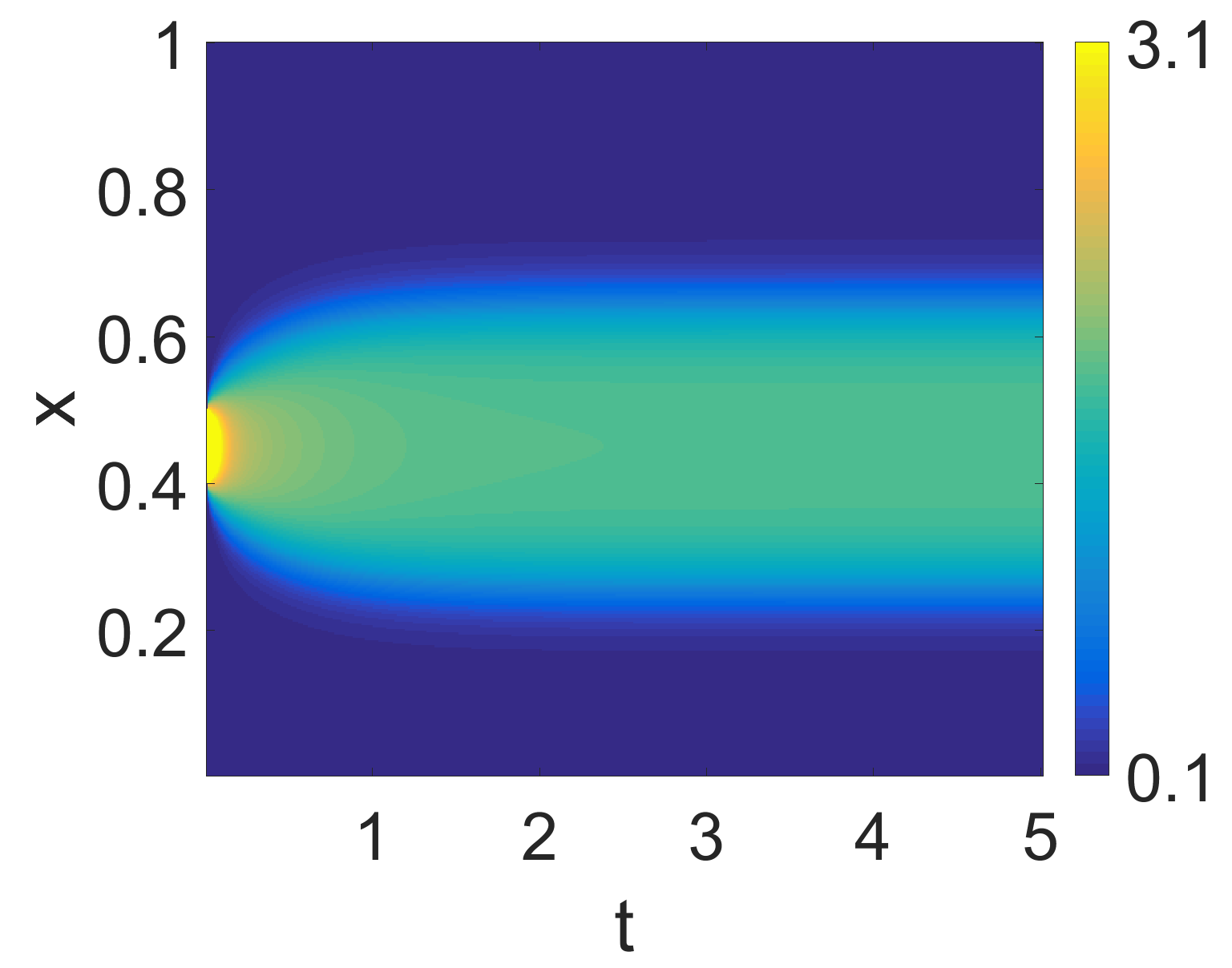}
    \end{subfigure}~
    \begin{subfigure}[h]{0.5\textwidth}
        \centering
        \caption{v}
        \includegraphics[width=\textwidth]{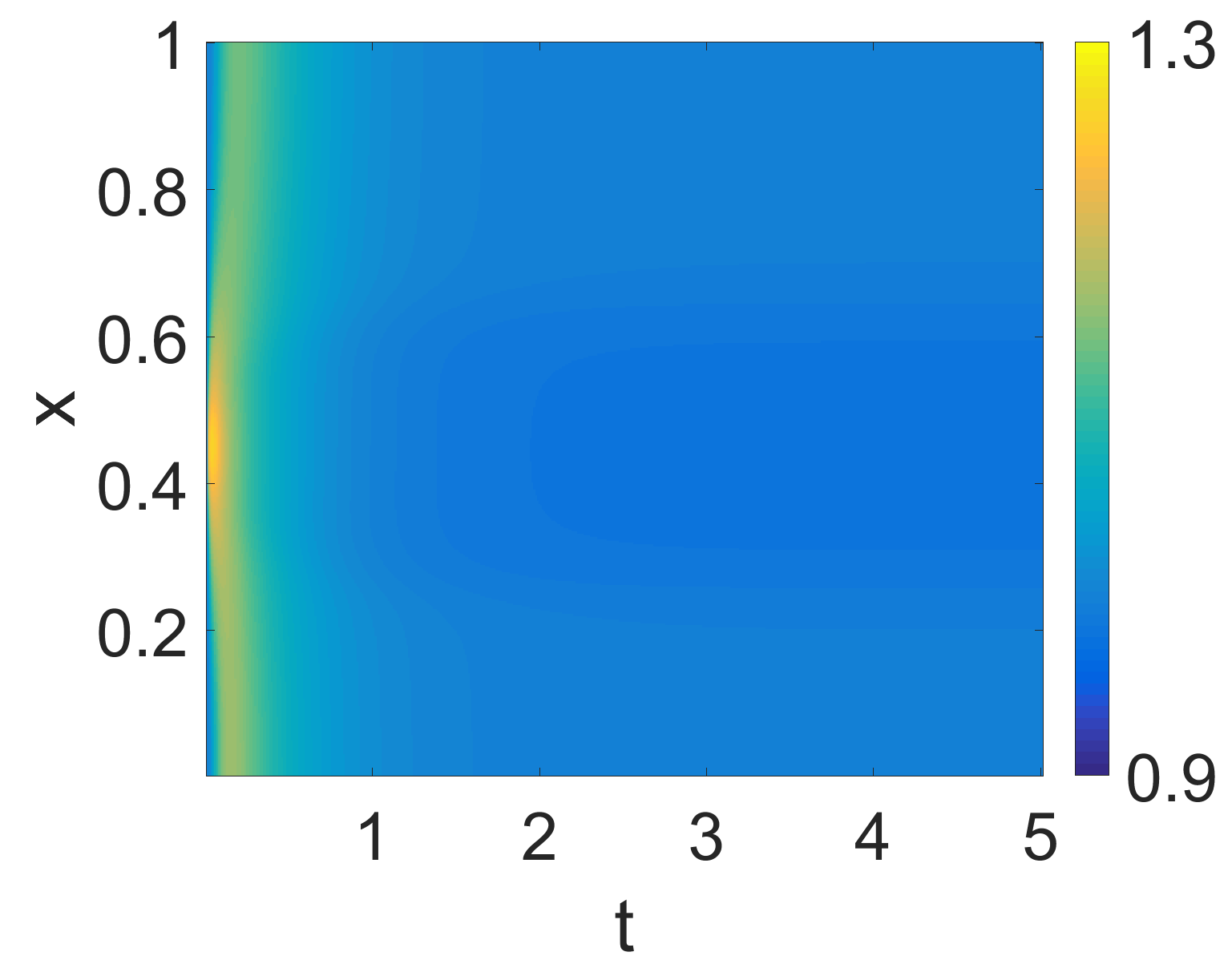}
    \end{subfigure}
    \begin{subfigure}[h]{0.5\textwidth}
        \centering
        \caption{u}
        \includegraphics[width=\textwidth]{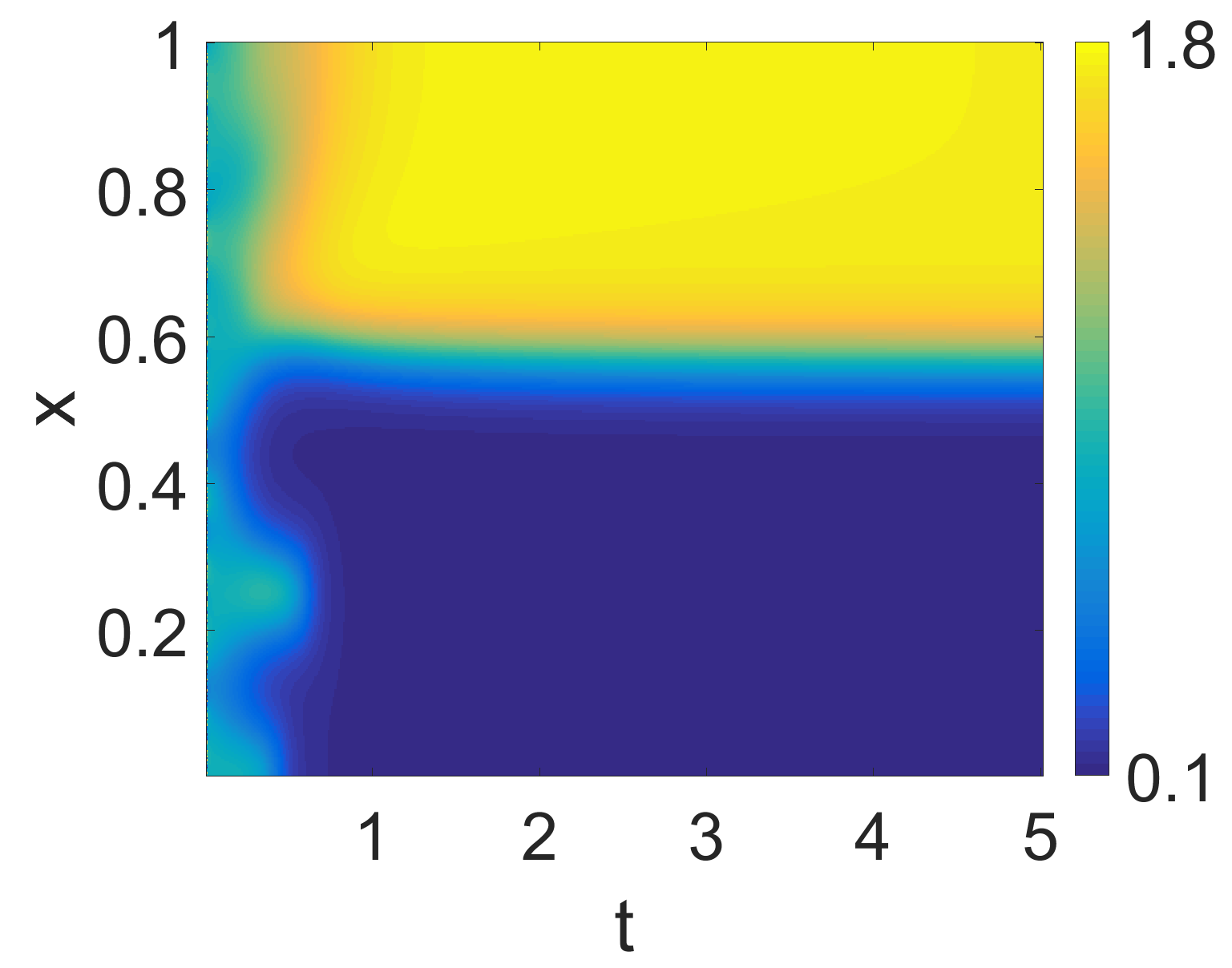}
    \end{subfigure}~
    \begin{subfigure}[h]{0.5\textwidth}
        \centering
        \caption{v}
        \includegraphics[width=\textwidth]{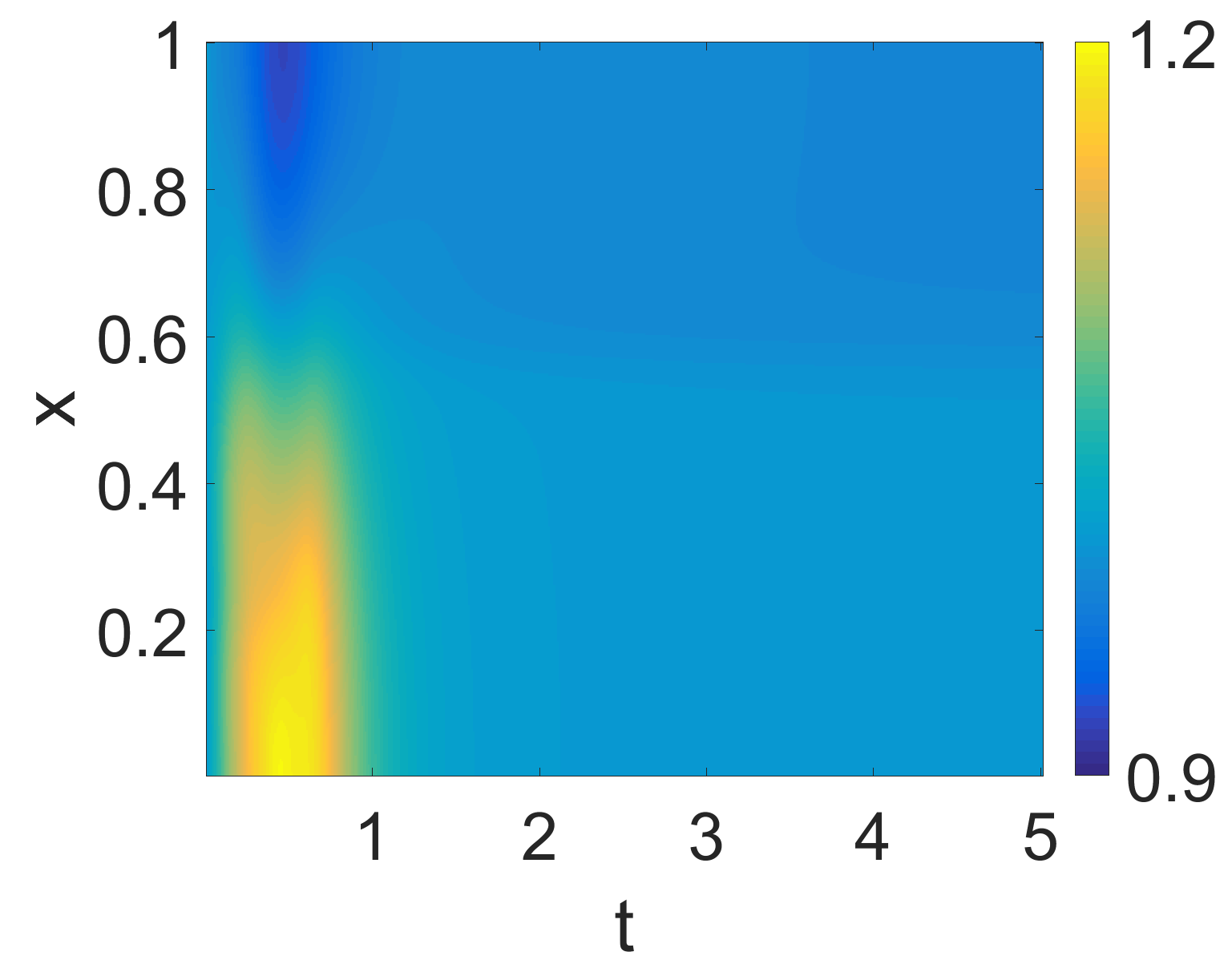}
    \end{subfigure}
    \vspace{-0.15cm}
	\caption[Wave pinning simulation]{Simulation of the wave pinning model (WP),
	%\eqref{sys:wavepin}, 
	with parameters from Table~\ref{tab:parameters}(WP). Initial condition:  $v=1$, $u=0.102$ with perturbation $u=6$ for (a,b) $0 \leq x \leq 0.1$; (c,d) $0.4 \leq x \leq 0.5$; (e,f) random noise, $u=0.834\cdot \epsilon(x)$. Note that not all initial conditions result in wave pinning: a small perturbation from the HSS will simply decay and no pattern forms. The behaviors shown in (a,b,e,f) correspond to solutions shown in Fig.~2 of \cite{mori2008wave}. }
	\label{fig:sim_wavepin}
\end{figure}

Fig.~\ref{fig:sim_wavepin} shows the results for the WP model. Observe that for the first two cases (a,c), the initial perturbation decays considerably, but nevertheless this results in formation of a pattern associated with polarization. The random initial condition (e) also results in a polarized steady state. In these simulations, $u$ can vary greatly across the domain while $v$ becomes nearly uniform, as expected given its much faster rate of diffusion.

\begin{figure}
    \centering
    \vspace{-0.25cm}
    \begin{subfigure}[h]{0.33\textwidth}
        \centering\caption{$u$, $k=1.5,s=18$}
        \includegraphics[width=\textwidth]{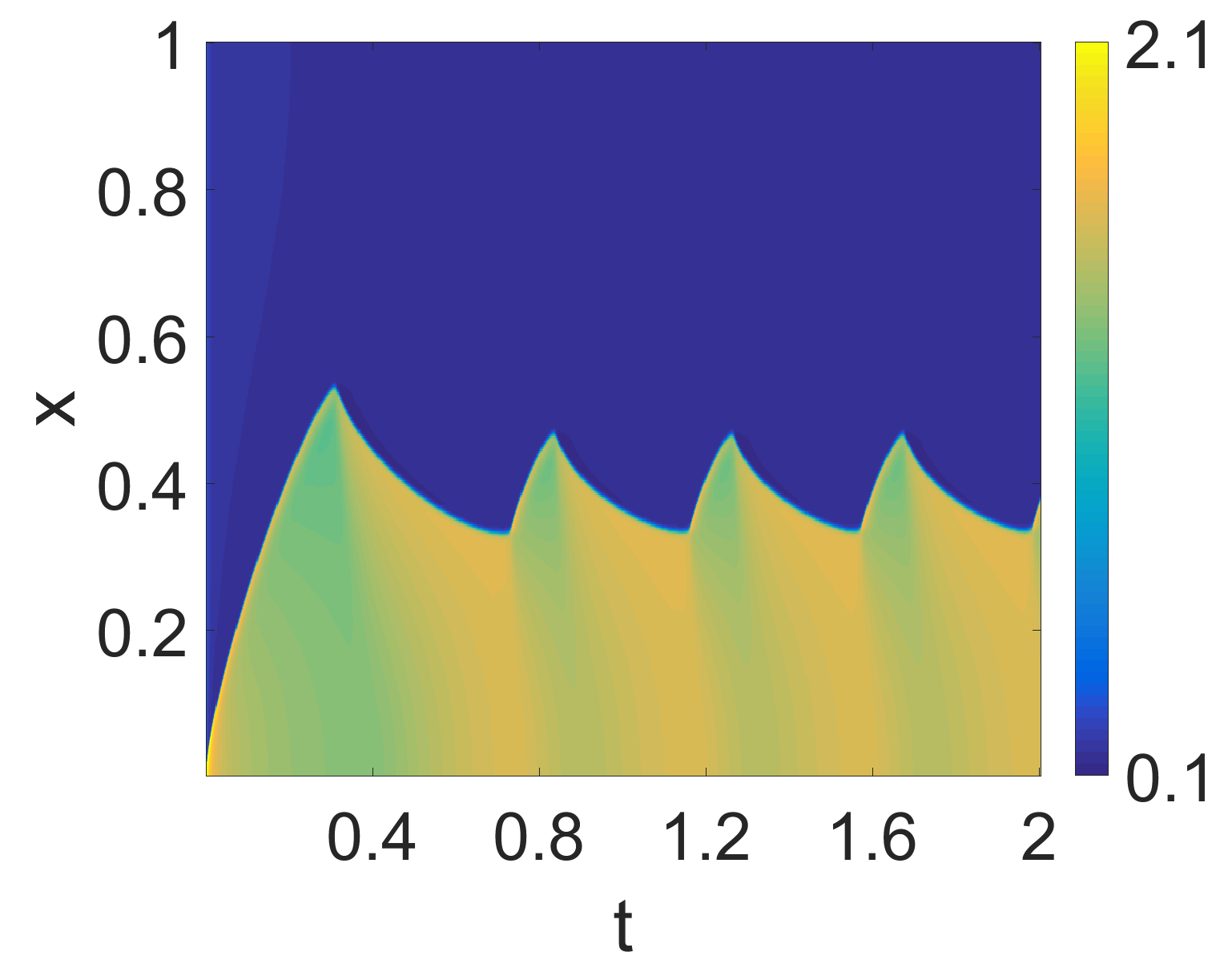}
    \end{subfigure}%
    \begin{subfigure}[h]{0.33\textwidth}
        \centering\caption{$v$, $k=1.5,s=18$}
        \includegraphics[width=\textwidth]{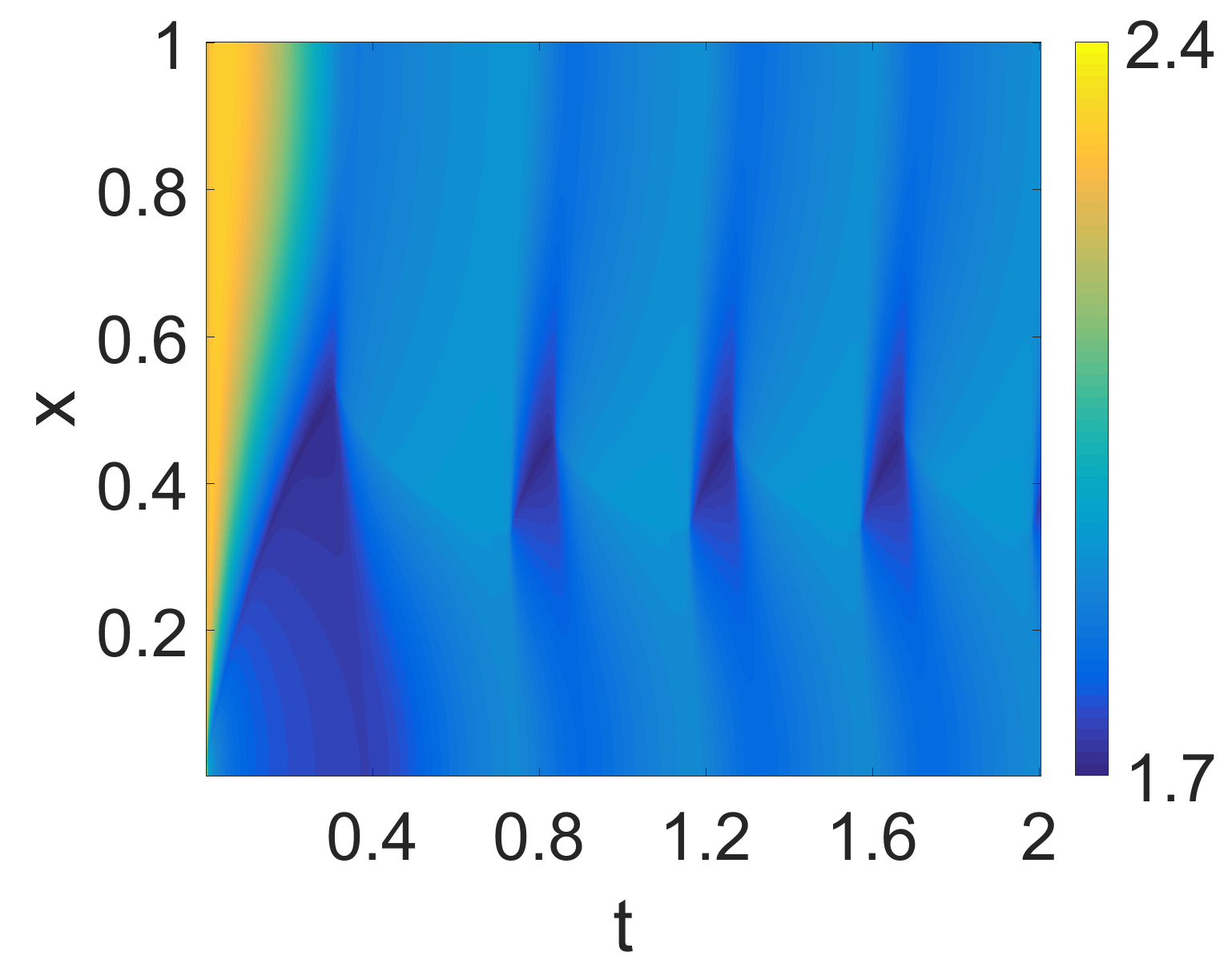}
    \end{subfigure}%
    \begin{subfigure}[h]{0.33\textwidth}
        \centering\caption{$F$, $k=1.5,s=18$}
        \includegraphics[width=\textwidth]{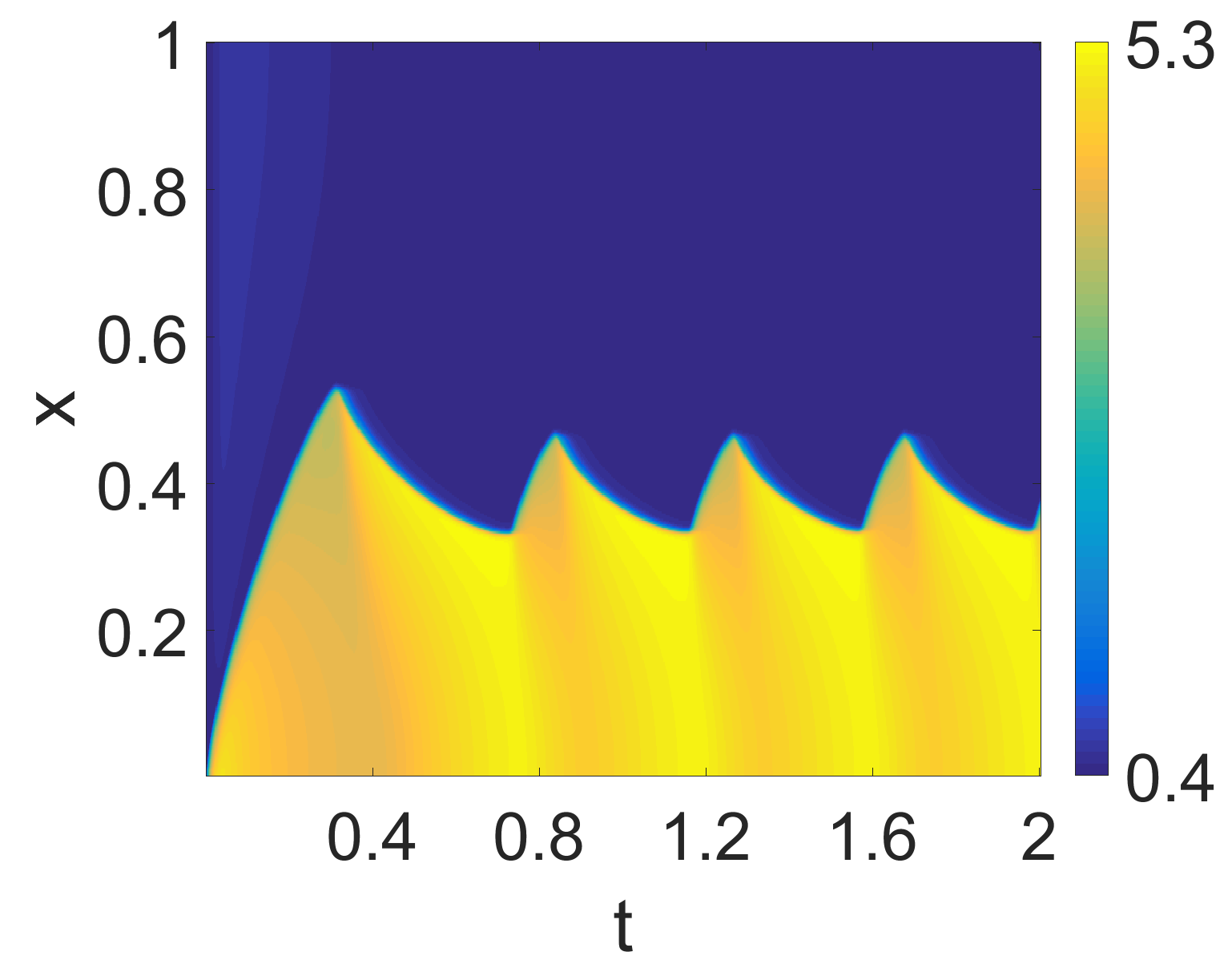}
    \end{subfigure}
    \begin{subfigure}[h]{0.33\textwidth}
        \centering\caption{$u$, $k=1.5,s=27$}
        \includegraphics[width=\textwidth]{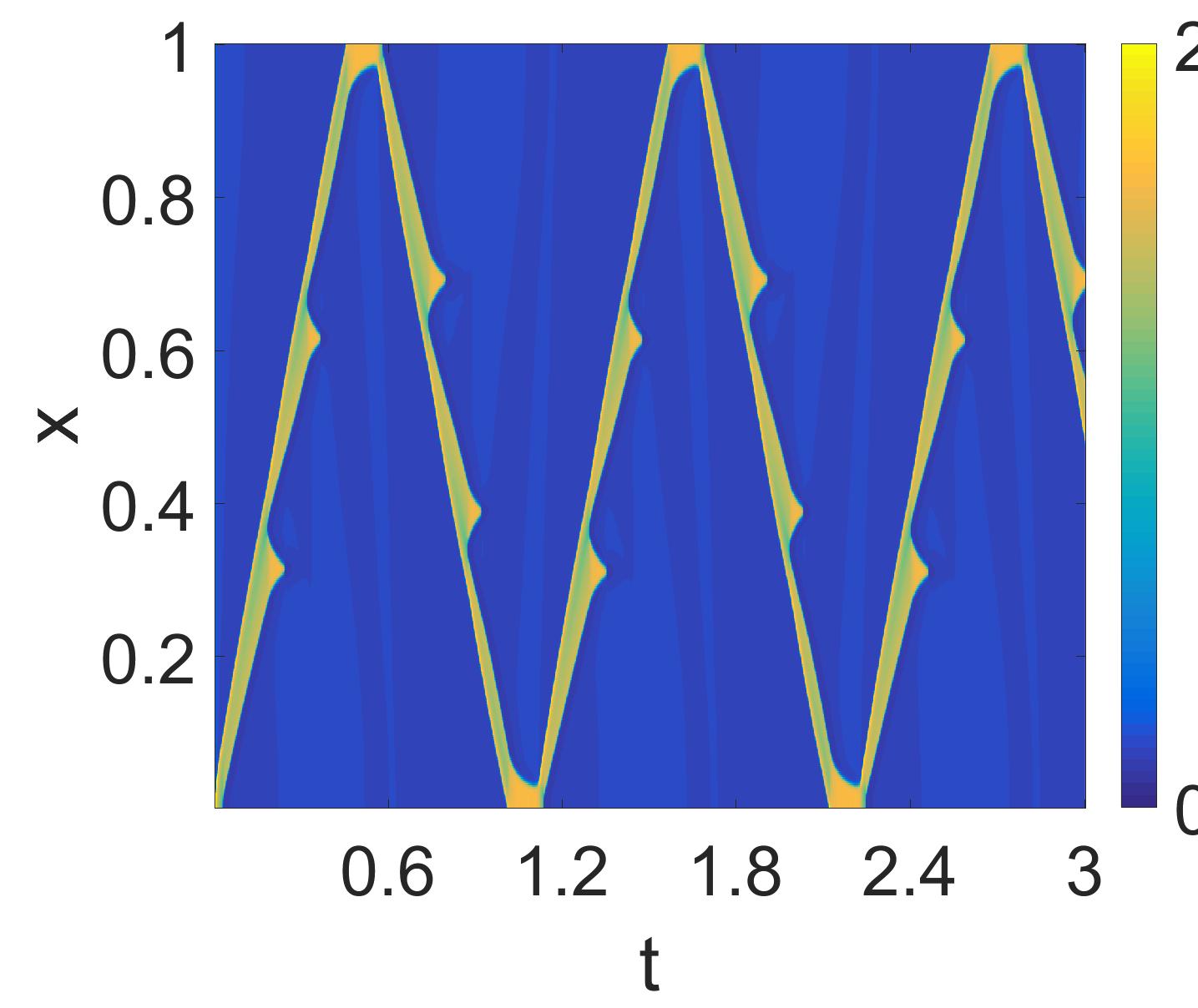}
    \end{subfigure}%
    \begin{subfigure}[h]{0.33\textwidth}
        \centering\caption{$v$, $k=1.5,s=27$}
        \includegraphics[width=\textwidth]{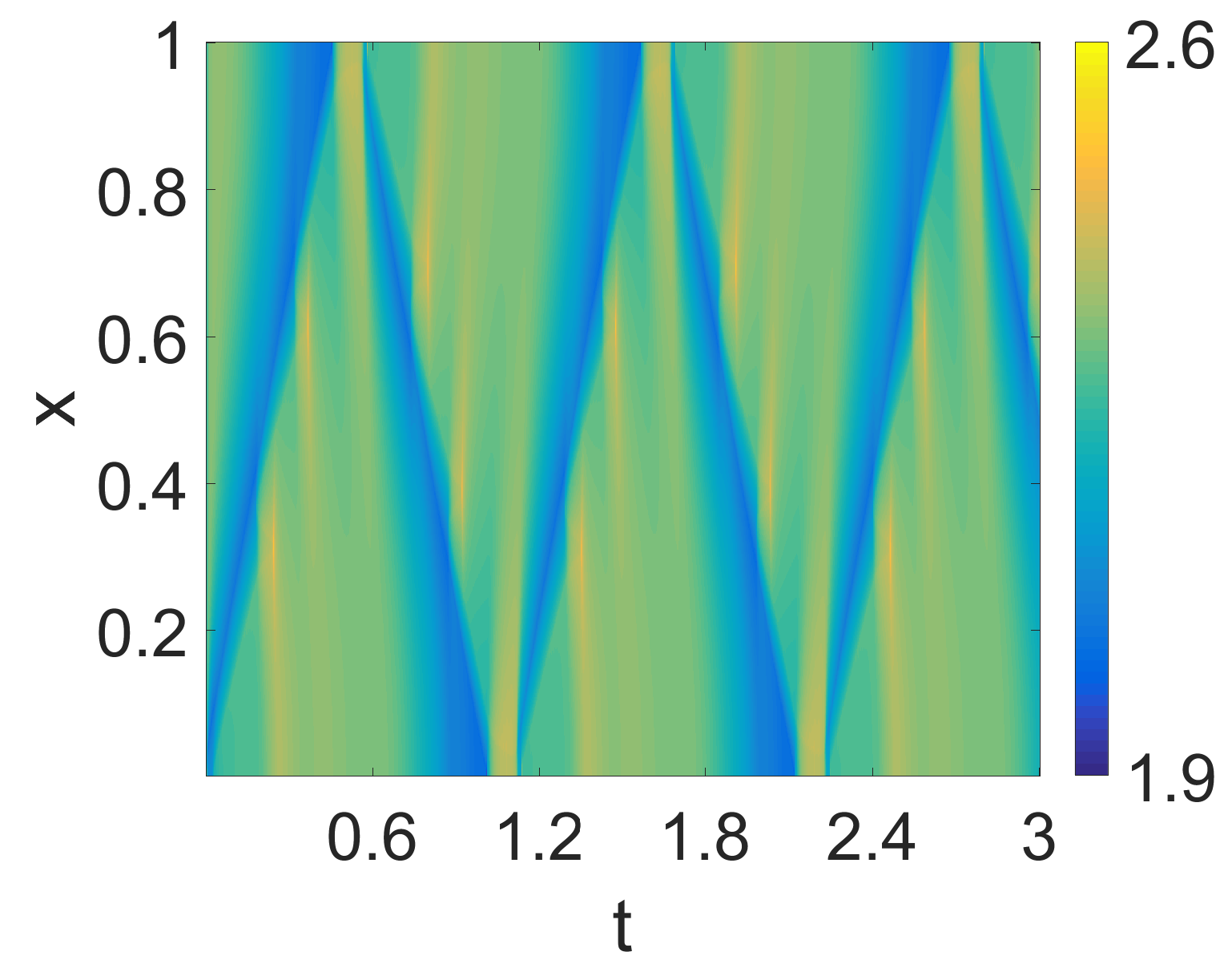}
    \end{subfigure}%
    \begin{subfigure}[h]{0.33\textwidth}
        \centering\caption{$F$, $k=1.5,s=27$}
        \includegraphics[width=\textwidth]{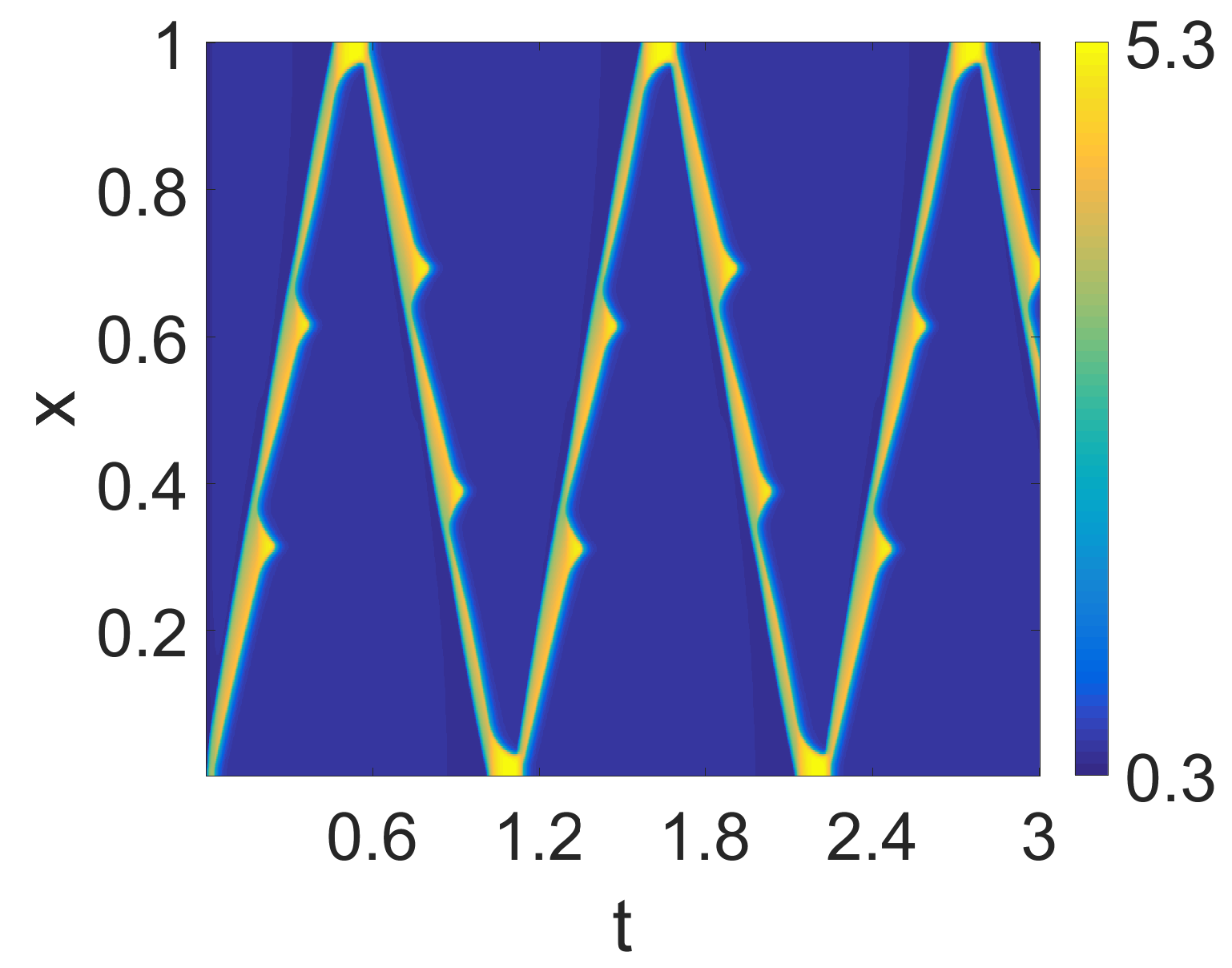}
    \end{subfigure}
    \begin{subfigure}[h]{0.33\textwidth}
        \centering\caption{$u$, $k=1.5,s=36$}
        \includegraphics[width=\textwidth]{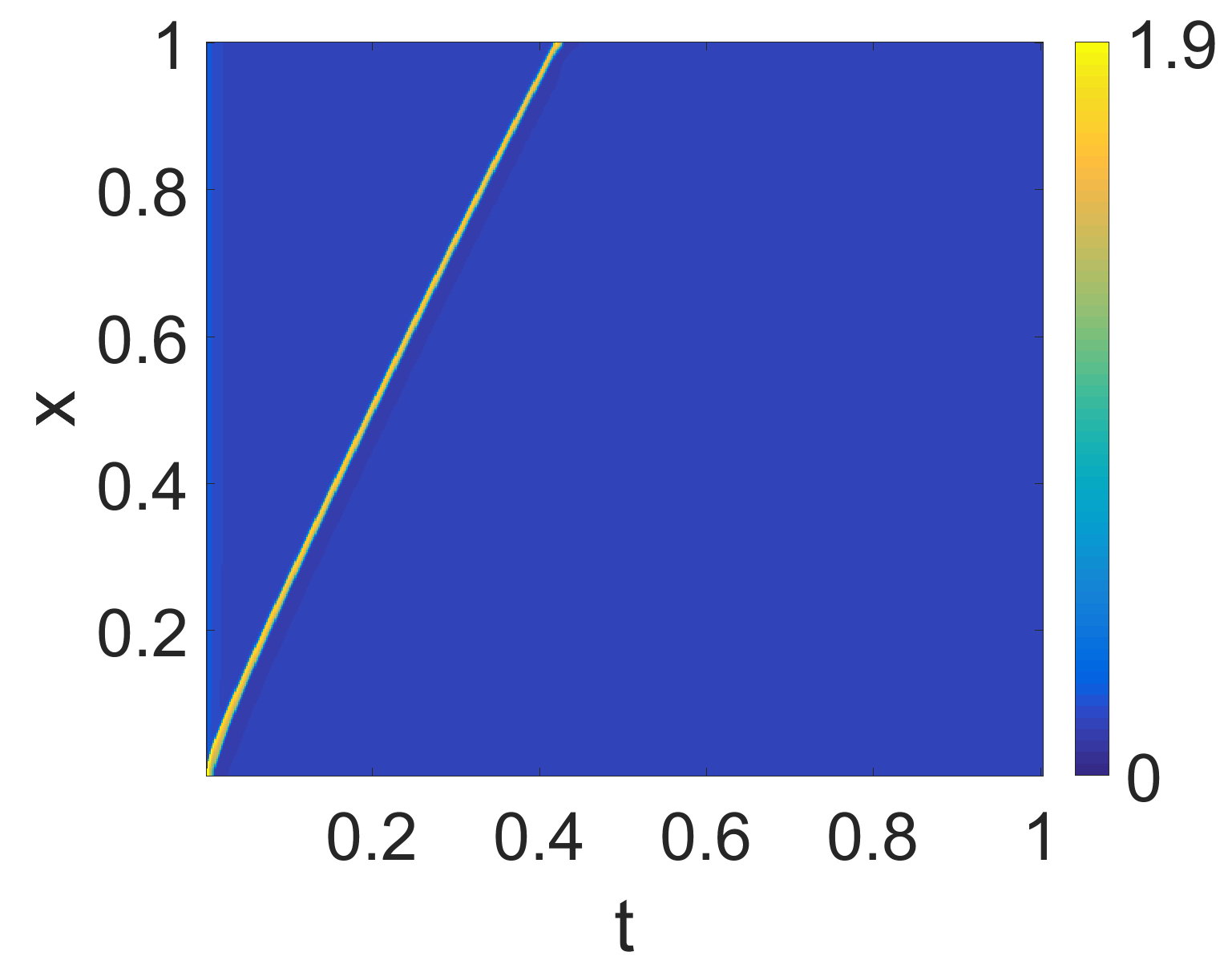}
    \end{subfigure}%
    \begin{subfigure}[h]{0.33\textwidth}
        \centering\caption{$v$, $k=1.5,s=36$}
        \includegraphics[width=\textwidth]{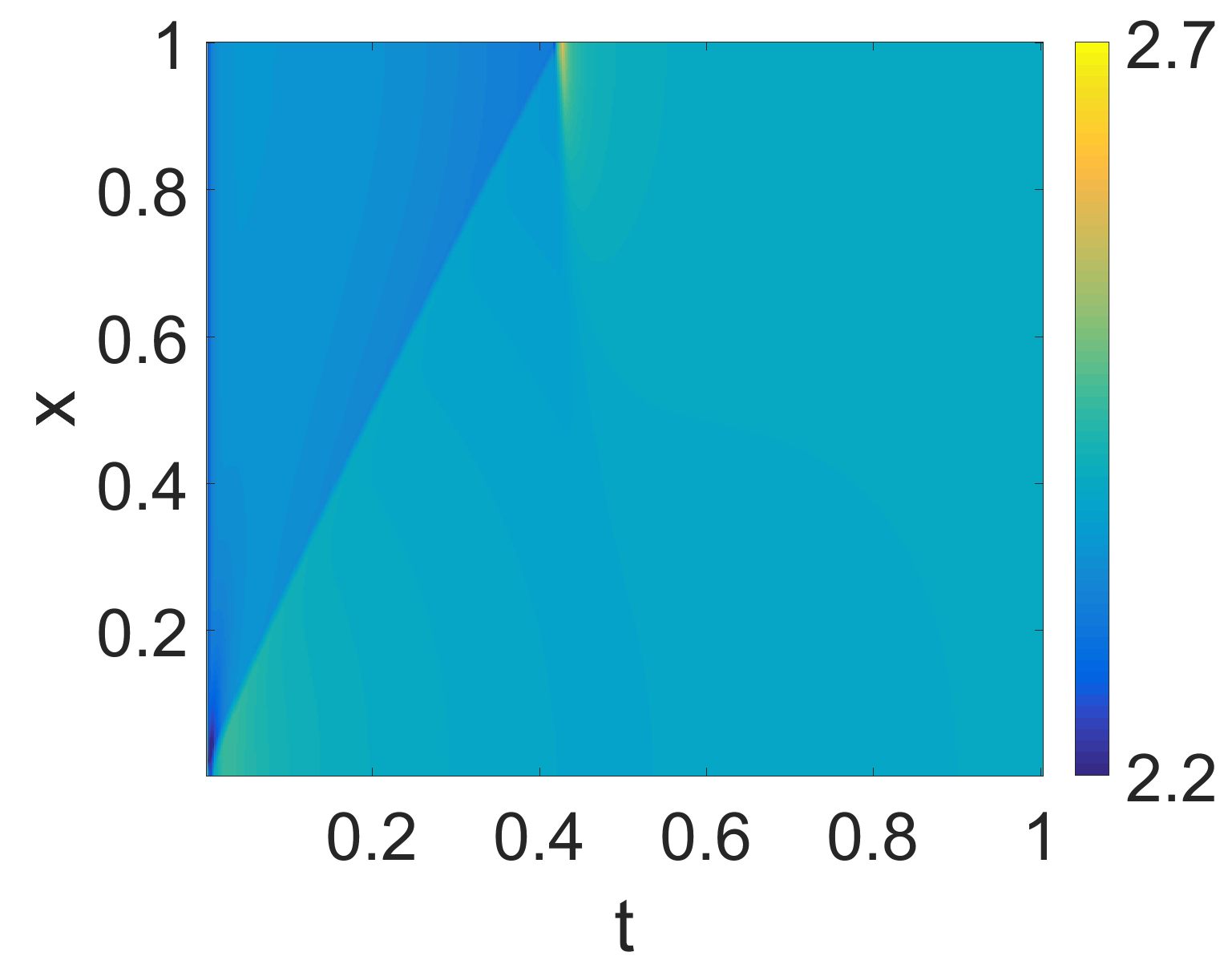}
    \end{subfigure}%
    \begin{subfigure}[h]{0.33\textwidth}
        \centering\caption{$F$, $k=1.5,s=36$}
        \includegraphics[width=\textwidth]{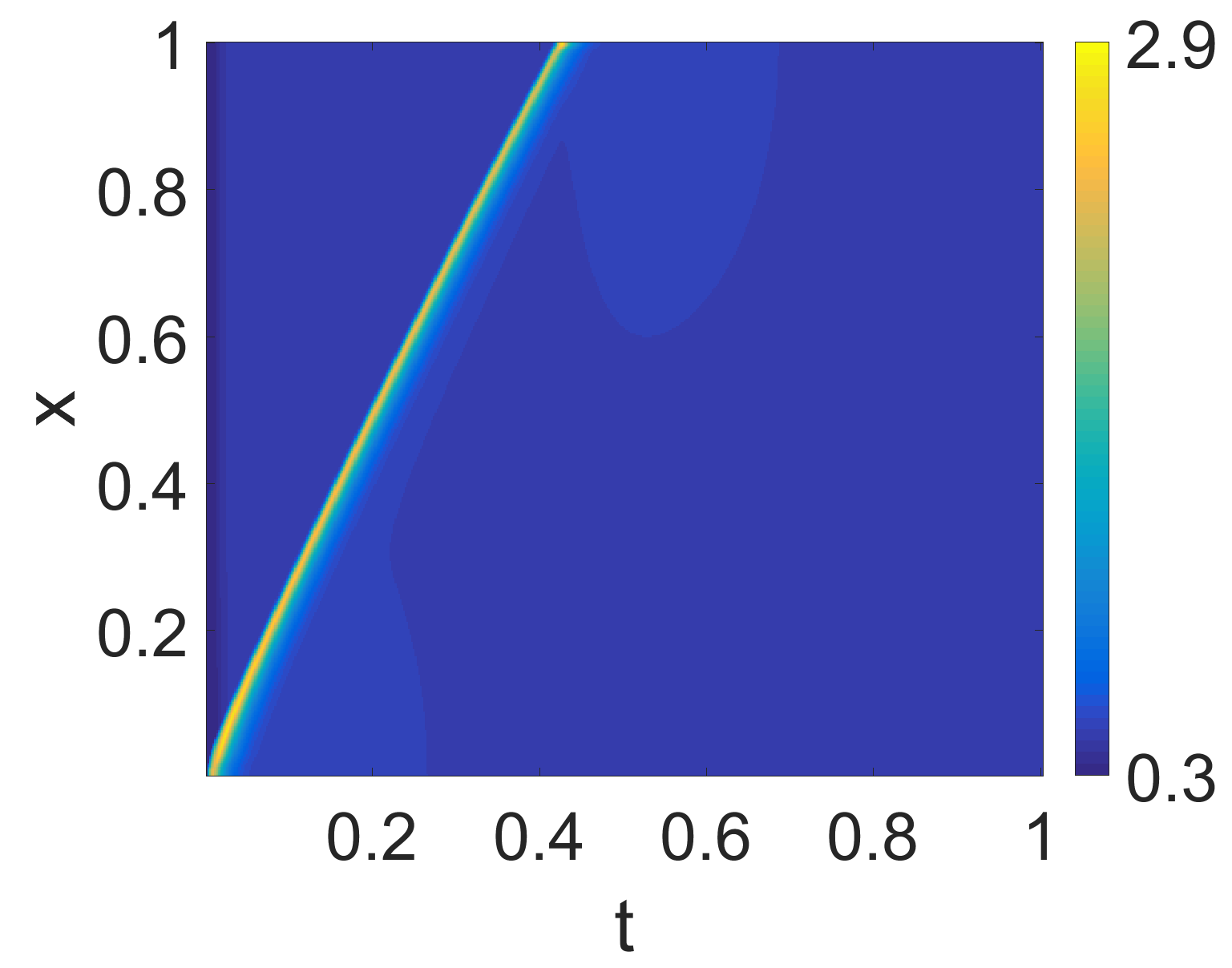}
    \end{subfigure}
    \begin{subfigure}[h]{0.33\textwidth}
        \centering\caption{$u$, $k=6,s=30$}
        \includegraphics[width=\textwidth]{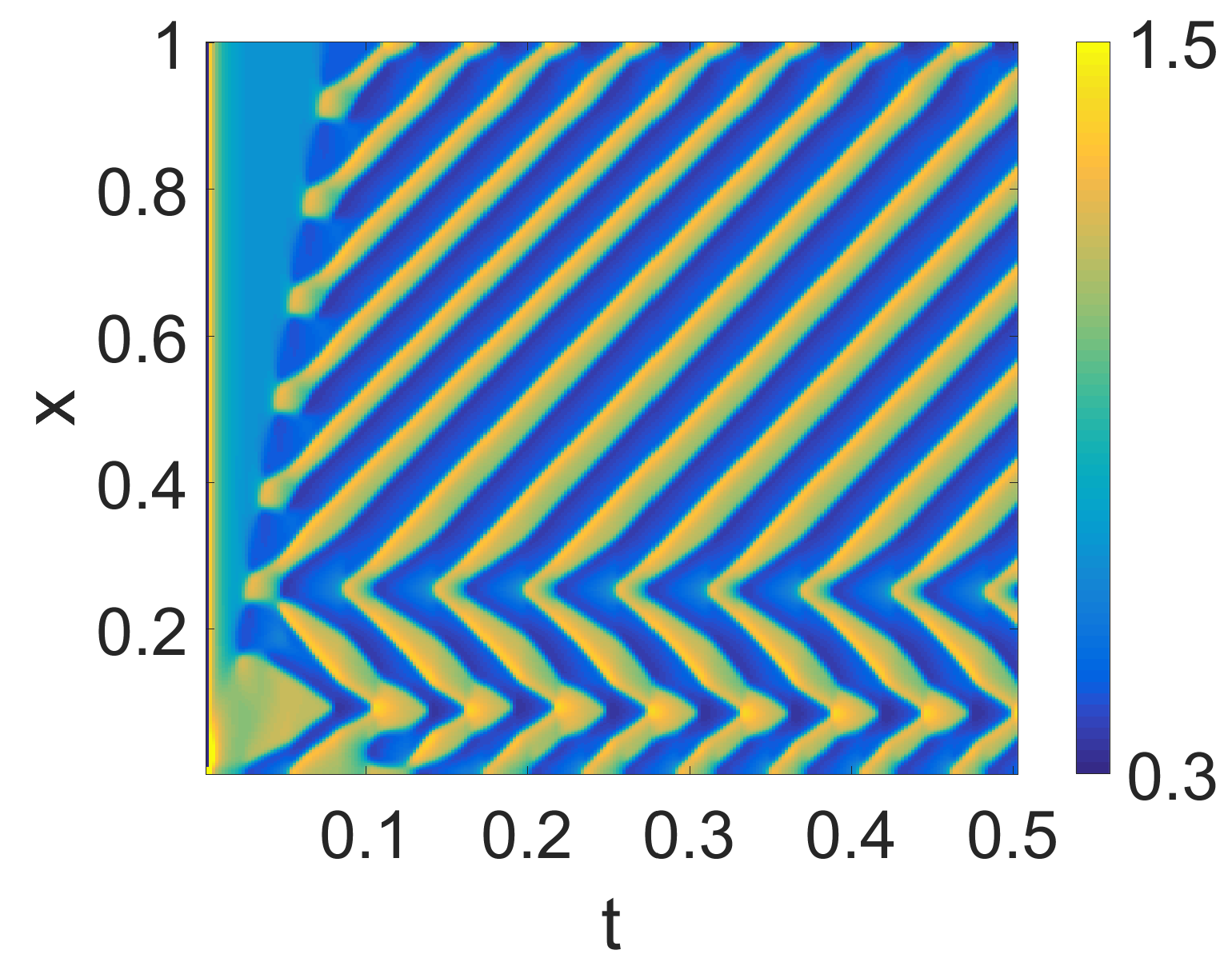}
    \end{subfigure}%
    \begin{subfigure}[h]{0.33\textwidth}
        \centering\caption{$v$, $k=6,s=30$}
        \includegraphics[width=\textwidth]{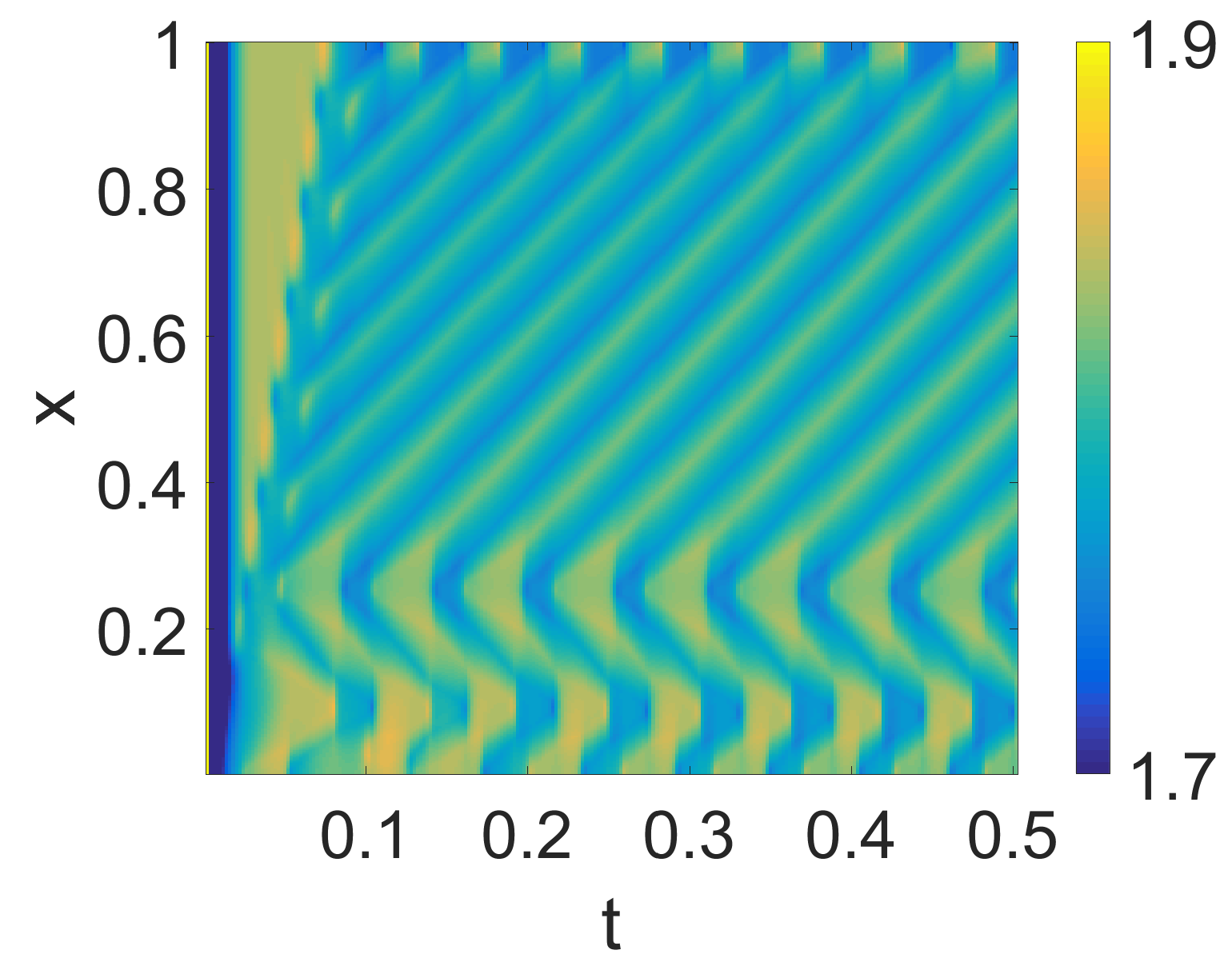}
    \end{subfigure}%
    \begin{subfigure}[h]{0.33\textwidth}
        \centering\caption{$F$, $k=6,s=30$}
        \includegraphics[width=\textwidth]{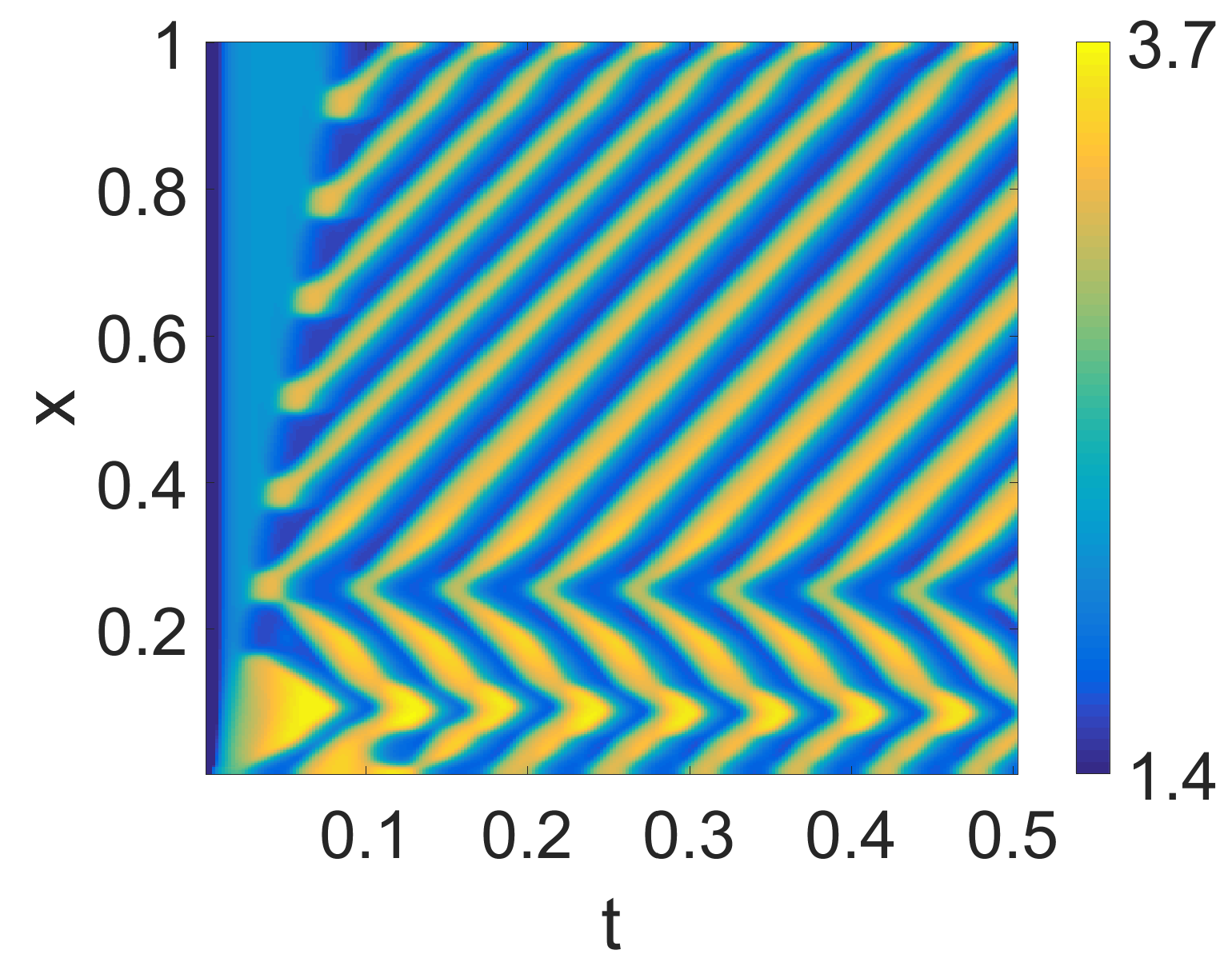}
    \end{subfigure}
    \vspace{-0.15cm}
    \caption[Numerical simulations of the actin feedback model]{Simulations of the actin feedback model (AF)
  %  \eqref{sys:actinwave} 
    with parameters from Table~\ref{tab:parameters}(AF) ($s,k$ as indicated on labels), and default initial conditions. Each row corresponds to one parameters set, showing $u,v,F$ (left to right). We observe four behaviors by varying $k $ and $s$: (a-c) Wave pinning with oscillating front (WPO); (d-f) Reflecting waves (RW); (g-i) Single pulse absorbed at boundary (SP); (j-l) Persistent wave trains (WT). We used a larger domain length than \cite{holmes2012regimes}, leading to a richer set of patterns. }
    \label{fig:sim_actin}
\end{figure}

Fig.~\ref{fig:sim_actin} shows the results for the actin feedback (AF) model. The default initial conditions are $u=0$ except $u=4$ for $0 \leq x \leq 0.01$, $v=2.5$, $F=0$. We are not initializing near a stable HSS because doing so usually does not result in patterning. The patterns observed are quite sensitive to initial conditions. In addition to simple wave pinning observed at low $s$ (not shown), the system displays four qualitatively different behaviors: (1) wave pinning with oscillating boundary (WPO), where polarization occurs as in wave pinning, but with an oscillating front position; (2) reflecting pulse (RW), where a single pulse traverses the domain at constant velocity and gets reflected back at the boundary; (3) a single pulse (SP) that is absorbed at a boundary, before the system returns to HSS; (4) a wave train (WT),  that originates either at a boundary or in the interior of the domain, propagates with constant velocity and gets absorbed at a boundary.

In general, the spatial profile of $F$ lags behind $u$, as expected, since it is a slow variable depending on $u$. The pattern in $v$ is usually opposite that of $u$, i.e, $v$ is high where $u$ is low, and vice versa. Moreover, the gradient of $v$ tends to be much shallower than $u$ due to the faster diffusion of $v$.

\begin{figure}[ht]
    \centering
    \begin{subfigure}[h]{0.33\textwidth}
        \centering
        \includegraphics[width=\textwidth]{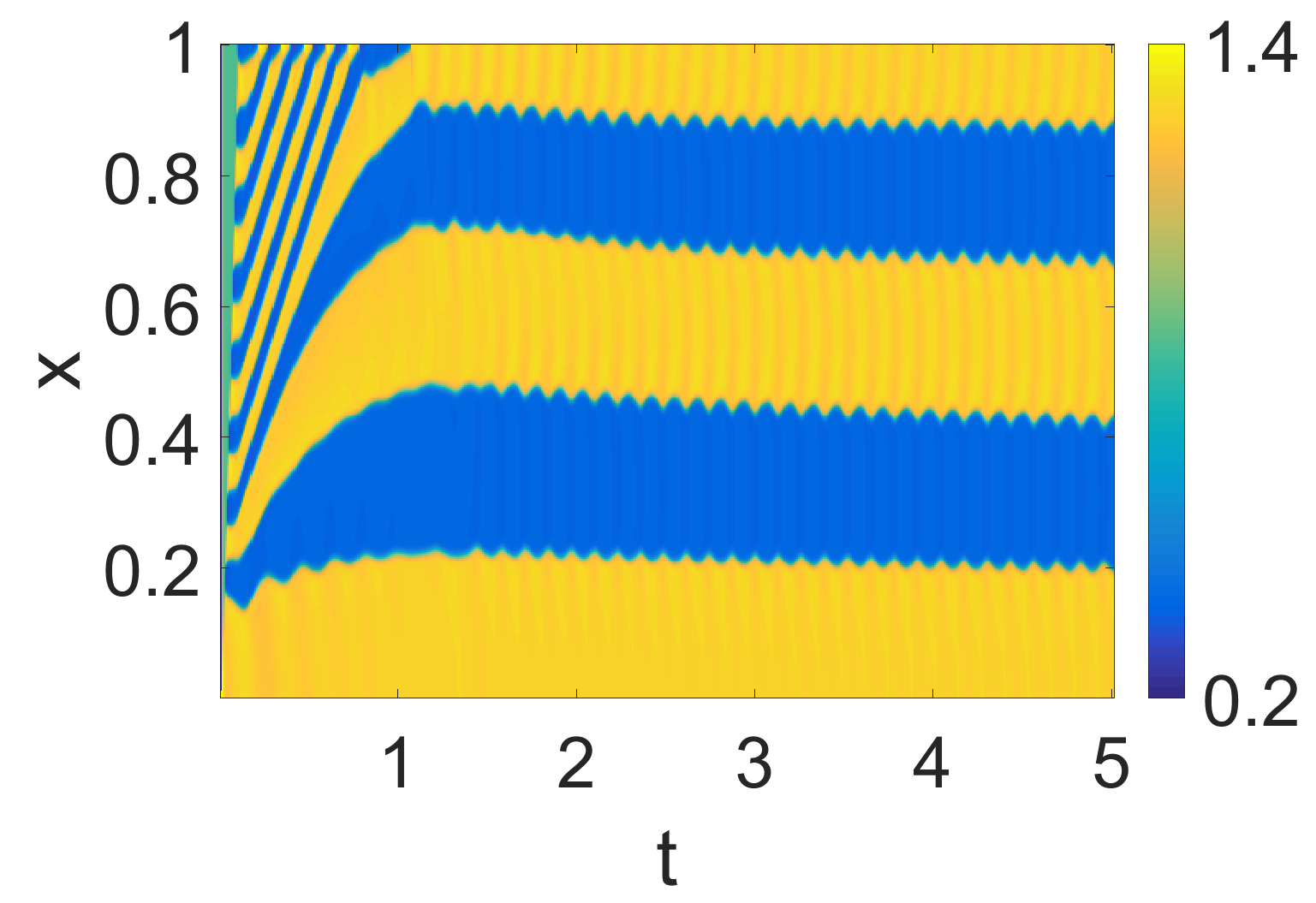}
        \caption{u}
    \end{subfigure}%
    \begin{subfigure}[h]{0.33\textwidth}
        \centering
        \includegraphics[width=\textwidth]{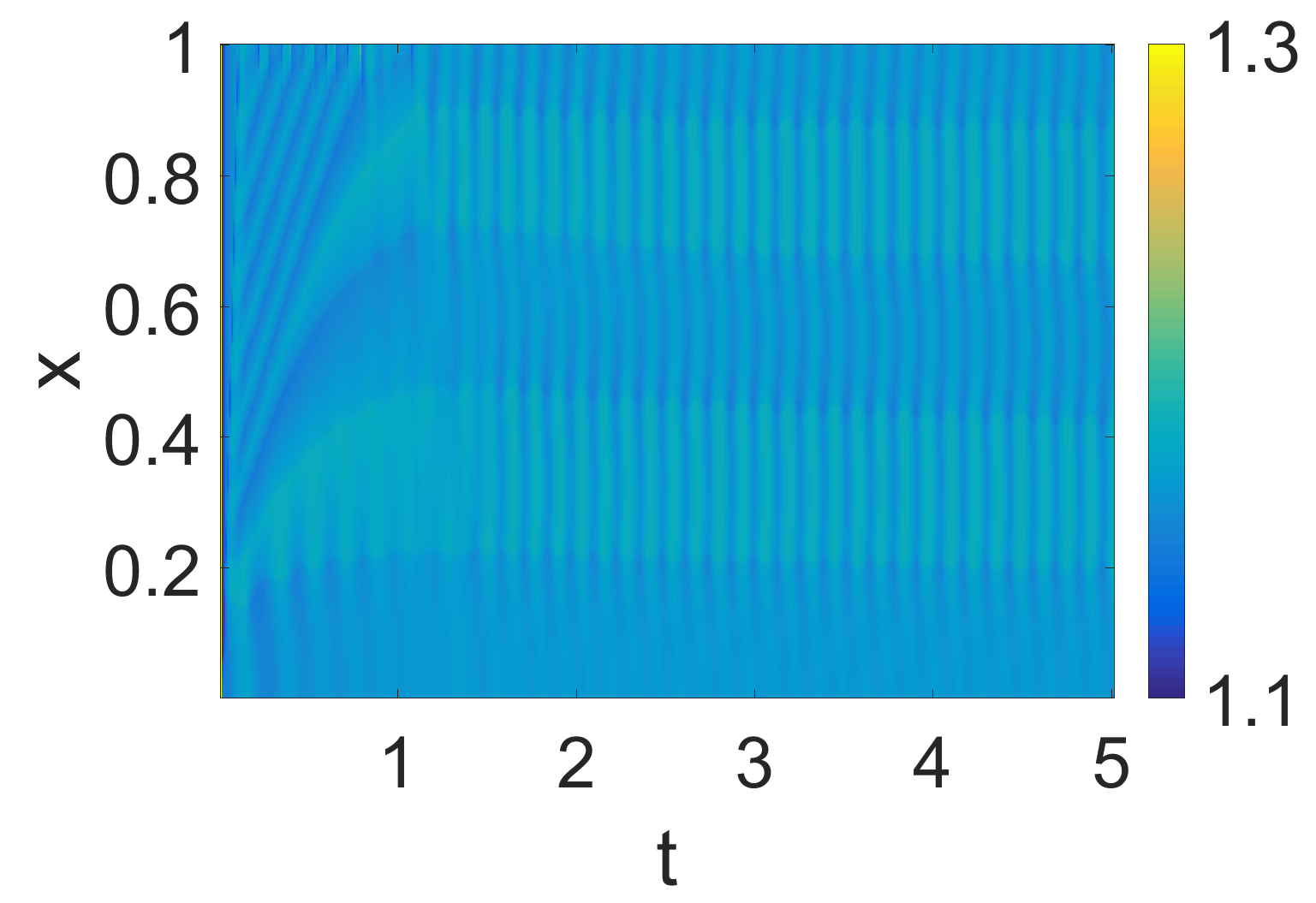}
        \caption{v}
    \end{subfigure}%
    \begin{subfigure}[h]{0.33\textwidth}
        \centering
        \includegraphics[width=\textwidth]{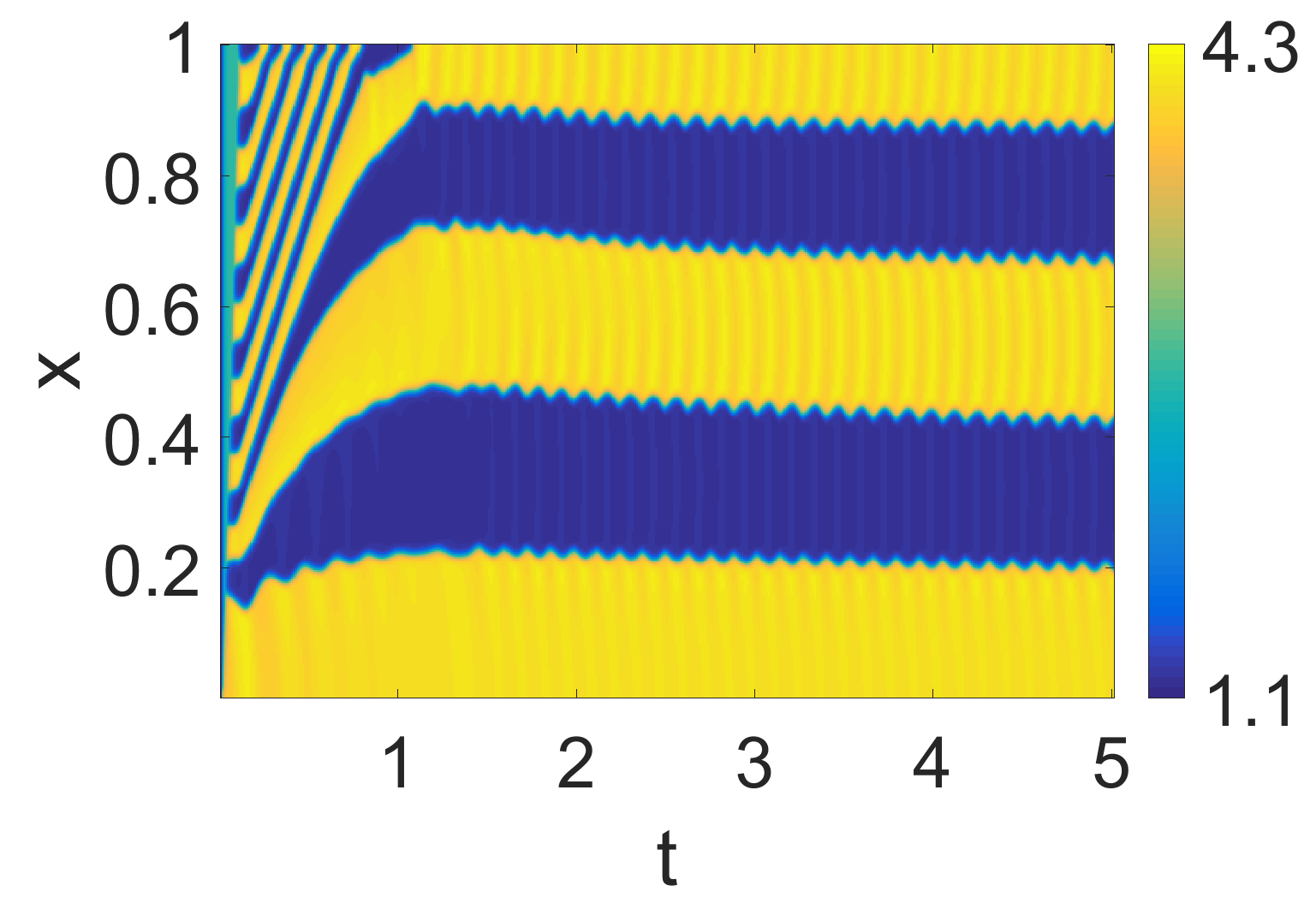}
        \caption{F}
    \end{subfigure}
    \begin{subfigure}[h]{0.33\textwidth}
        \centering
        \includegraphics[width=\textwidth]{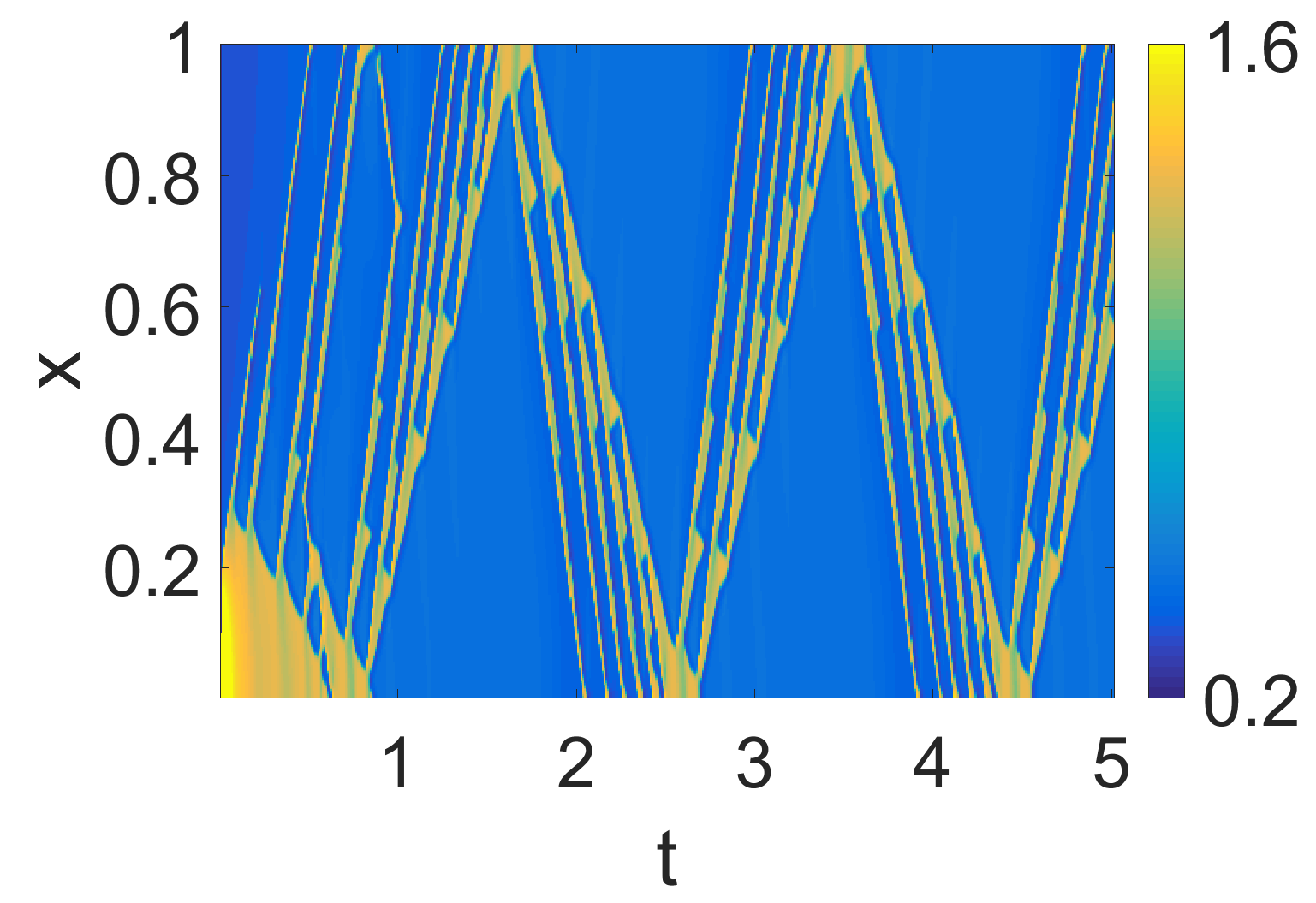}
        \caption{u}
    \end{subfigure}%
    \begin{subfigure}[h]{0.33\textwidth}
        \centering
        \includegraphics[width=\textwidth]{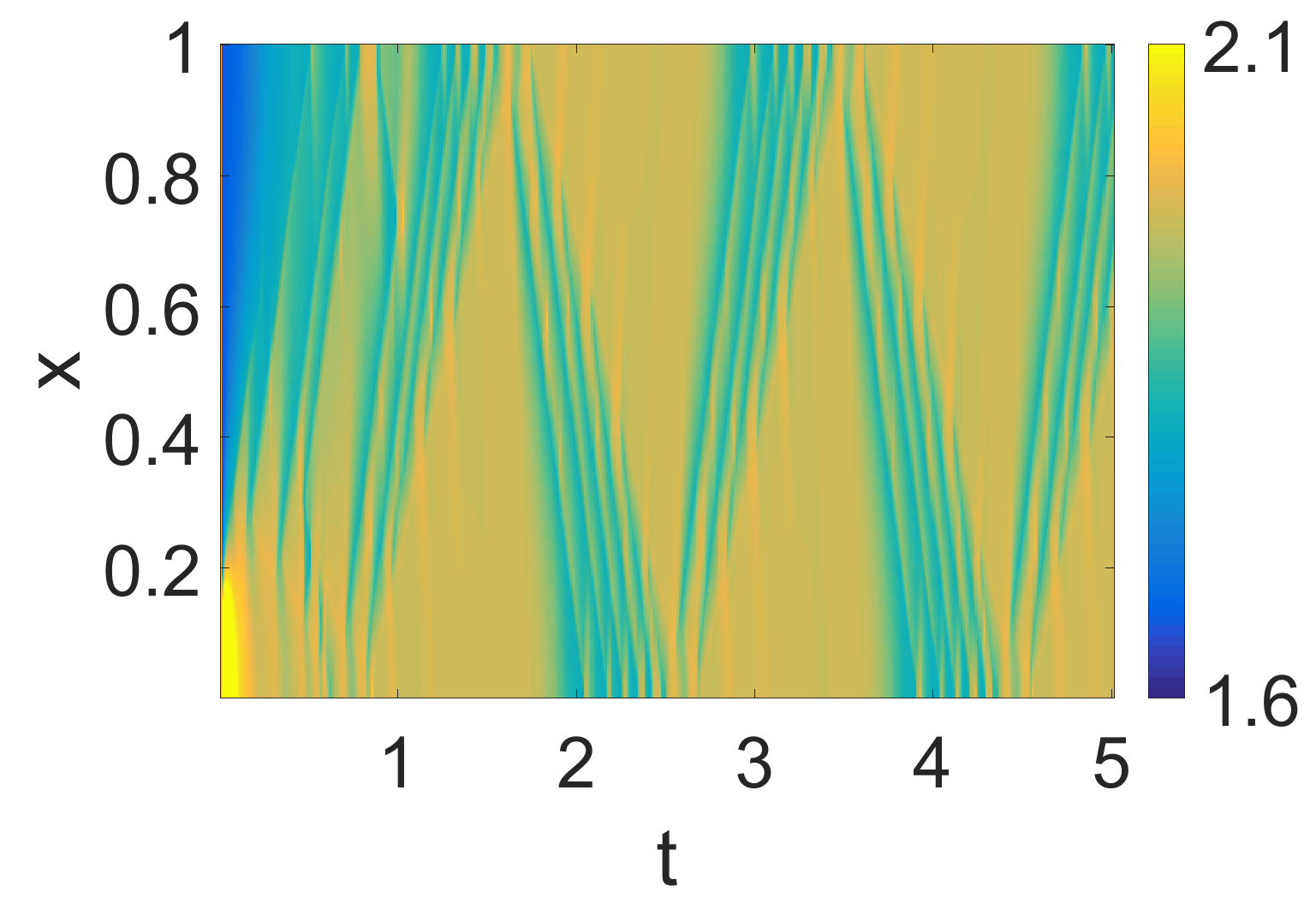}
        \caption{v}
    \end{subfigure}%
    \begin{subfigure}[h]{0.33\textwidth}
        \centering
        \includegraphics[width=\textwidth]{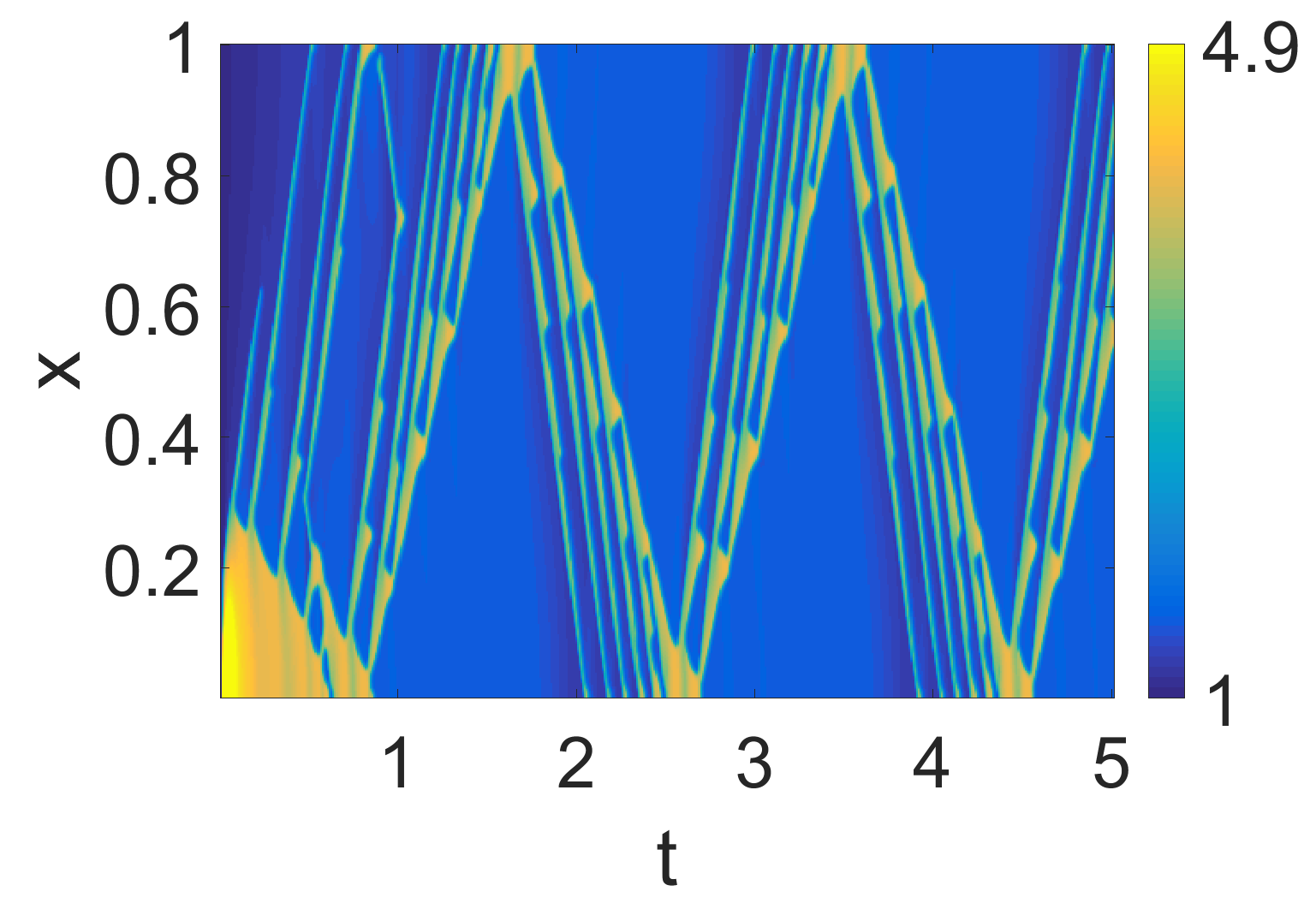}
        \caption{F}
    \end{subfigure}
    \caption[Exotic patterns observed under the actin feedback model]{Exotic patterns observed in the actin feedback (AF) model. %\eqref{sys:actinwave} 
    Parameters as in Table~\ref{tab:parameters}(AF) but varying $k,s$. (a-c) $k=5, s=10$, default initial conditions. The pattern resembles WPO but with several subregions; (d-f) $k=5, s=30$, default initial conditions with excitation region $0\leq x \leq 0.1$. The resulting pattern is similar to RW but with a group of four pulses traversing the domain. }
    \label{fig:sim_actin_weird}
\end{figure}

Some other more complex patterns are shown in Fig.~\ref{fig:sim_actin_weird}. These share some characteristics with the simpler patterns. The patterns shown in (a-c) are similar to  (WPO), but the domain is divided into five regions instead of two, with an initial transient reminiscent of (WT). The patterns in (d-f) can be seen as a group of four reflecting pulses (similar to RW) rather than one. 
Compared to \cite{holmes2012regimes}, we find a richer range of patterns using a similar parameter set (with different scaling). The main difference is that the larger domain used here, $L=10$, allows more space for pattern to develop. (In \cite{holmes2012regimes}, $L=1$, so patterns are more confined and boundary effects are prominent.)

\begin{figure}[htbp]
    \centering
    \begin{subfigure}[h]{0.5\textwidth}
        \centering
        \caption{u}
        \includegraphics[width=\textwidth]{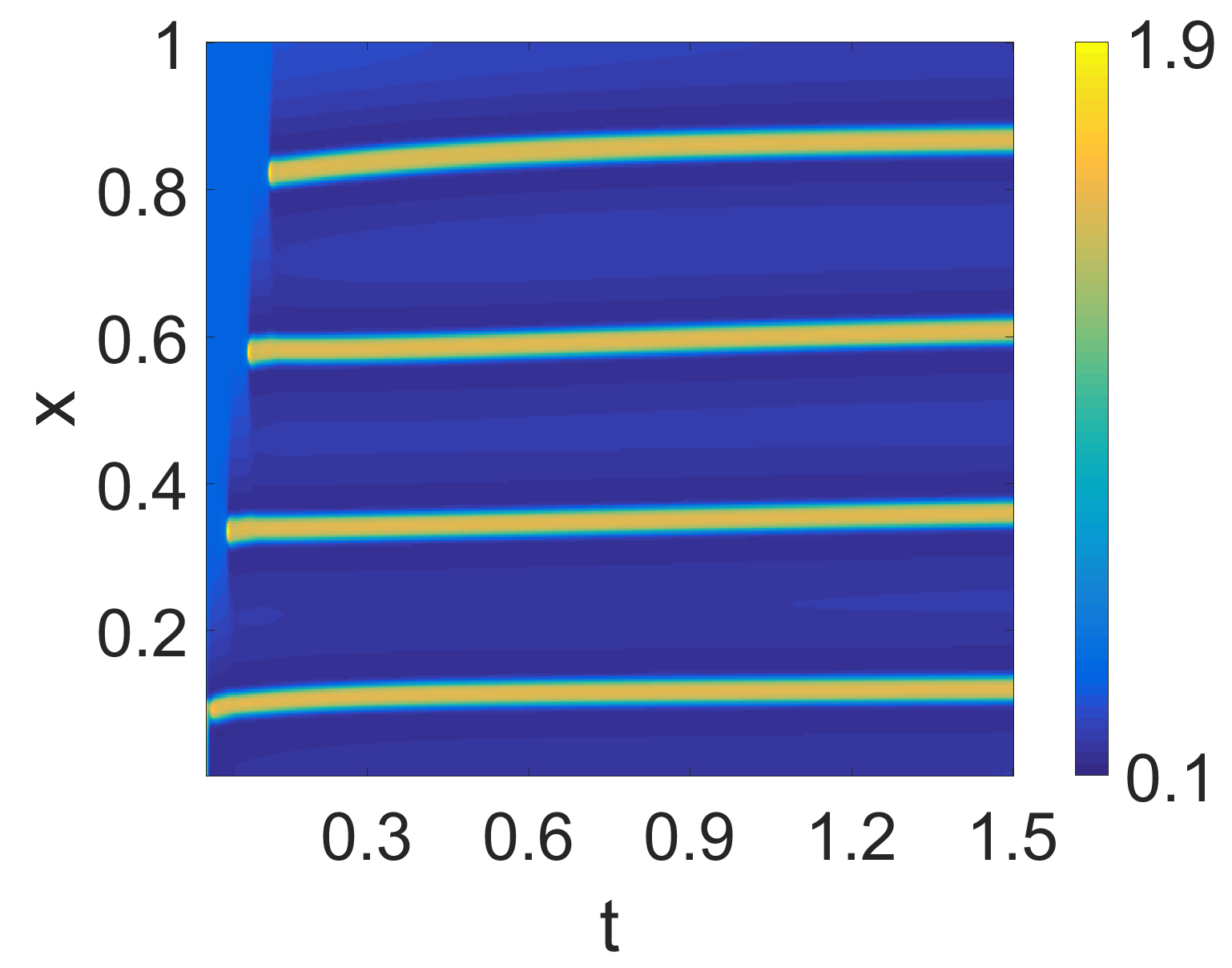}
    \end{subfigure}~
    \begin{subfigure}[h]{0.5\textwidth}
        \centering
        \caption{v}
        \includegraphics[width=\textwidth]{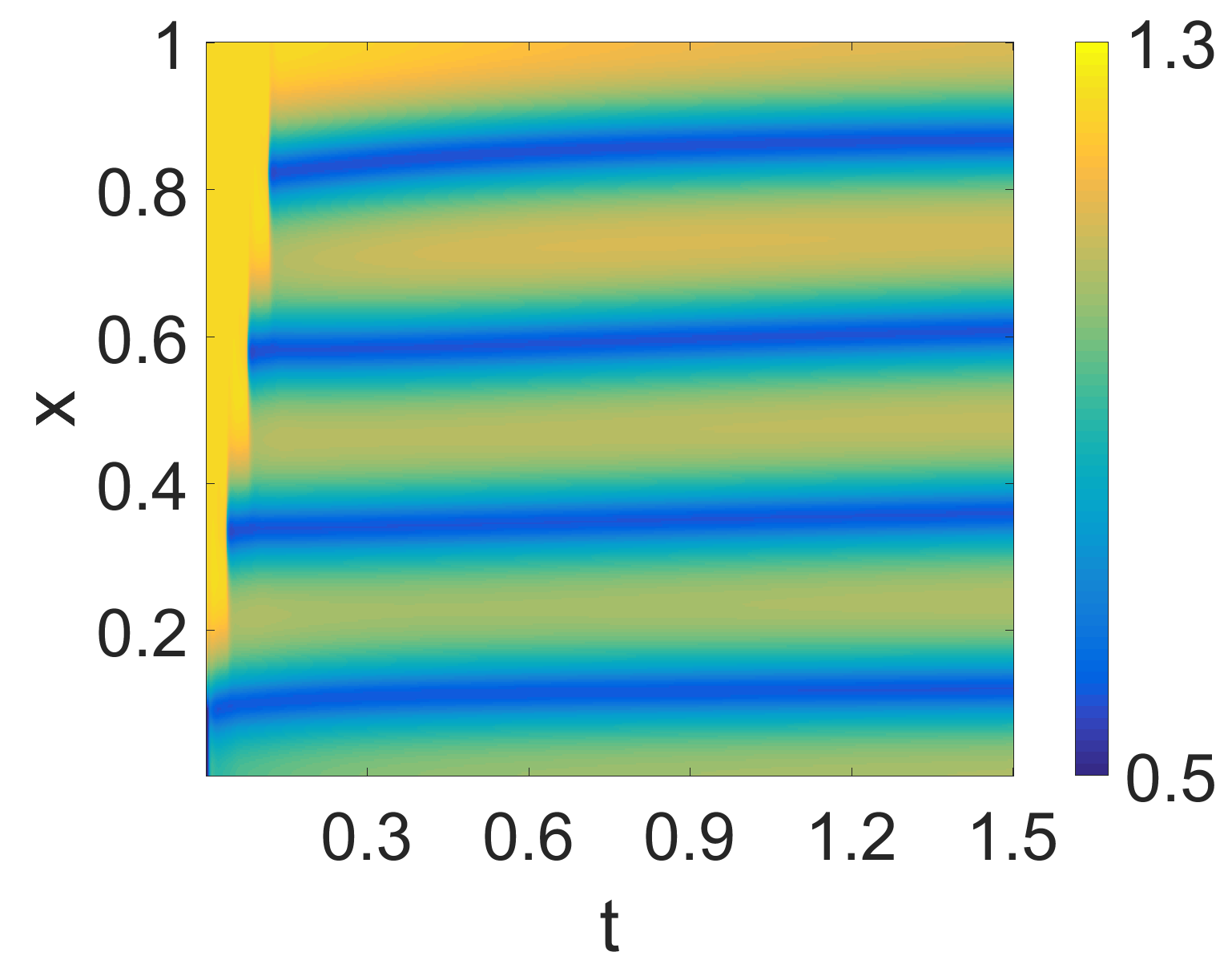}
    \end{subfigure}
    \begin{subfigure}[h]{0.5\textwidth}
        \centering
        \caption{u}
        \includegraphics[width=\textwidth]{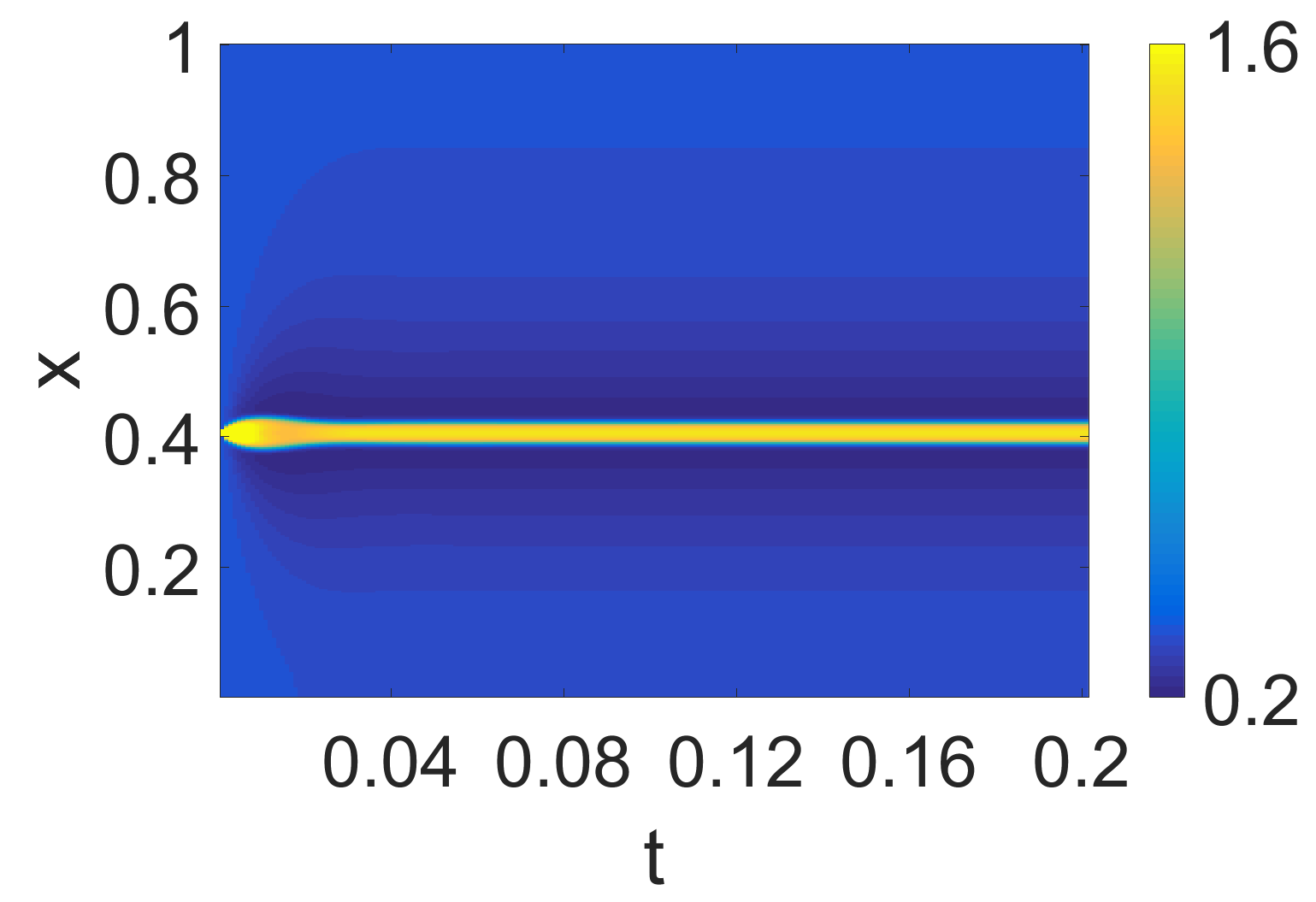}
    \end{subfigure}~
    \begin{subfigure}[h]{0.5\textwidth}
        \centering
        \caption{v}
        \includegraphics[width=\textwidth]{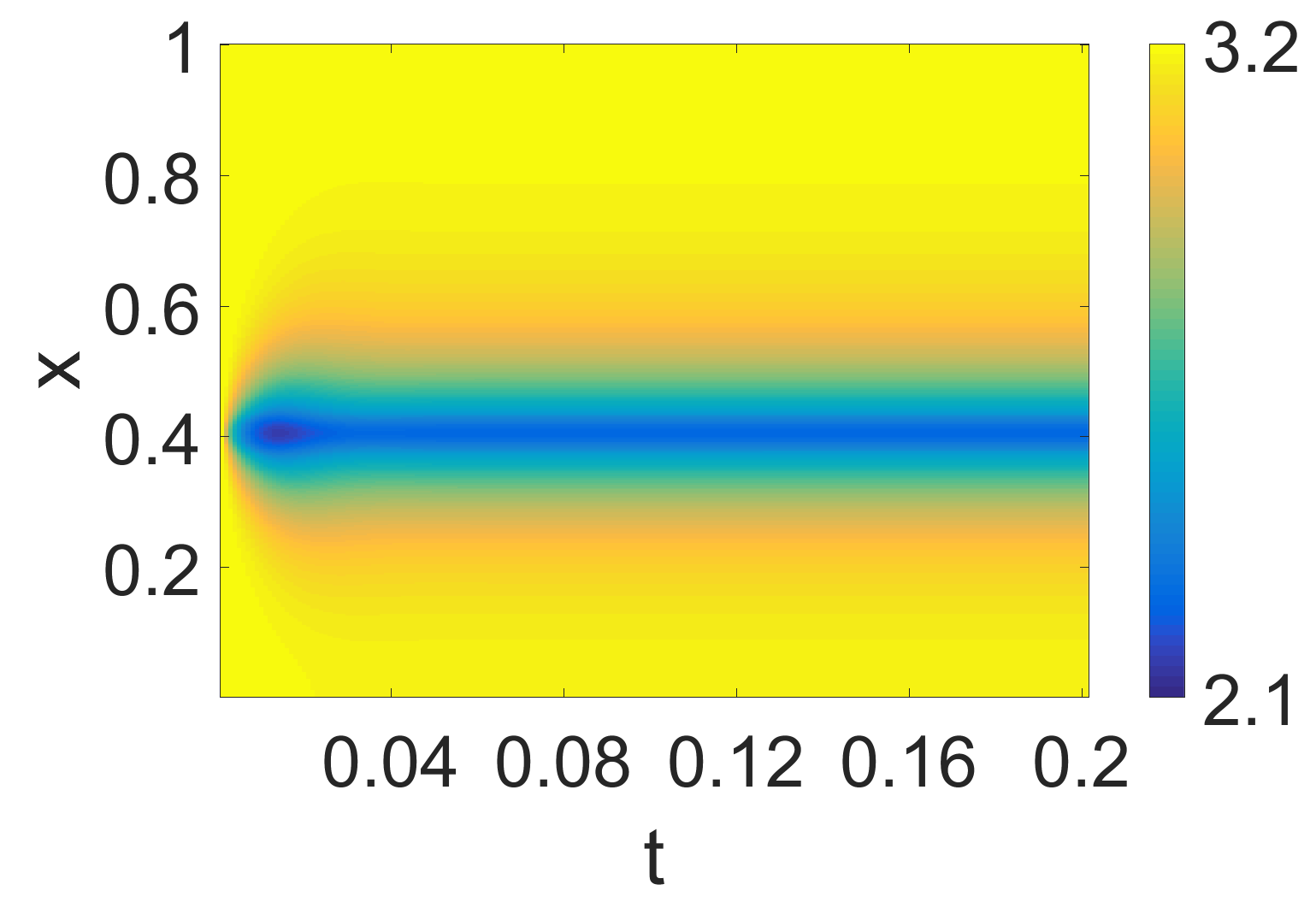}
    \end{subfigure}
	\caption[Numerical simulations of the non-conservative model]{Simulation of the non-conservative model (NC)
	%\eqref{sys:champ} 
	with (a,b) default parameters (Table~\ref{tab:parameters}(NC)); (c,d) $\gamma=15L^2, \eta=15L^2$. Initial condition: (a,b) $u=u_*$ except $u=1$ on $0 \leq x \leq 0.1$, $v=v_*$. (c,d) $u=u_*=0.33333$ except $u=10u_*$ on $0.4\leq x \leq 0.41$, $v=v_*=3.19298$. In (a,b), the formation of a peak on the left  triggers some new peaks farther away, until space runs out. Once all peaks form, they shift slightly to be evenly spaced. In (c,d), the single initial peak persists, without triggering new peaks. We refer to this as the soliton solution. }
	\label{fig:sim_champ}
\end{figure}

\begin{figure}[ht]
    \centering
    \begin{subfigure}[h]{0.5\textwidth}
        \centering
        \caption{Default parameters}
        \includegraphics[width=\textwidth]{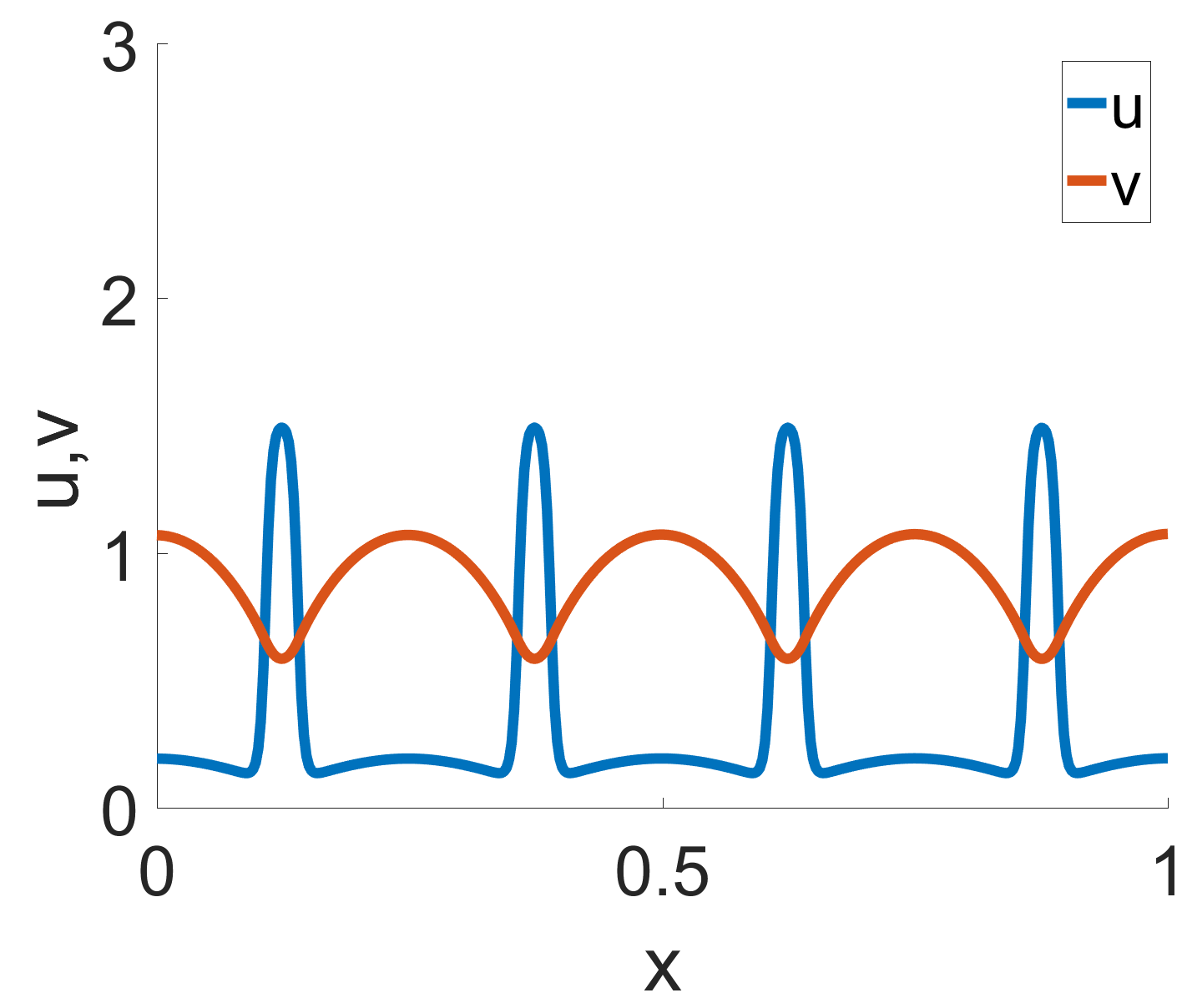}
    \end{subfigure}~
    \begin{subfigure}[h]{0.5\textwidth}
        \centering
        \caption{$\gamma=15L^2$}
        \includegraphics[width=\textwidth]{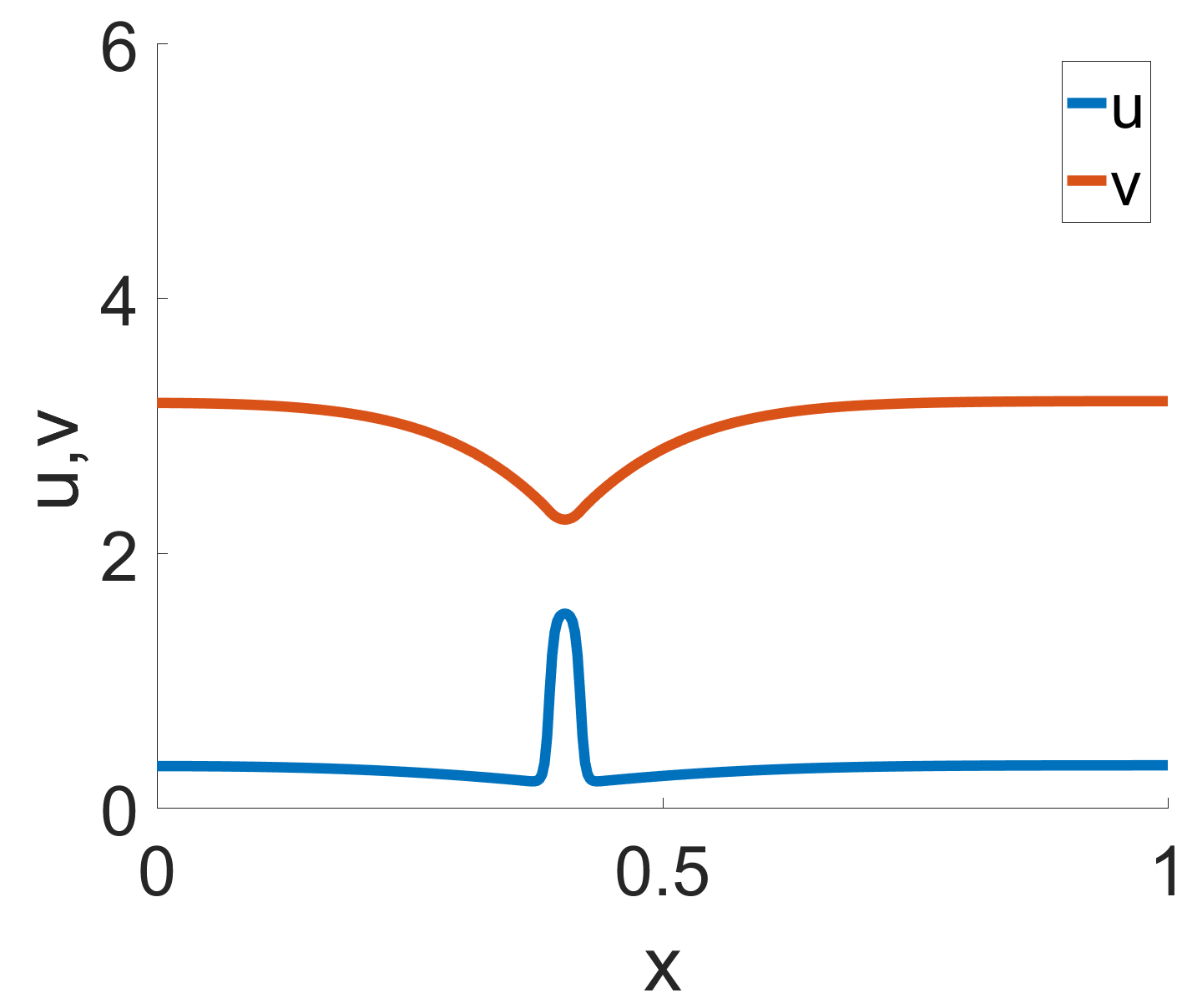}
    \end{subfigure}
    \begin{subfigure}[h]{0.5\textwidth}
        \centering
        \caption{$L=1$}
        \includegraphics[width=\textwidth]{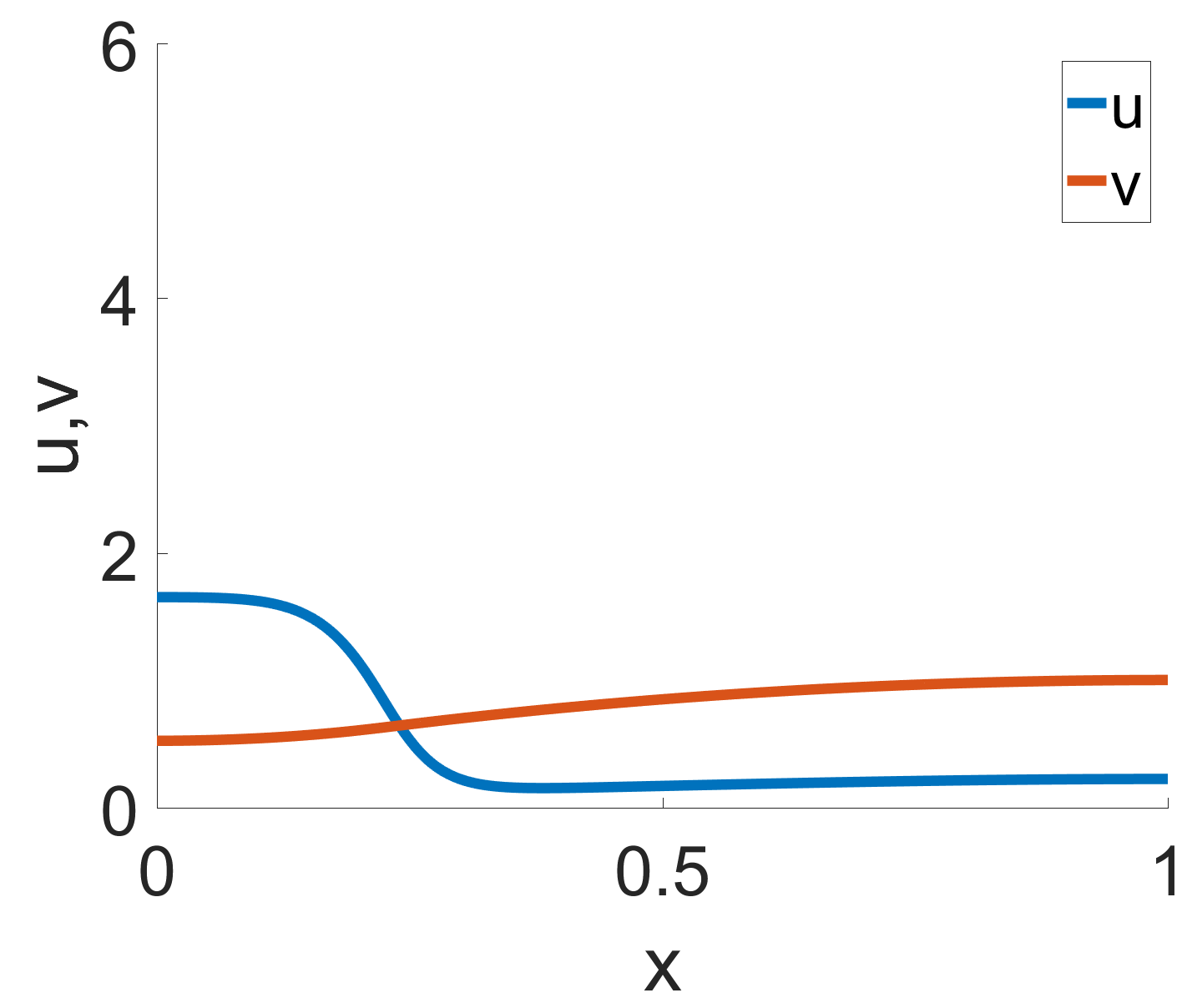}
    \end{subfigure}~
    \begin{subfigure}[h]{0.5\textwidth}
        \centering
        \caption{$\eta=15$}
        \includegraphics[width=\textwidth]{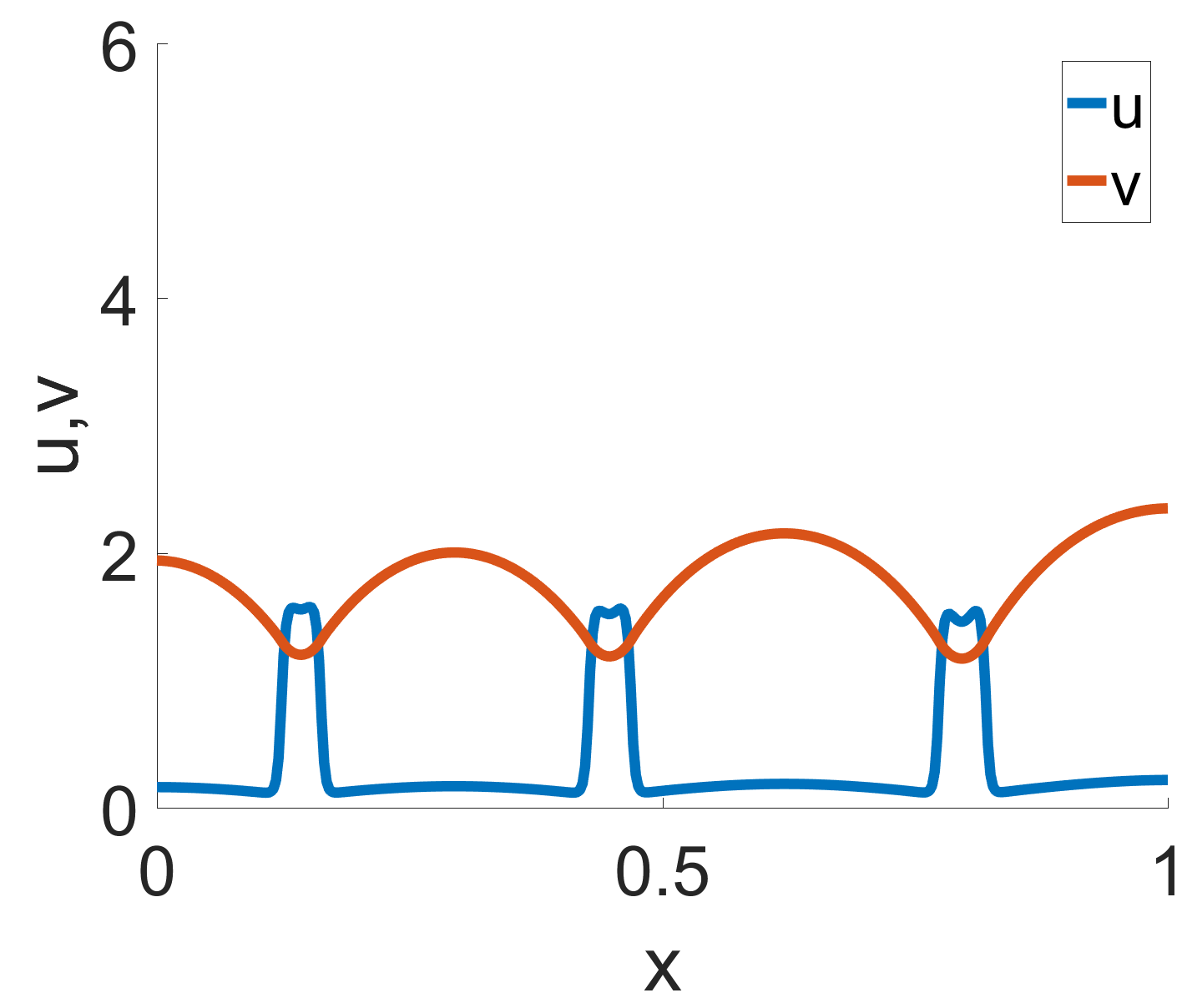}
    \end{subfigure}
	\caption[Steady state pattern of the non-conservative model]{Final steady state pattern of the non-conservative model (NC) %\eqref{sys:champ} 
	with most parameters from Table~\ref{tab:parameters}(NC), except the parameters indicated on the labels. (a) and (b) correspond to the steady state of Fig.~\ref{fig:sim_champ}(a,b) and (c,d) respectively. In (c) the shortened domain results in wave pinning; (d) Higher inactivation rate $\eta=15$ results in bifurcating peaks. (a,b) corresponds to Fig.~5 (a,d) of \cite{Champneys}, respectively.}
	\label{fig:sim_champ2}
\end{figure}

Fig.~\ref{fig:sim_champ} shows two typical patterns in the NC model: a static, Turing-type pattern consisting of a series of evenly spaced spikes, and a single spike ``soliton" pattern. The final profiles of these two patterns are shown in Fig.~\ref{fig:sim_champ2}(a,b). The domain length, $L$, must be large enough to support such patterns. If $L$ is too small to support a full period of the pattern, the result would be  simple polarization similar to wave pinning (Fig.~\ref{fig:sim_champ2}(c)). Using a higher rate of inactivation $\eta$, or a smaller diffusion ratio $\delta$ can result in spikes that split into two, as shown in Fig.~\ref{fig:sim_champ2}(d).

\begin{figure}[ht]
    \centering
    \begin{subfigure}[h]{0.5\textwidth}
        \centering
        \caption{$s=5$}
        \includegraphics[width=\textwidth]{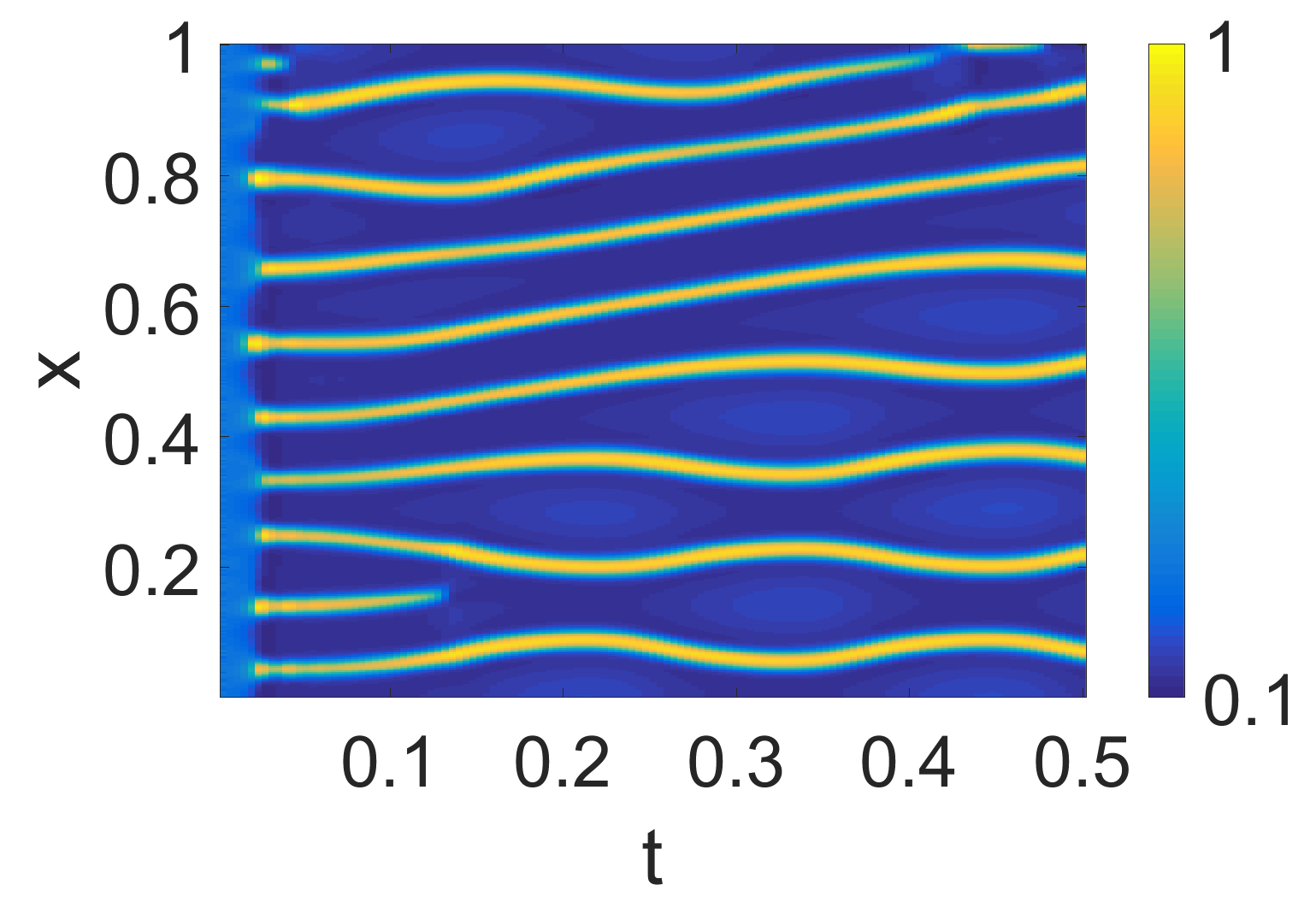}
    \end{subfigure}~
    \begin{subfigure}[h]{0.5\textwidth}
        \centering
        \caption{$s=8$}
        \includegraphics[width=\textwidth]{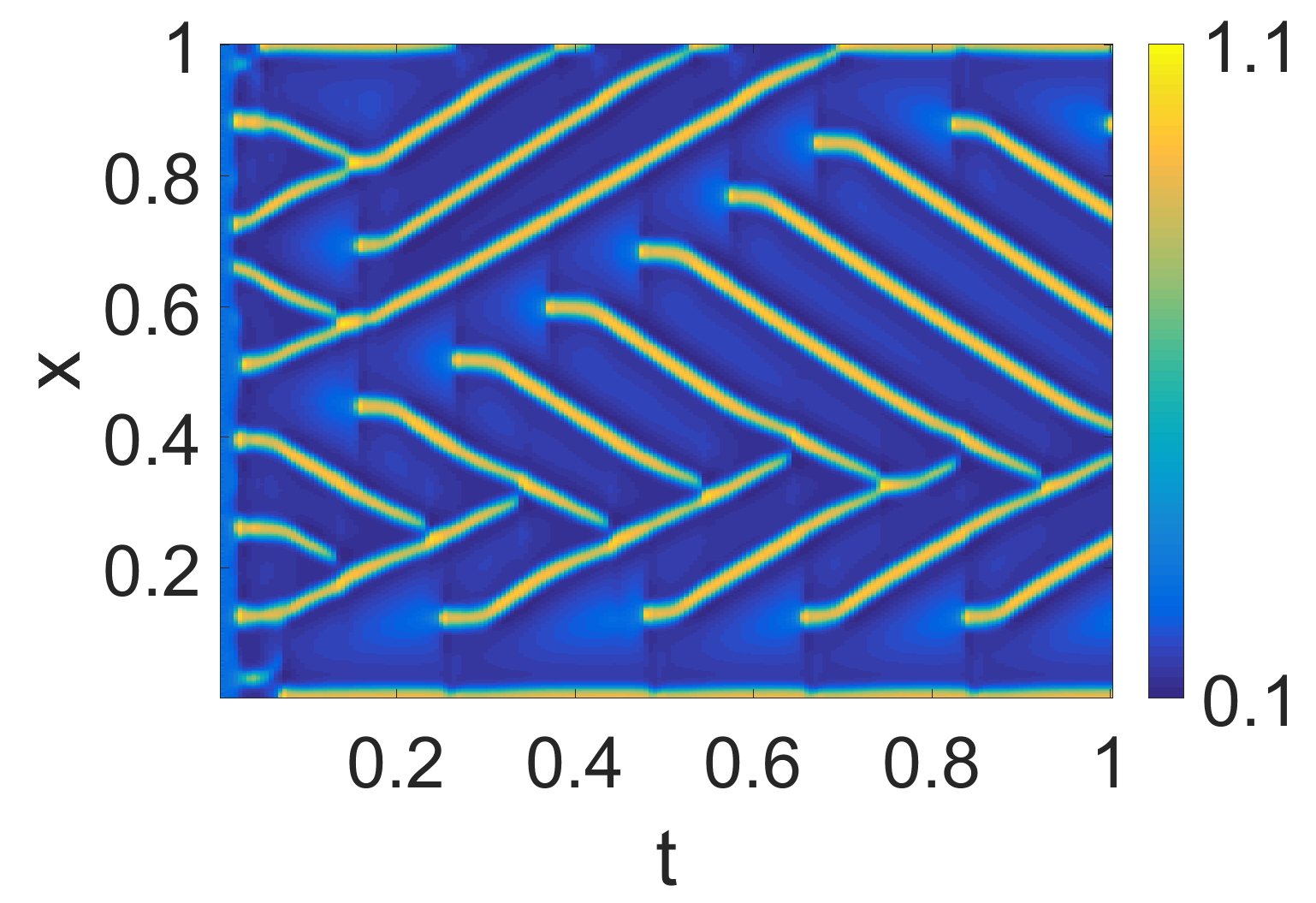}
    \end{subfigure}
    \begin{subfigure}[h]{0.5\textwidth}
        \centering
        \caption{$s=18$}
        \includegraphics[width=\textwidth]{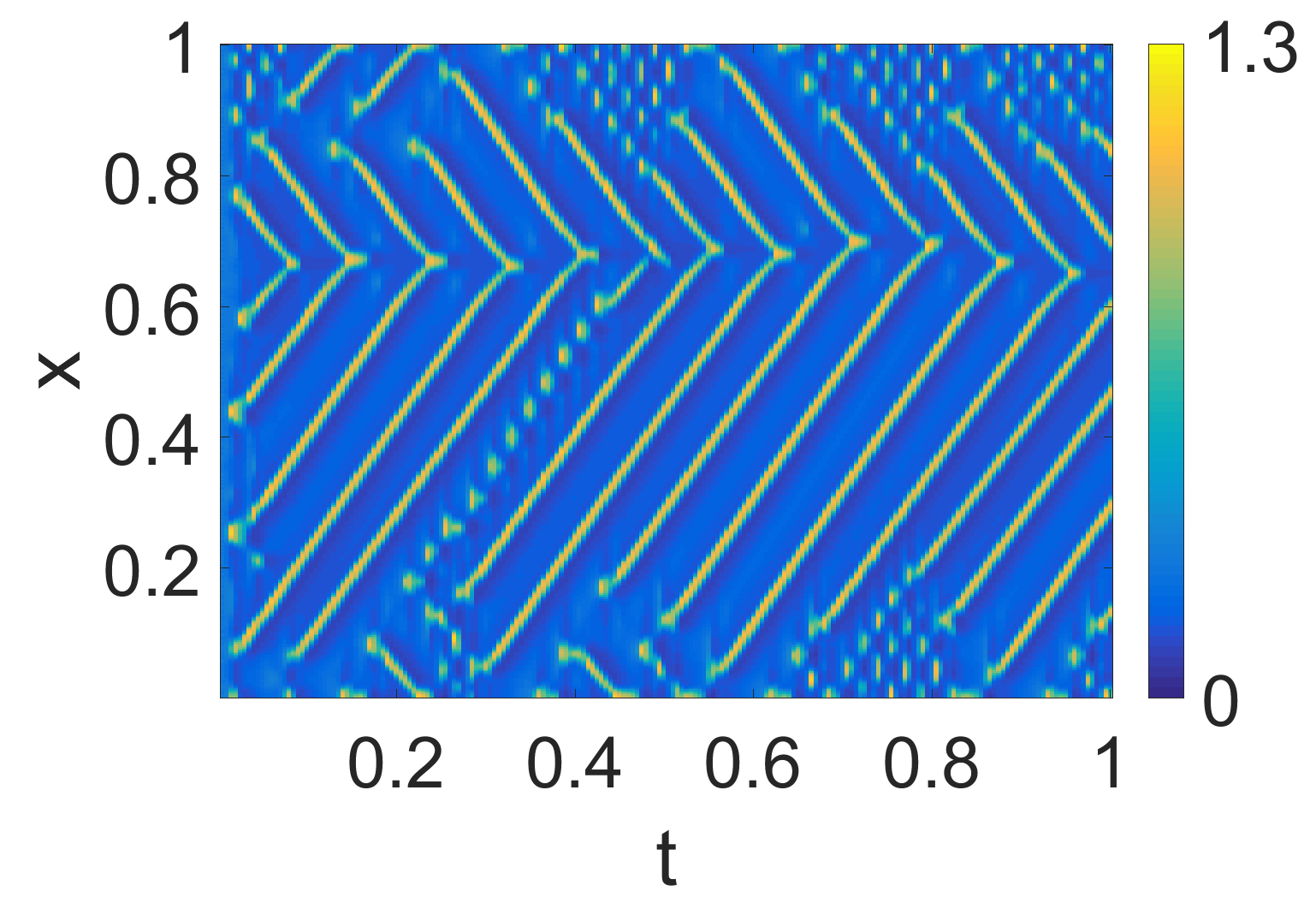}
    \end{subfigure}~
    \begin{subfigure}[h]{0.5\textwidth}
        \centering
        \caption{$s=35$}
        \includegraphics[width=\textwidth]{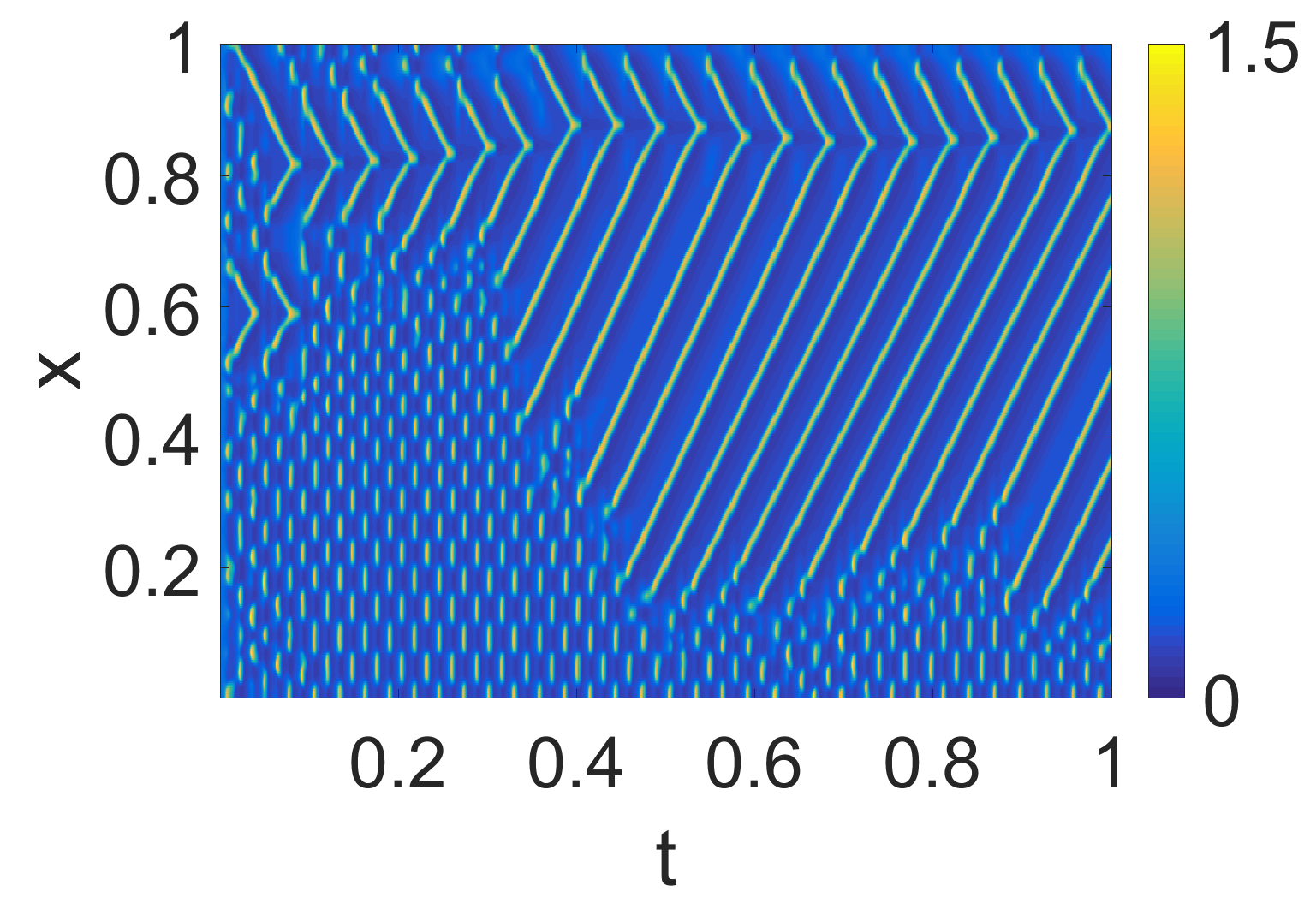}
    \end{subfigure}
	\caption[Numerical simulations of the combined model (2)]{Simulations of the combined model (CM). 
	%\eqref{sys:combined} 
	Parameters as in Table~\ref{tab:parameters}(CM2) but varying $s$, and HSS+noise initial condition as described in text. As we increase the actin feedback strength $s$, the behavior transitions from slowly moving, repelling peaks to colliding peaks. At higher $s$, there is a rapidly oscillating standing wave pattern in some parts of the domain.}
	\label{fig:sim_combined2}
\end{figure}

For the combined model, we use parameters from Table~\ref{tab:parameters}(CM2), mostly similar to the NC case. In Fig.~\ref{fig:sim_combined2}, we show the effect of increasing $s$ (strength of actin feedback) on system behavior. 
With $s$ low enough, the system behavior resembles the $s=0$ case of a static, spatially periodic pattern, as in the NC case. For increasing $s$, the peaks begin to move with constant velocity by themselves, repelling one another when too close. For moderate values of $s$, the peak repulsion is strong enough that peaks reverse their direction of motion if on a collision course (Fig.~\ref{fig:sim_combined2}(a)). At higher $s$, they collide (Fig.~\ref{fig:sim_combined2}(b)). At even higher $s$, we observe a localized standing wave pattern that oscillates rapidly in Fig.~\ref{fig:sim_combined2}(c), and even more prominently in (d).

%%%%%%%%%%%%%%%%
%%%%%%%%%%%%%%%%%
\subsection{Simulations in a fixed 2D domain}

%\textcolor{red}{Put these two links into the appropriate figure captions}
%\mycomment{Actin waves 2D animations: \url{https://imgur.com/a/61GwiA9}}

%\mycomment{Combined model 2D animations: \url{https://imgur.com/a/a0u57GQ}}

\begin{figure}[ht]
    \centering
    \begin{subfigure}[h]{0.74\textwidth}
        \centering
        \caption{Wave pinning, $t=1.25$}
        \includegraphics[width=\textwidth]{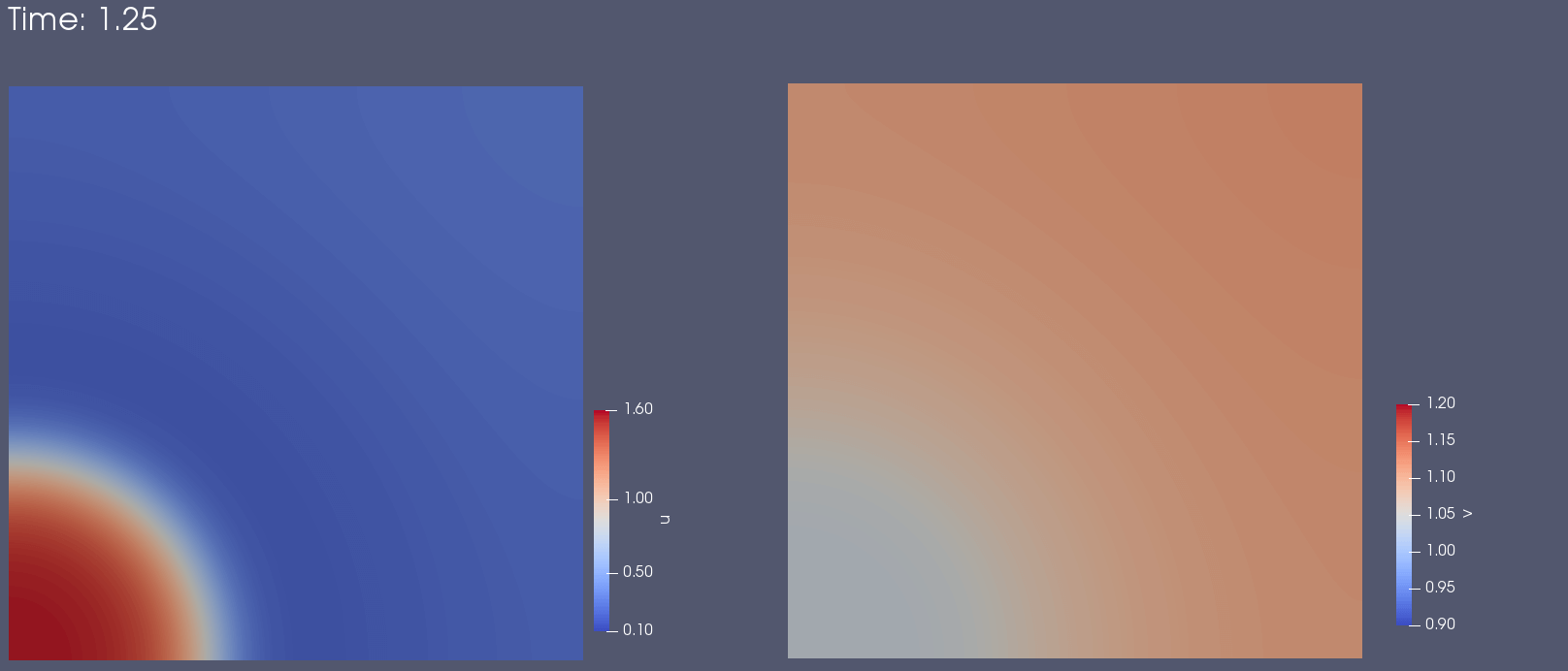}
    \end{subfigure}
    \begin{subfigure}[h]{0.74\textwidth}
        \centering
        \caption{Wave pinning, $t=30$}
        \includegraphics[width=\textwidth]{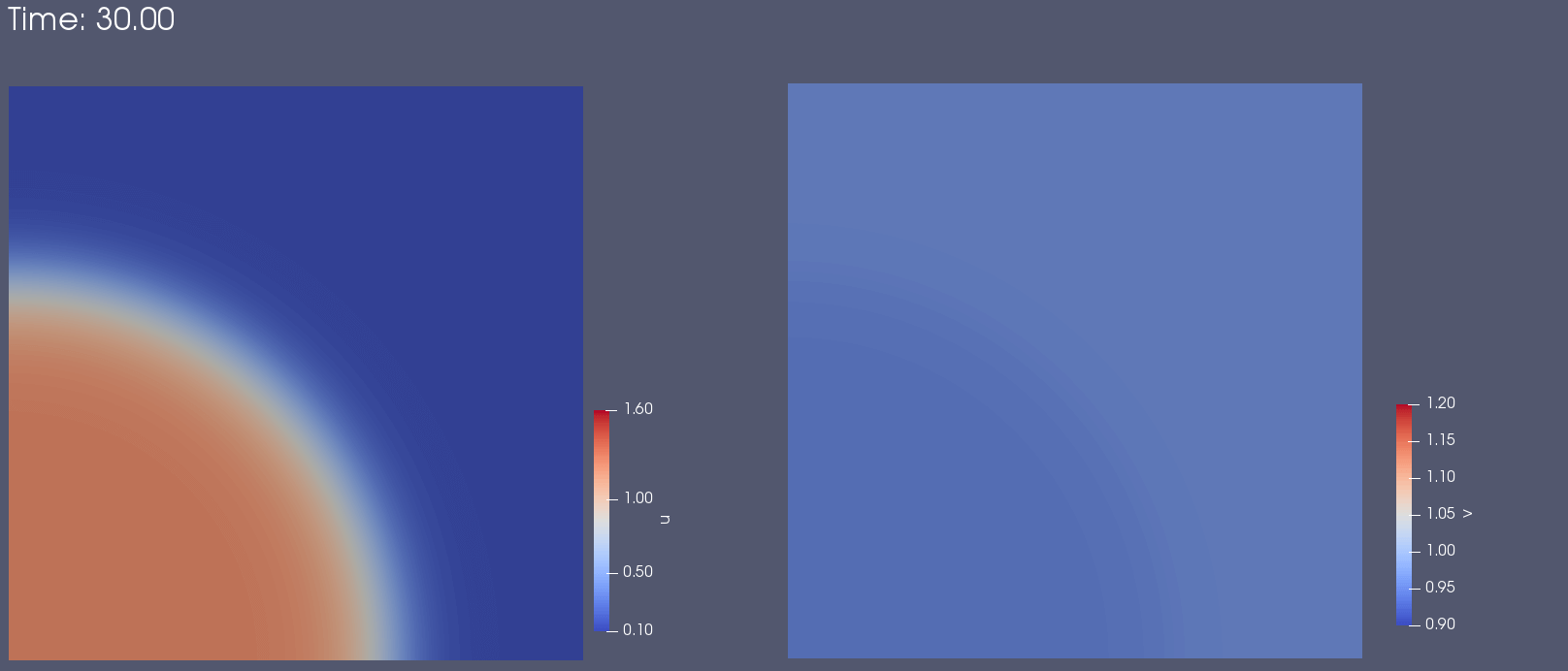}
    \end{subfigure}
    \begin{subfigure}[h]{0.74\textwidth}
        \centering
        \caption{Non-conservative extension, $t=0.145$}
        \includegraphics[width=\textwidth]{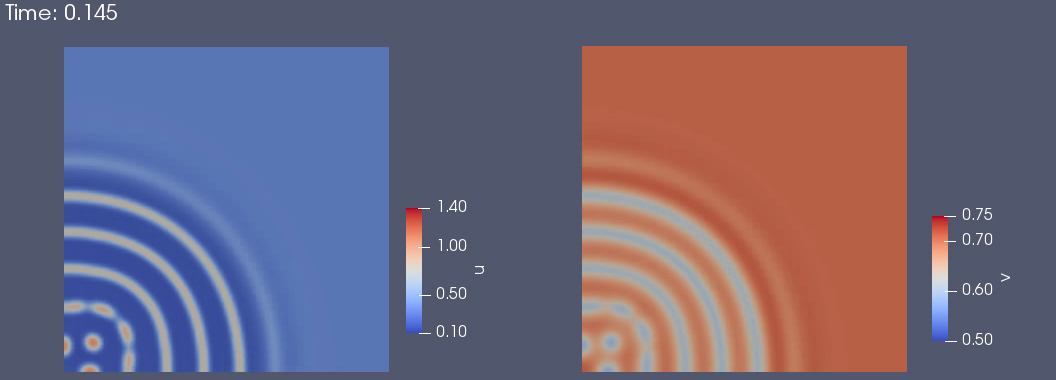}
    \end{subfigure}
    \begin{subfigure}[h]{0.74\textwidth}
        \centering
        \caption{Non-conservative extension, $t=0.994$}
        \includegraphics[width=\textwidth]{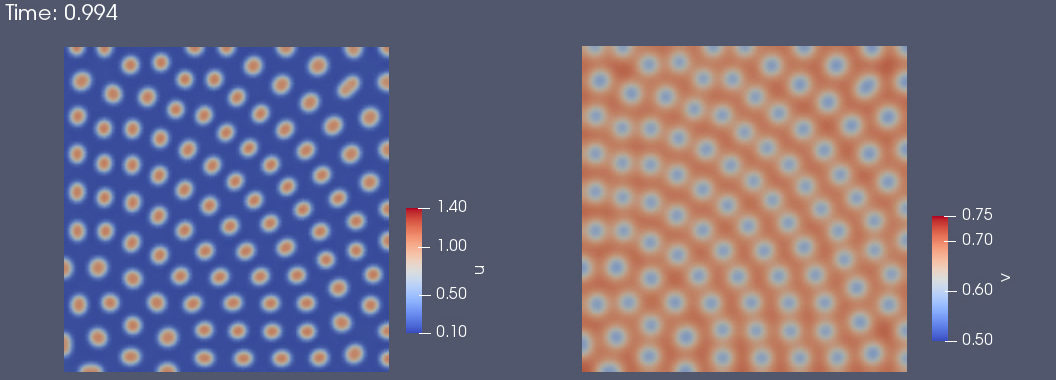}
    \end{subfigure}
	\caption[2D simulations of the wave pinning and non-conservative model]{2D simulations of the wave pinning (a,b) and non-conservative (c,d) models, using the same parameters as in 1D (Table~\ref{tab:parameters}(WP) and (NC)). Left: $u$ , Right: $v$. For each model, two snap shots are shown: one when the pattern begin to take shape, and another after the system reached steady state.}
	\label{fig:sim_2d}
\end{figure}

In two spatial dimensions, we use the same parameters as in 1D. For the WP model, we start at HSS and perturb one corner of the domain. The pattern we observe (Fig.~\ref{fig:sim_2d}(a,b)) is a direct analogue to the 1D case (compare to Fig.~\ref{fig:sim_wavepin}(a,b)): $u$ initially spreads out from the corner as a 2D travelling wave, and that is eventually pinned along a front determined by the initial conditions.

We use a similar initial condition for the NC model. Based on 1D simulations, we expect evenly-spaced stripes to form around the corner as concentric rings, as happens initially (Fig.~\ref{fig:sim_2d}(c)). However, these rings quickly break up into spots (Fig.~\ref{fig:sim_2d}(d)). The spots spread out, and then settle into a steady state. We have not found any parameter sets for stable ring patterns.
The patterns are insensitive to the shape of the domain. Simulations on circular, rectangular and other domains with simple shapes produced patterns with the same qualitative characteristics (not shown).

%\textcolor{red}{I am not sure if this is a generic issue or something that you had issues with that we don't understand}
%For the AF model, we found that the patterns are not very robust in 2D and it is difficult to obtain patterns starting from simple initial conditions. To solve this problem, we start our simulations with a initial condition of HSS plus global noise and $c=1$ (i.e. the combined model, which is much more robust), wait long enough for a pattern to form, then gradually decrease $c$ to 0. In this case, the patterns persists. Fig.~\ref{fig:sim_2d_actin} shows a few snapshot of the simulations. For low $s$, the pattern resembles slowly drifting and deforming blobs. As $s$ increases, the pattern transitions into spiral waves with decreasing width. 

For CM, we initialize the system at HSS and perturb with noise. Fig.~\ref{fig:sim_2d_combined} shows the simulation results. With a low $s$, the pattern is indistinguishable from the static spots under the non-conservative model. As $s$ increases, the spots become mobile and repel each other as in the 1D case. In 2D, as $s$ is increased further, the spots transitions to spiral waves.

\begin{figure}[ht]
    \centering
    \begin{subfigure}[h]{\textwidth}
        \centering
        \caption{$s=12$}
        \includegraphics[width=\textwidth]{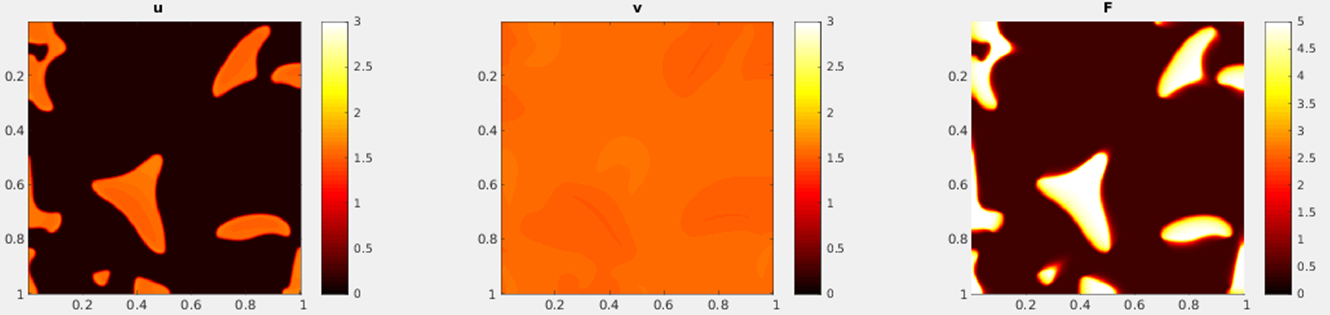}
    \end{subfigure}
    \begin{subfigure}[h]{\textwidth}
        \centering
        \caption{$s=18$}
        \includegraphics[width=\textwidth]{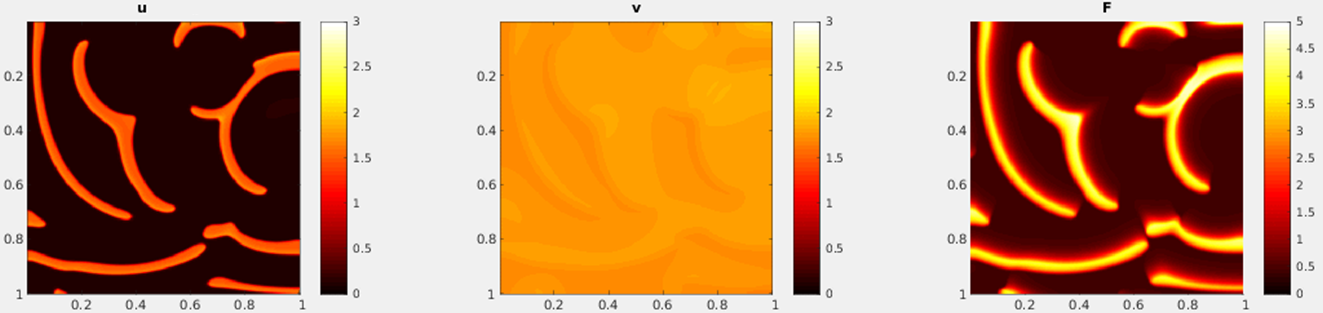}
    \end{subfigure}
    \begin{subfigure}[h]{\textwidth}
        \centering
        \caption{$s=27$}
        \includegraphics[width=\textwidth]{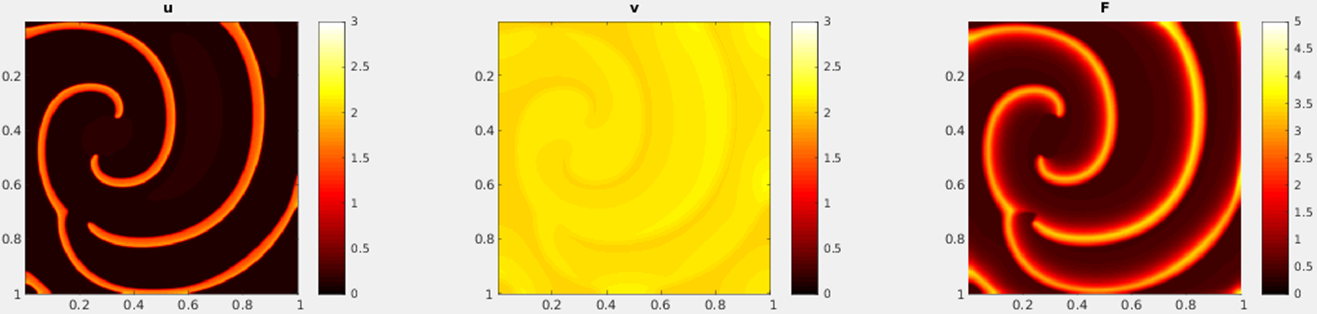}
    \end{subfigure}
    \begin{subfigure}[h]{\textwidth}
        \centering
        \caption{$s=36$}
        \includegraphics[width=\textwidth]{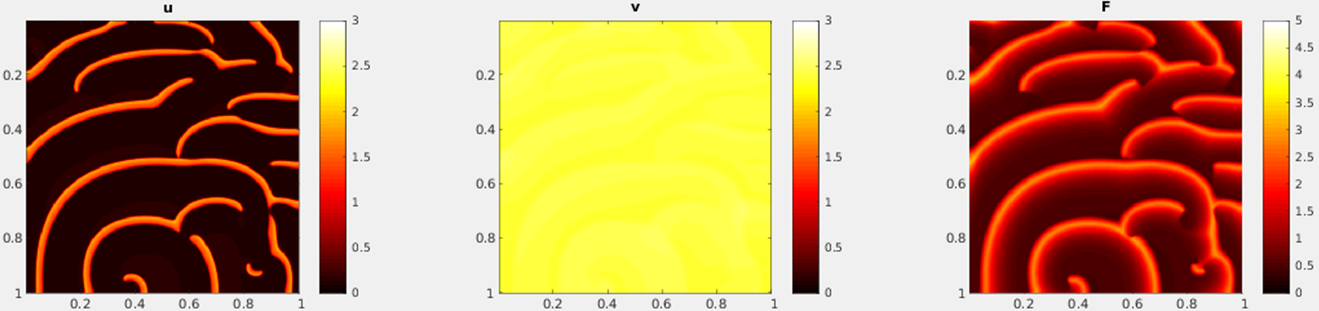}
    \end{subfigure}
	\caption{2D simulations of the actin feedback (AF) model, with parameters from Table~\ref{tab:parameters}(AF) and initial conditions described in the text. These snapshots are taken after the patterns have fully developed. As $s$ increases, blobs transitions into thinner and thinner spiral waves. See movies at \url{https://imgur.com/a/61GwiA9}.}
	\label{fig:sim_2d_actin}
\end{figure}

\begin{figure}[ht]
    \centering
    \begin{subfigure}[h]{\textwidth}
        \centering
        \caption{$s=8$}
        \includegraphics[width=\textwidth]{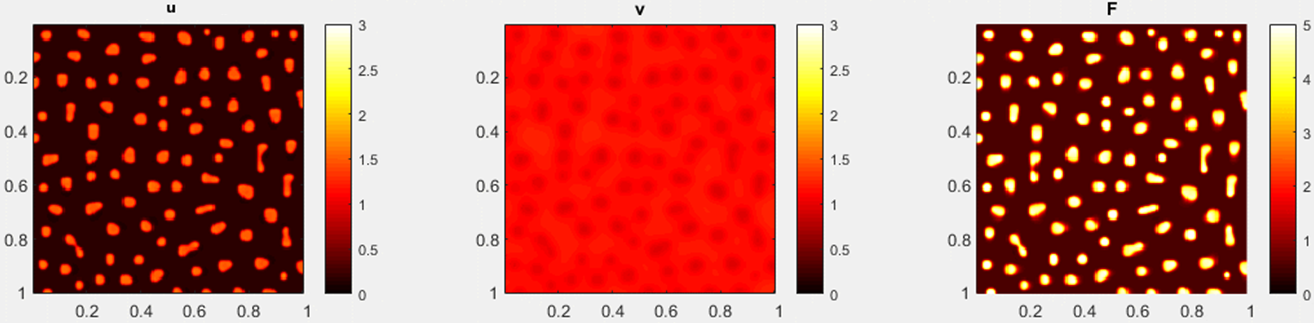}
    \end{subfigure}
    \begin{subfigure}[h]{\textwidth}
        \centering
        \caption{$s=12$}
        \includegraphics[width=\textwidth]{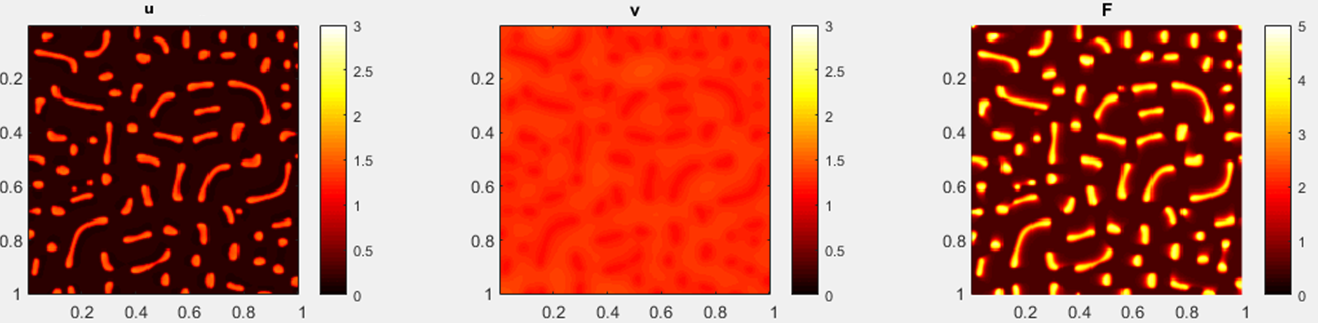}
    \end{subfigure}
    \begin{subfigure}[h]{\textwidth}
        \centering
        \caption{$s=18$}
        \includegraphics[width=\textwidth]{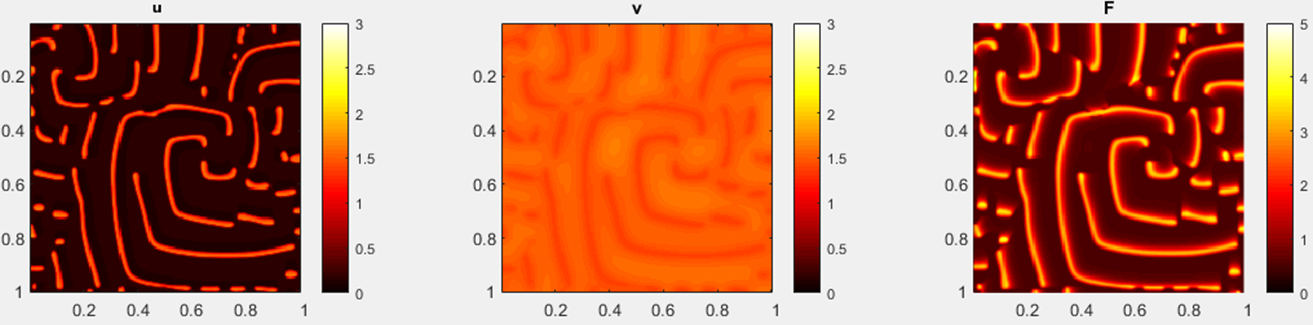}
    \end{subfigure}
    \begin{subfigure}[h]{\textwidth}
        \centering
        \caption{$s=4$}
        \includegraphics[width=\textwidth]{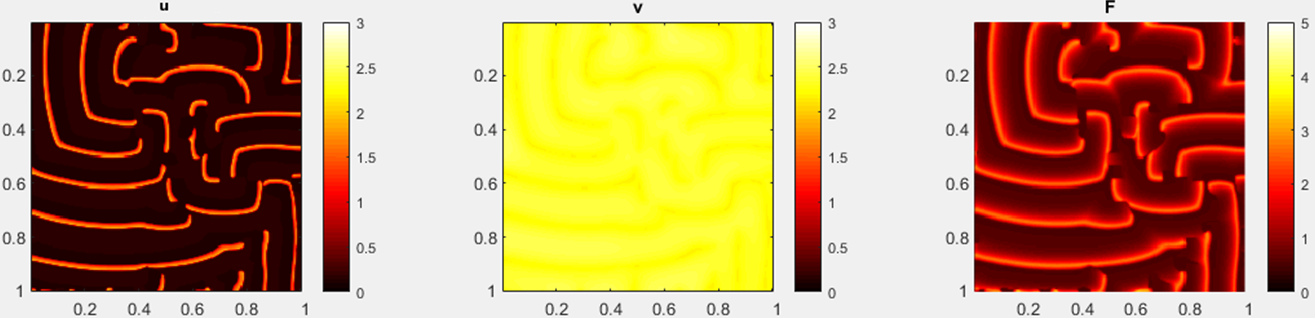}
    \end{subfigure}
	\caption{Simulations of the combined model (CM) in 2D, with parameters from Table~\ref{tab:parameters}(CM2) and HSS + noise initial condition. There is a transition from spots to spiral waves near $s=12$. See movies at \url{https://imgur.com/a/a0u57GQ}.}
	\label{fig:sim_2d_combined}
\end{figure}

We also arrived at the AF model by initializing the CM model at HSS plus global noise and $c=1$. After a pattern starts to form, we gradually decreased $c$ to 0 to arrive at the AF model. (In our hands, this produced more robust results, with patterns that persisted.) 
Fig.~\ref{fig:sim_2d_actin} shows a few snapshot of the simulations. For low $s$, the pattern resembles slowly drifting and deforming blobs. As $s$ increases, the pattern transitions into spiral waves with decreasing width.

%%%%%%%%%%%%%%%
%%%%%%%%%%%%%%%
\subsection{Simulations in a 2D deforming domain}\label{sec:cpm}

As a final set of numerical experiments, we simulate the models in an evolving 2D domain. The boundaries deform in response to the chemical levels close to the boundary. We use the Cellular Potts Model (CPM) for these examples.

Full details of the CPM can be found elsewhere \cite{maree2007cellular} and are briefly summarized in the Appendix. The essential feature of the CPM is its ability to track an evolving shape such as morphology of a motile biological cell \cite{scianna2013cellular}. (In 2D, the cell is ``viewed from above'' as it migrates on a flat 2D surface.) The neighbourhood of each point inside the shape represents a 2D projection of some small cylinder in 3D, containing both membrane and cytosol. Hence, active and inactive GTPases ($u$ and $v$) coexist at every point inside the given shape, as they do in our fixed domain 2D simulations.)

Commonly, for the CPM, a scalar Hamiltonian, analogous to a potential is assumed to depend on the area and perimeter of the cell, as well as  the interface contact with other cells or empty space. Changes to the boundary of the cell are accepted or rejected stochastically, according to the net changes in the Hamiltonian, as described in the Appendix. Our simulations include the following additional features: (1) solving the reaction-diffusion PDEs inside the evolving domain with Neumann boundary conditions at the cell boundaries and (2) modifying the Hamiltonian to depend on the local RD variables.

In real cells, actin polymerizes into  F-actin, and promotes protrusion of a cell edge. Hence, we link the F-actin variable $F$ in the model to forces on the cell boundary, (by superimposing a chemically-dependent potential $H_0=\pm \beta F$ for retractions(+) vs extensions(-) on the basic Hamiltonian, see Appendix). In variants of the model that do not explicitly track F-actin, we assume that the GTPase $u$ plays a similar role (i.e., that $u$, like the GTPase Rac, locally promotes cytoskeleton assembly, creating a protrusive force at the cell edge).

\begin{figure}
    \centering
    \includegraphics[scale=1.2]{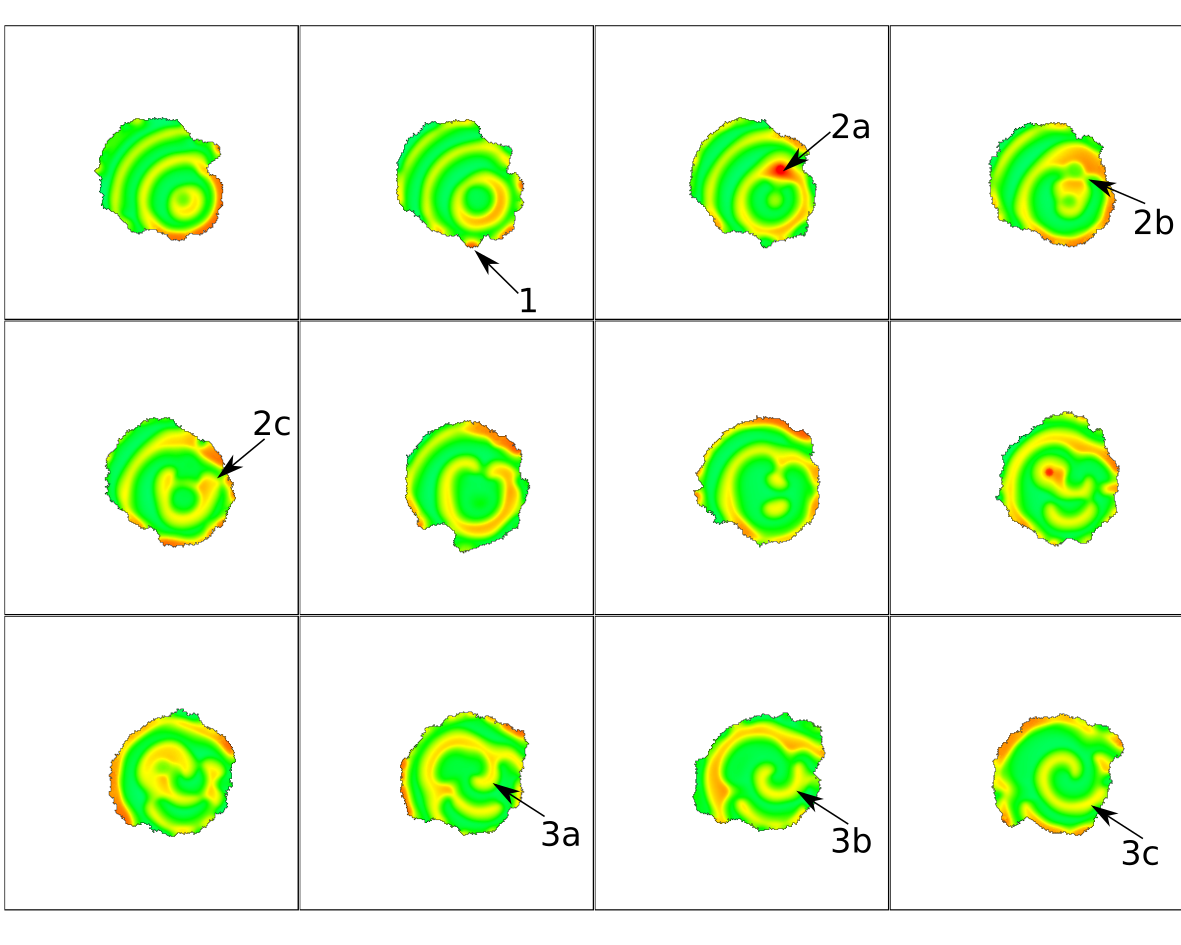}
    \caption{Snapshots of 2D CPM simulation with parameters from the absorbing waves in a static domain. Visualized is F-actin ($F$) that promotes protrusions ($H_0=\pm \beta F$). Arrows indicate examples of interesting dynamics. Snapshots are 20 MCS apart. Parameters are: $D_a = 0.06$, $\eta=15$, $k=6$, $n=3$, $\gamma=30$, $\epsilon=0.1$, $k_s=7.5$, $k_n=24$, $s=30$. CPM Parameters are: $a=12000$, $\lambda_a=2$, $p=500$, $\lambda_p=20$, $J=50$, $r=3,\xi(r)=18$, $\beta=150$, $T=100$. Movie link \url{https://imgur.com/a/7OmgctR}. }
    \label{fig:CPMabsorbing}
\end{figure}

Simulations are initiated with a circular cell and internal variables close to HSS but with a randomly placed peak of active GTPase, $u$ somewhere inside the cell.
Figure \ref{fig:CPMabsorbing} shows a time series of a CPM simulation with parameters that produced the absorbing wave simulations in the static domain. We observe three new types of dynamics (indicated by arrows 1, 2 and 3) resulting from cell movement. An initial burst in the lower right of the cell splits into two waves, moving to the lower right and upper left. The initial burst continues to produce additional waves that split and move towards the cell edge. Waves that impinge on the cell boundary push it outwards. The waves break and smaller protrusions are formed (see Arrow 1).

A new random burst appears (Arrow 2a) and produces waves in two directions (Arrow 2b). When waves collide, they break, amplify, move 
left and right (Arrow 2c) and eventually give rise to a spiral wave. Because the resulting spiral wave has a lower magnitude, there is weaker effect on the boundary at this time. 
% \begin{itemize}
%     \item three interesting dynamics, indicated by arrow 1,2,3
%     \item initial burst in the lower right of the cell makes waves to the lower left and upper right 
%     \item arrow 1 shows that wave hits membrane and pushes it out, by which the wave breaks and smaller protrusions are formed
%     \item arrow 2 shows a random burst appearing (arrow 2a)
%     \item this wave breaks into two waves traveling upwards and downwards (arrow 2b)
%     \item the wave traveling down then hits another wave, which causes it to break into two and amplify (traveling to the left and right)
%     \item arrow 3 shows a spiral wave occurring \mycomment{interesting to note that the waves are absorbed without extending the boundary too much}
% \end{itemize}

\begin{figure}
    \centering
    \includegraphics[scale=1.2]{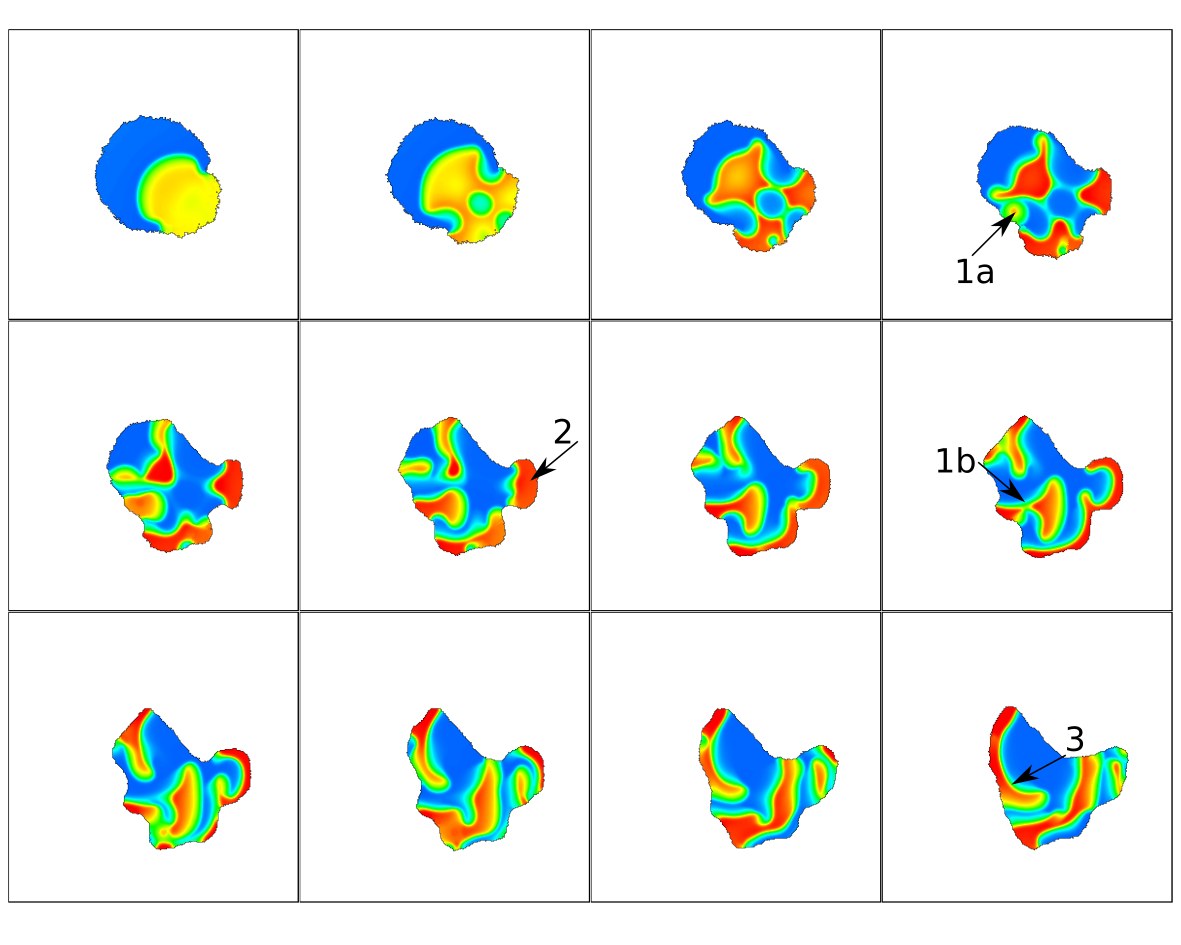}
    \caption{Snapshots of 2D CPM simulation with parameters from the oscillating waves in a static domain. Visualized is F-actin ($F$) that promotes protrusions ($H_0=\pm \beta F$) Arrows indicate examples of interesting dynamics. Snapshots are 20 MCS apart. Parameters are as in Fig. \ref{fig:CPMabsorbing}, but with $k=1.5$, $s=18$. CPM parameters are as in Fig. \ref{fig:CPMabsorbing}, but with $\beta=50$. Movie link \url{https://imgur.com/a/eIAjr59}}
    \label{fig:CPMoscillating}
\end{figure}

In Figure \ref{fig:CPMoscillating}, we show a time series for  parameters that produced oscillating waves in the static domain. As before, the initial burst is in the lower right, and a new burst (Arrow 1a) breaks apart into two waves that broaden. We find a protrusion that is much broader than in Figure \ref{fig:CPMabsorbing} 
(Arrow 2). Wave absorption is lower than in Figure \ref{fig:CPMabsorbing}, so the cell edge is pushed further out.

% \begin{itemize}
%     \item again an initial burst at the lower left of the cell
%     \item arrow 1a shows an additional burst appearing
%     \item arrow 1b shows how this new burst breaks apart into two waves, where the wave towards the inside of the cell flattens out
%     \item arrow 2 shows a fat protrusion (much fatter than in figure \ref{fig:CPMabsorbing}). I think it's fatter because the wave is not being absorbed as much in compared to parameter settings of figure \ref{fig:CPMabsorbing} \mycomment{the waves are also much ``blobier" (better word?) that the previous case, where it's quite thin. This is also observed in the static boundary case}
% \end{itemize}

\begin{figure}
    \centering
    \includegraphics[scale=1.2]{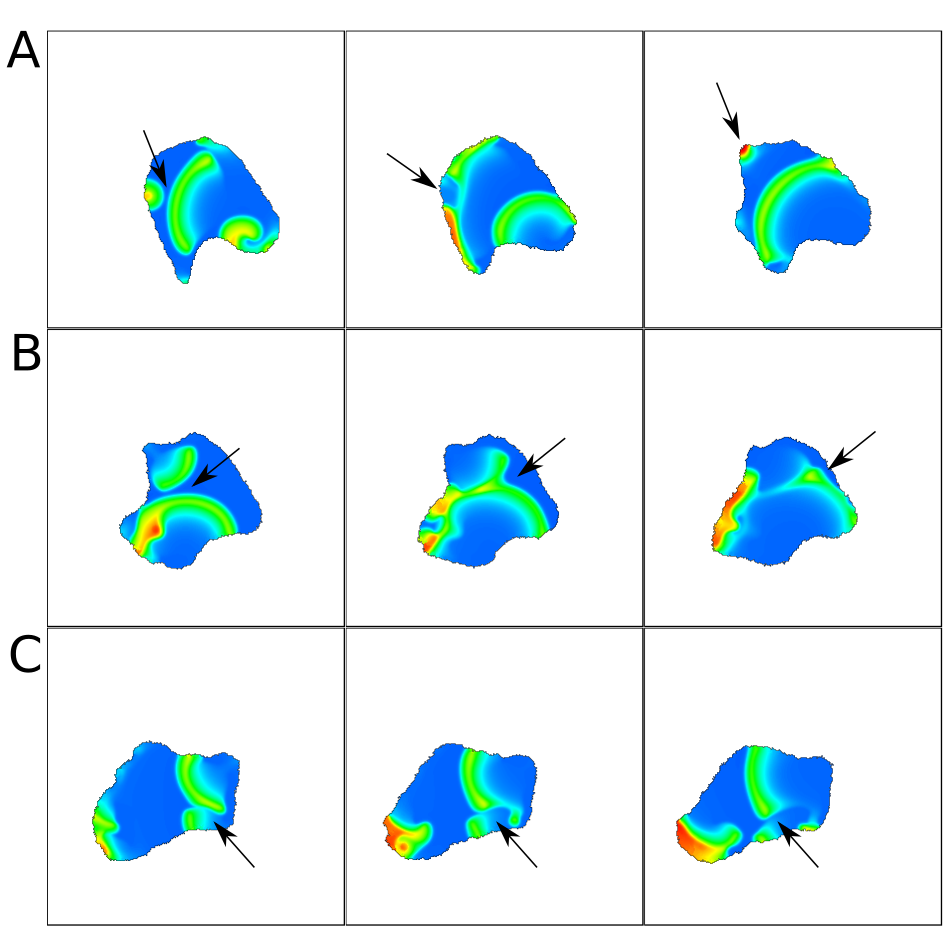}
    \caption{Reflecting wave parameter set. Visualized is F-actin ($F$) that promotes protrusions ($H_0=\pm \beta F$). Snapshots in A,B,C are 10,20,20 MCS apart respectively. Parameters are as in Fig. \ref{fig:CPMabsorbing}, but with $k=1.5$, $s=27$. CPM parameters are as in Fig. \ref{fig:CPMabsorbing}, but with $\beta=50$. Movie link \url{https://imgur.com/a/FDCn3NY}}
    \label{fig:CPMreflecting}
\end{figure}

Figure \ref{fig:CPMreflecting} shows results for parameters corresponding to reflecting waves in a static domain. Here, because the cell boundary moves outwards, the waves are usually absorbed, rather than reflected. Occasionally, if the wave hits the cell edge tangentially, it is reflected (e.g. at 19 sec in the movie, upper left corner). We furthermore observe three new wave dynamics in a moving cell with random bursts. A wave can break apart when it hits a burst (A), waves can merge (B), or avoid each other (C).
% \begin{itemize}
%     \item usually waves are not reflecting, because the CPM cell is moving out \mycomment{can talk about this case shows how the moving boundary really does change the pattern}
%     \item they can reflect when they approach the boundary tangentially and curve back \mycomment{an example is at 19 sec into the movie, upper left corner}
%     \item Fig.A shows a wave breaking apart when it hits a burst
%     \item Fig.B shows to waves merging
%     \item Fig.C shows to waves avoiding each other
% \end{itemize}

\begin{figure}
    \centering
    \includegraphics[scale=1.2]{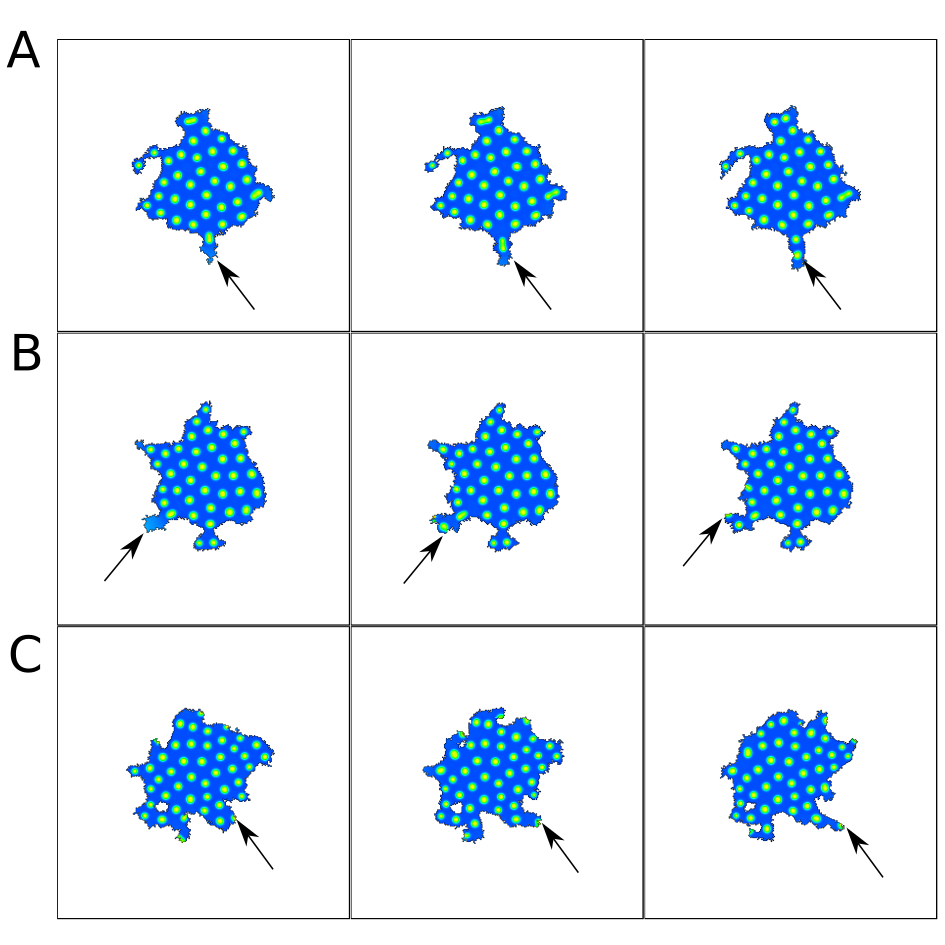}
    \caption{Spots. Visualized is Rac ($u$) that promotes protrusions ($H_0=\pm \beta u$). Snapshots in A,B,C are 10,5,25 MCS apart respectively. Parameters are: $D_a = 0.1$, $\eta=60$, $k=6$, $n=3$, $\gamma=30$, $c=1$, $\theta=18$, $\alpha=6$. CPM Parameters are: $a=10000$, $\lambda_a=0.02$, $p=1000$, $\lambda_p=0.04$, $J=40$, $r=3,\xi(r)=18$, $\beta=200$, $T=20$. Movie link \url{https://imgur.com/dADHOnS}.}
    \label{fig:CPMspots}
\end{figure}

As a last experiment, we simulate the formation of spots in the NC model (Figure \ref{fig:CPMspots}). Since this model has no F-actin variable, we base the edge protrusion on the active GTPase $u$ (assumed to act like Rac in promoting local cytoskeleton assembly). The spots are highly dynamic and, as expected, lead to the formation of small protrusions (``filopodia'') (C). Furthermore, edge deformation also causes the spot pattern to change. When a protrusion forms (stochastically or by locally elevated $u$), the spot in a region close to the protrusion can split into two, one of which moves into the protrusion (A). We also see formation of new spots inside protrusions (B).

% \begin{itemize}
%     \item Fig.A shows a protrusion forming and a spots breaking into two and the spot traveling into the protrusion
%     \item Fig.B shows new spots being made a the membrane
%     \item Fig.C shows a spot leading a protrusion \mycomment{interesting to see the thin, long protrusions. Also a spot at the boundary will quickly carve out a protrusion, so in a metastable state all spots are in the interior}
% \end{itemize}

\section{Discussion}

In summary we have explored extensions of the wave-pinning model (WP) \cite{mori2008wave,mori2011asymptotic} that coupled the non-conservative variant proposed by \cite{Champneys} (NC) and the actin feedback (AF) model of \cite{holmes2012modelling}. We found that the combined model (CM)  borrows features from both, with moving peaks and wave trains, as well as more complex hybrid dynamics. At the same time, we were unable to find blinking localized spots as observed experimentally in \cite{robin2016excitable}. Despite the fact that the work of \cite{robin2016excitable} points to interactions of F-actin with the GTPase Rho, other unknown factors, missing in our model, should be considered to explain such behavior. 

We used the local perturbation analysis (LPA) on each model variant.
As noted before \cite{holmes2015_lpa}, LPA recovers Turing analysis. States that are LPA stable are also Turing stable. LPA helps to identify potentially interesting parameter ranges, including polarizable regimes that cannot be detected by Turing analysis. 
At the same time, LPA does not predict details of patterns that emerge, nor accurate bifurcation points in the full PDE system. We also encountered examples where
 LPA identified apparent bifurcations that did not materialize as true regimes of behavior in the PDEs.
 
%    \item without source/sink terms, the actin wave patterns are not very robust (at least in 2D)

As a second innovation, we simulated all model variants in 2D on both a static and a deforming domain. Previous work \cite{Champneys,holmes2012regimes} was concerned with fixed 1D domains for the PDEs. We hence showed that the patterns for the nonconservative (NC) model were primarily spots, not bands, whereas the AF model, while appearing to be less robust, produced a variety of moving peaks, bands, and waves, including spiral waves. 

Finally, our simulations of the models in a deforming domain mimicking a motile cell allowed us to consider the connection to experimentally observed waves of actin \cite{inagaki2017}. We showed that the internal dynamics of the models (and in particular the actin feedback model) have an interesting consequence on the motility of a ``model cell''. Indeed, the waves of high and low signaling levels led to formation of cellular protrusions, and gave rise to nontrivial motion in the deforming cell.
 
 \bigskip
 \noindent{\bf Acknowledgements: } LEK is funded by a Natural Sciences and Engineering Research Council of Canada (NSERC) Discovery Grant. YL was also supported by an NSERC postgraduate Fellowship. We are grateful to Zachary Pellegrin for his development of a CPM reaction-diffusion solver. We thank members of the Feng-Keshet groups for feedback.

\section{Appendix: LPA diagram for the NC model} \label{apd:NC_lpa}

\begin{figure}[!hp]
    \centering
    \includegraphics[width=0.7\textwidth]{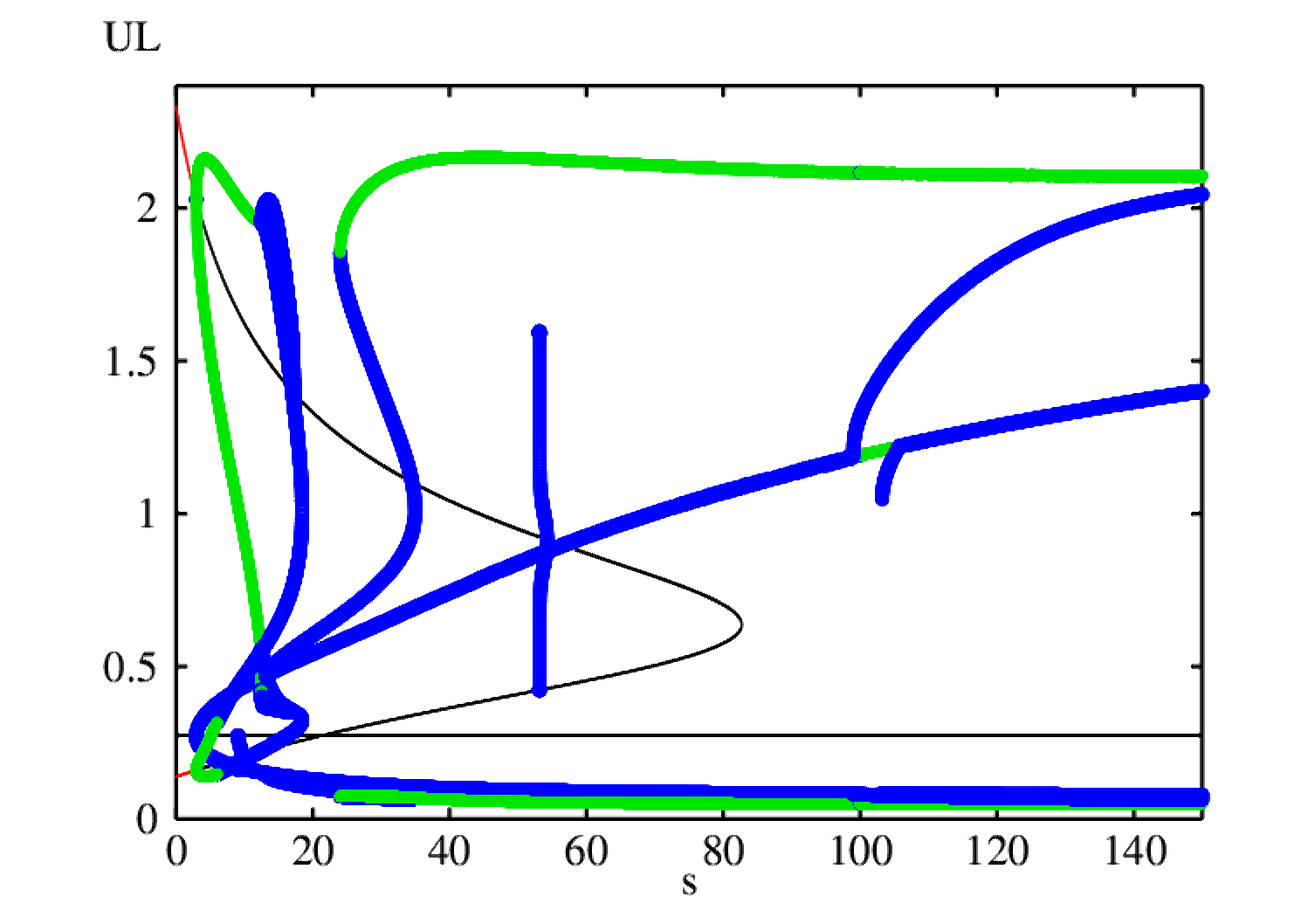}
    \caption[Bifurcation diagram of the combined LPA system]{LPA bifurcation diagram for the combined model (CM) with respect to the parameter $s$. Other parameters as in Table~\ref{tab:parameters}(CM2). There are many apparent branches of periodic solutions. In a parameter range around $s=20$, there are no stable equilibria nor stable periodic solutions even though the system remains bounded, which suggests the presence of chaos.}
    \label{fig:lpa_combined}
\end{figure}

The LPA diagram for the nonconservative model (NC) is given in Figure~\ref{fig:lpa_combined}. Due to the many intertwined bifurcations, it is difficult to interpret this diagram, and we present it here only to demonstrate the limitations of LPA.

%\mycomment{YL: will write a bit here soon}

\section{Appendix: Relation between Turing and LPA} \label{apd:turing_vs_lpa}

Linear stability analysis (i.e. Turing analysis) \cite{turing1952chemical} is a more traditional method of analyzing the stability of reaction-diffusion systems which focuses on the stability of a HSS in response to perturbations in the form of a global noise with infinitesimal height.

Here we examine the relation between Turing and LPA. This was previously done by \cite{mata2013model} for the two-variable case in terms of eigenvalues, and for the general case by Theorem~4.1 of \cite{holmes2014efficient}. (The proof for this theorem is quite involved.) Here we present a much more elementary argument for the two-variable case.

Suppose in a general reaction-diffusion PDE with a slow-diffusing quantity $u$ and a fast-diffusing quantity $v$. Assume that time has been rescaled so that the diffusion coefficient of the fast quantity is $1$. 
\begin{align*}
\pd{u}{t} &= \delta \nabla^2 u +f(u,v),\\
\pd{v}{t} &=  \nabla^2 v +g(u,v) .
\end{align*}
The corresponding well-mixed (WM) system is
\begin{align*}
\pd{u}{t} &= f(u,v),\\
\pd{v}{t} &= g(u,v) .
\end{align*}
The LPA system is the above plus an additional equation for the local variable:
\begin{align*}
\pd{u_L}{t} &= f(u_L,v) .
\end{align*}
Suppose the PDE system has a homogeneous steady state (HSS) $(u_*,v_*)$.
The Jacobian of the well-mixed system and LPA system at the corresponding equilibrium is given by
\[J_{WM} = \begin{bmatrix}
\partial_u f & \partial_v f \\
\partial_u g & \partial_v g 
\end{bmatrix}, \quad J_{LPA} = \begin{bmatrix}
\partial_u f & \partial_v f & 0\\
\partial_u g & \partial_v g & 0 \\
0 & \partial_v g & \partial_u f \\
\end{bmatrix} = \begin{bmatrix}
J_{WM} & 0 \\
* & \partial_u f
\end{bmatrix} ,\]
where the partial derivatives are understood to be evaluated at the HSS, and $*$ denote entries that are unimportant for later analysis.
Notice that eigenvalues of $J_{LPA}$ are the two eigenvalues of $J_{WM}$, plus $\partial_u f$. This is due to $J_{LPA}$ being block-lower-triangular.
We will show that saying the HSS is Turing-unstable in the limit $\delta \to 0$ is equivalent to saying it is a stable equilibrium for WM but unstable for LPA.

Define the relevant matrices for Turing analysis:
\[M(q^2) = J_{WM} - Dq^2, \quad D = \begin{bmatrix}
\delta & 0 \\
0 & 1
\end{bmatrix} \xrightarrow{\delta \to 0} \begin{bmatrix}
0 & 0 \\
0 & 1
\end{bmatrix}\]
Suppose the HSS satisfy the condition for Turing instability, which has three conditions (see, for example, \cite[Ch.~11.4]{leahsbook}):
\begin{subequations}
\begin{align}
\Tr(J_{WM}) &<0 , \label{eqn:turingcond1}\\
\det(J_{WM}) &>0 , \label{eqn:turingcond2}\\
\det(M(q^2)) &<0 \text{ \ for some \ } q^2>0 \label{eqn:turingcond3} \,.
\end{align}
\end{subequations}
Conditions \eqref{eqn:turingcond1}, \eqref{eqn:turingcond2} are equivalent to saying that the HSS is a stable equilibrium for WM. 
Next, in the limit of $\delta \to 0$, we compute
\[\det(M) = \det(J_{WM} - Dq^2) = \det\lb J_{WM} - \begin{bmatrix}
0 & 0 \\
0 & q^2
\end{bmatrix} \rb = \det(J_{WM}) - q^2 \partial_u f\]
Notice that by setting $\delta = 0$, this equation is linear in $q^2$ instead of quadratic. This means \eqref{eqn:turingcond3} is equivalent to $\partial_u f >0$ .
But since $\partial_u f$ is an eigenvalue for $J_{LPA}$, this means the HSS is unstable in the LPA system.

Conversely, suppose that the HSS is stable in WM and unstable in LPA. This means the eigenvalues of $J_{WM}$ all have negative real part, and $\partial_u f >0$, which as shown above is equivalent to Turing-unstable.

\begin{figure}[!hp]
    \centering
    \begin{subfigure}[h]{0.5\textwidth}
        \centering
        \caption{LPA}
        \includegraphics[width=\textwidth]{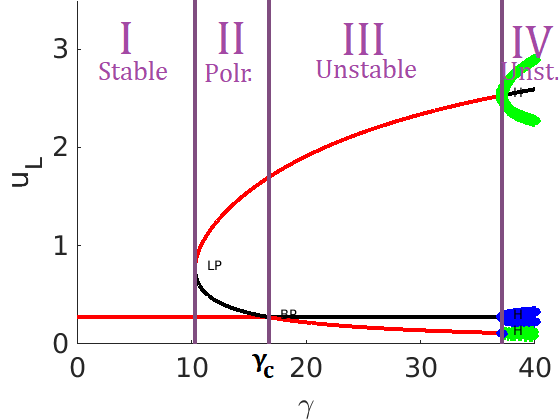}
    \end{subfigure}~
    \begin{subfigure}[h]{0.5\textwidth}
        \centering
        \caption{Turing}
        \includegraphics[width=\textwidth]{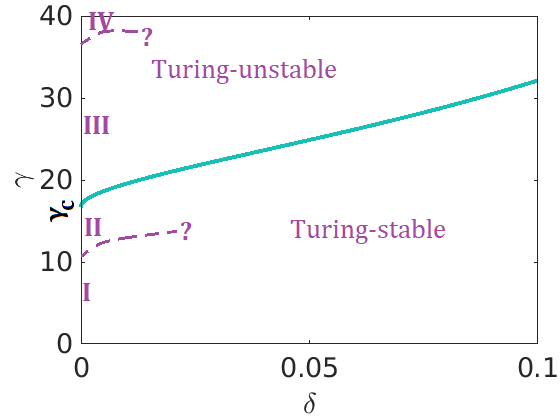}
    \end{subfigure}
	\caption[Comparison of LPA and Turing regimes for the non-conservative model]{Comparison of LPA and Turing bifurcation diagrams for the non-conservative model. (a) is a zoom of the LPA diagram from Fig.~\ref{fig:lpa_champ}(b). (b) is the Turing bifurcation diagram reproduced from Fig.~5 of \cite{Champneys}, using the same parameters. Observe that both the LPA-stable (I) and the LPA-polarizable (II) regimes in (a) located to the left of $\gamma_c=16.765$ correspond to the Turing-stable regime below the blue curve in (b). The LPA-unstable regimes (III, IV) correspond to the Turing-unstable regime above the curve. The curve passes through $\delta = 0, \gamma=\gamma_c$. The bifurcation boundary between Regimes I and II, and between III and IV cannot be detected by Turing analysis. Given that numerical simulations have shown that the PDE produces the same behavior (Fig.~\ref{fig:sim_champ2}(a)) in both Regimes III and IV, it is possible that these are not distinct regimes for the PDE. Overall, the LPA diagram (a) can be seen as a vertical slice of the Turing diagram (b) at $\delta=0$, with additional bifurcation boundaries that separates the LPA-stable and LPA-polarizable regimes.}
	\label{fig:lpa_turing_comparison}
\end{figure}

Compared to Turing analysis, LPA has several advantages and disadvantages. 
LPA  is essentially a ``zeroth-order" expansion in $\delta$, so it is only valid in the limit of $\delta \to 0$ and offers no information on the effect of $\delta>0$, as opposed to Turing analysis. However, in this limit, LPA contains the Turing stability of the system. In particular, LPA-unstable is the same as Turing-unstable, whereas LPA-polarizable and LPA-stable regimes are Turing-stable. This follows the analysis above, and also \cite{mata2013model,holmes2014efficient}. This correspondence is illustrated in Fig.~\ref{fig:lpa_turing_comparison}, where we show how the LPA regimes from Fig.~\ref{tab:lpa_champ_regimes}(b) lines up with Turing regimes.

\begin{figure}
    \centering
    \includegraphics[width=0.7\textwidth]{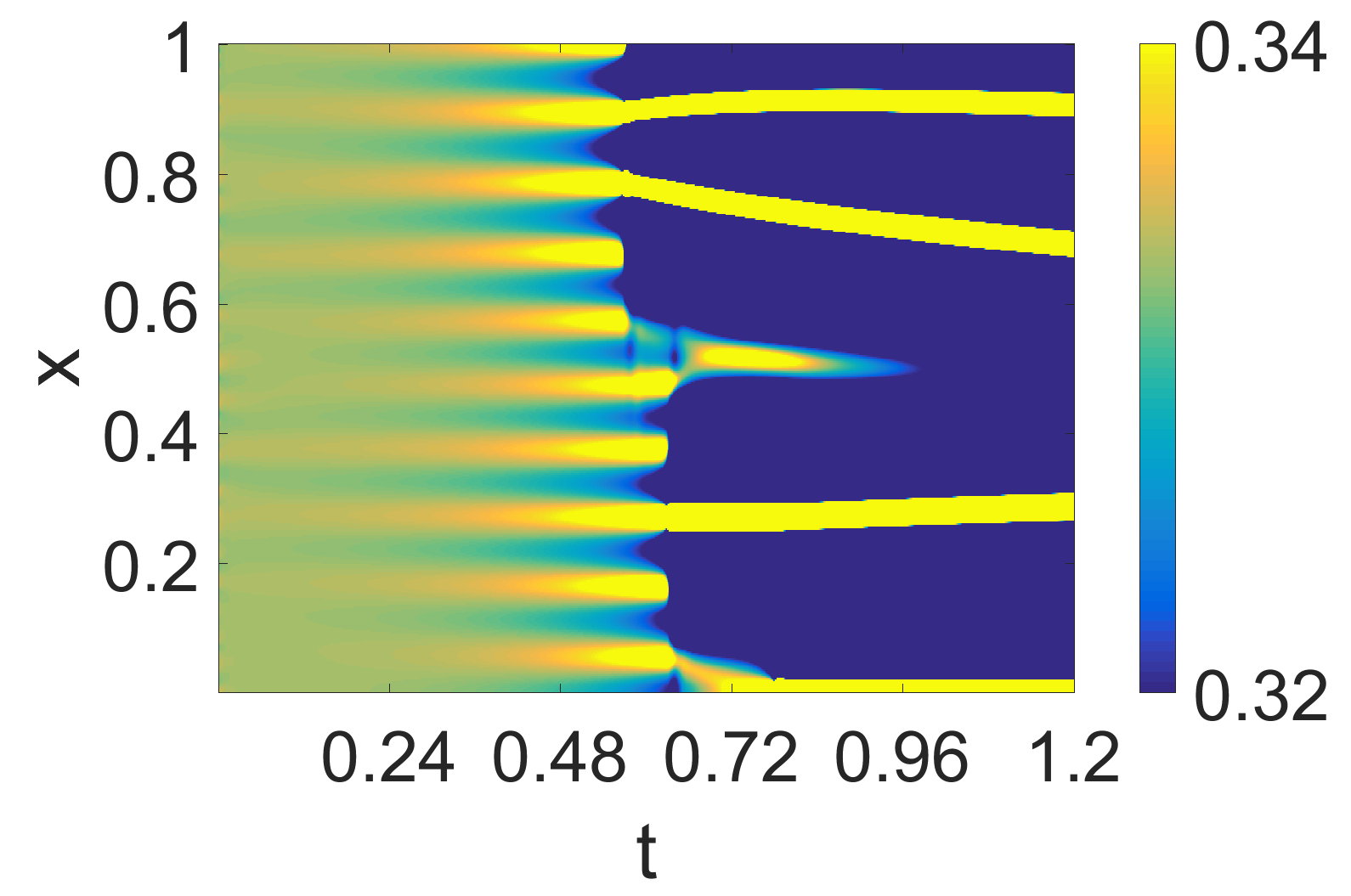}
    \caption{Simulations for the non-conservative model with random initial conditions $u=u_* (1+\mbox{Unif}(-0.01,+0.01)), v=v_*$ with default parameters from Table~\ref{tab:parameters}(NC) except $\gamma=25$. The color range is chosen so that the precursor pattern is more visible. The rapid transition from the shallower, higher frequency precursor pattern to the final pattern can be clearly seen.}
    \label{fig:turing_zoom}
\end{figure}

In pattern-forming regimes, LPA does not help predict the exact form of the pattern. This is most relevant to the actin feedback model, when there are many different possible patterns and a large number of parameter regimes.
Turing analysis also cannot predict the final pattern, but it does allow us to predict the initial precursor pattern that forms and exists only for small $t$. In Fig.~\ref{fig:turing_zoom}, we simulated the non-conservative model with parameters chosen such that only a single wave number is unstable. This results in a periodic, shallow precursor pattern which has the exact frequency as the unstable wave number. 

As the system continue to evolve, once a peak of the precursor pattern reaches a certain amplitude, it very rapidly grow to the full size of the final pattern while suppressing nearby peaks. Other precursor peaks farther away from the grown one might survive longer and eventually transition to full size, or be suppressed by another nearby peak which has transitioned sooner. These non-linear interactions cannot be captured by Turing analysis. 

In the special case that a static, periodic pattern forms, such as in the non-conservative model, the ``minimal patch size" idea from \cite{painter2011spatio}, which is based on Turing analysis, can give an upper bound on the number of periods the pattern can have. 

\section{Appendix: Methods}
Bifurcation plots were produced with XPPAUT \cite{xppaut,auto} and Matcont \cite{matcont}. PDEs in 1D were solved using finite difference methods with Crank–Nicolson time stepping in Matlab with $\Delta x= 0.005, \Delta t= 0.0002$. Plots were produced with Matlab. PDEs in fixed 2D domain were solved using the FEniCS \cite{fenics} package in Python, plots and movies produced by Paraview. The codes for both are published at \url{https://github.com/liuyue002/Wave-pinning-model}.
%\mycomment{YL: I guess we should add in a bit for CPM?}

\section{Appendix: Cellular Potts Model simulations}

In the CPM, a biological cell is represented by a set of contiguous
lattice site in 2D (or 3D) all assigned an index $\sigma$. (For a single cell, the index is 1 and the surrounding medium is given an index of 0.) Here we focused on a 2D CPM model cell, representing a top-down view of a ``biological cell''  attached to a flat surface. We use the classic Hamiltonian 
\begin{equation}
H=\lambda_a(A-a)^2 +\lambda_p (P-p)^2 + J P. 
\label{eq:Hamiltonian} 
\end{equation}
Here the three terms represent the energetic cost for change of area $A$ (cell contraction/expansion) away from the preferred ``rest area'' $a$, a cost for elongation or shortening of the cell perimeter $P$ away from a preferred  ``rest perimeter'' $p$, and a term that describes an interfacial energy associated with the cell-medium interface. The weighting factors, $\lambda_a, \lambda_p, J$ are adjusted to set the relative importance of the various energy terms. 

The perimeter $P$ is approximated as in \cite{magno_cpm}. For each cell site, we calculate the number of lattice sites within a certain radius $r$ (here $r=3$) in contact with the medium 
%(this works for a single cell simulation, for a multiple cell simulation it's slightly different).
Then we take the sum over all boundary sites and rescale by $\xi(r)$ (here $\xi(r)=18$)  to obtain a perimeter approximation:

\begin{equation}
P = \frac{1}{\xi(r)} \sum_{x : \sigma(x)=1} \; \; \; \sum_{y : \lbrace |x-y|^2<r^2 \wedge \sigma(y)=0 \rbrace } 1
\end{equation}

At each Monte Carlo Step (MCS) in the simulation, points along the cell edge may protrude or retract with some probability. Formally, this is achieved by
$N$ so-called ``copy attempts''. A copy attempt consists of selecting a random site (``source'') on the lattice and copying the index into a random neighbouring site (``target'') from its Moore neighbourhood. The change/copy is accepted with probability 
\begin{equation}
P(\Delta H)= 
\begin{cases}
1 & \textrm{if}\;\Delta H + H_0 <0, \\
e^{-(\Delta H+ H_0)/T} & \textrm{if}\;\Delta H + H_0\geq 0.
\end{cases}
\label{eq:boltzmann}
\end{equation}
Here $T\geq 0$ is denoted  a cellular ``temperature'' and sets the intensity of random edge fluctuations. $H_0$ is a yield energy (force) to be overcome.

We start with a circular cell (12000 pixels). Assigning a 
nondimensional size $\Delta x = 0.005$ to each pixel implies that cell area is 0.3. We rescaled the parameters $\eta,k,\gamma,\alpha,\theta,k_n,k_s$ with $L^2$, where $L=10$.  CPM parameters $a, p$ are in terms of pixels (see figure captions).  

The initial conditions are $v=2.5, u=0$ and $F=0$ everywhere. We introduce a randomly placed circular spot (radius 3 pixels) of $u=5$ to represent an initial random burst of active GTPase. Then every 100 MCS we add another randomly placed spot of elevated GTPase $u += 15$ (higher than the initial burst to prevent decay) to depict stochastic bursts of GTPase activation in the cell.

For the model without F-actin, we have slightly different initial GTPase fields: $v = 1.1113, u=0.3333$, except in the upper left corner of the cell, with an area of about 1/9 of the cell, we set $u = 0.5454$.

After every MCS we solve the RD equations for 0.001s, using $dt=1e-6$, so 1000 iterations per MCS. Within a MCS, after every accepted membrane extension or retraction we update the GTPase fields as follows. After an extension we set $u$(target)=$u$(source) and subsequently rescale the level of $u$ throughout the cell to conserve mass. After retraction we set u(target)=0 and then u(target) is equally distributed throughout the whole cell. The same operations are carried out for $v$. For F-actin, we do the same but do not redistribute, i.e. $F$(target)=$F$(source) for extensions and $F$(target)=0 for retractions.

%\printbibliography
\bibliographystyle{spbasic}
\bibliography{ref}

\end{document}